\documentclass[]{JFM-FLM_Au}

\usepackage{graphicx}
\usepackage[dvipsnames]{xcolor}
\usepackage{epstopdf, epsfig}
\usepackage{color,soul}
\usepackage{physics}
\usepackage{hyperref}
\usepackage{subcaption}
\usepackage{pdflscape}
\usepackage{amssymb}
\usepackage{changes}
\usepackage{amsmath}

\lefttitle{X.-B. Li, Y.-F. Chen, C.-Y. Wen, and P.-X. Guo}
\righttitle{Journal of Fluid Mechanics}

\title{Nonlinear dynamics involving multiple modes in high-speed transitional boundary layer}

\author{Xiao-Bai Li\aff{1}, Yifeng Chen\aff{1}, Chihyung Wen\aff{1} \and Peixu Guo\aff{1}}

\affiliation{\aff{1}Department of Aeronautical and Aviation Engineering, The Hong Kong Polytechnic University, Kowloon, Hong Kong SAR, China}

\corresau{Peixu Guo, \email{peixu.guo@polyu.edu.hk}}

\begin{document}
\maketitle

\begin{abstract}
Extensive studies have investigated the transition mechanism of boundary layers initiated by a single primary instability, which can generate the secondary instability. In a real-world scenario, however, multiple primary instabilities of different physical nature would coexist and generate more complicated stages of mode--mode interactions. For this scenario, conventional secondary stability analysis may not be applicable. In this work, a general framework is established to decompose the input--output system and to quantify the transfer of energy involving various modes. The linearized governing equation with nonlinear forcings is applied in a Mach 6 boundary layer, where two different types of primary instabilities are added simultaneously. As the primary-wave amplitudes increase to certain threshold, the nonlinear effect gives rises to triadic forcings acting on multiple spatio/temporal scales. This causes the saturation of the second mode and secondary growth of the first mode. In the generation stage of each higher-order mode, a specific leading triadic forcing term can be identified. These higher-order waves manifest solely in response to the identified dominant forcing during their generation. At the moderate and late transitional stages, the forcings are, however, not equally transferred to the response via the resolvent operator. In other words, the base-flow-associated resolvent operator exerts different levels of `leverage' to transfer different forcings to responses. The nonlinear energy transfer via triadic forcings also drives the higher-order instability to inherit physical signature from the associated lower-order instability. Another noteworthy observation is that the acoustic signature of the Mack second mode attenuates due to the end of linear energy support, and later revives as a consequence of nonlinear re-supply. The re-supply comes from primary, secondary and tertiary modes, is active away from the wall and is not detectable by experimental wall sensors. Finally, the interplay between secondary/tertiary waves and primary waves occurs notably earlier then one may expect, namely before transition onset or in the early transitional region. This differs from the traditional secondary instability analysis that a large-amplitude primary wave is developed first to perform the bi-global analysis in the distorted base flow.\par
\end{abstract}


%{\bf MSC Codes }  {\it(Optional)} Please enter your MSC Codes here

\section{Introduction}\label{sec:intro}
Laminar-turbulent transition in boundary layer flow has attracted growing interest for decades, given its importance in both fundamental physics and engineering applications. In the natural transition with a low noise level, primary instabilities can be triggered to undergo the linear modal or non-modal growth \citep{morkovin_1994_transition}. As the instability wave increases to certain amplitudes, secondary instabilities can be formed due to nonlinear interactions \citep{herbert_1988_secondary}, which eventually leads to turbulent breakdown. Therefore, primary waves are key factors to determine the downstream transition path and breakdown mechanisms. In the low-speed flow, the viscous Tollmien–Schlichting (T--S) wave usually triggers the transition to turbulence. As the Mach number ($\mbox{\textit{Ma}}$) grows, the oblique first mode emerges and dominates, which is the supersonic counterpart of the T--S wave. Further, the inviscid Mack modes with acoustic signature become unstable \citep{mack_1984_boundary,mack_1990_inviscid},  where the two-dimensional (2-D) Mack second mode usually possesses the highest growth rate at $\mbox{\textit{Ma}}>4$. The possible existence of different primary waves  may therefore lead to different transition mechanisms in hypersonic states, i.e. $\mbox{\textit{Ma}}>5$.\par
Due to the prominent linear amplification rate, the Mack second mode with hundreds of kilohertz is usually assumed, or found to be the dominant primary wave in the controlled hypersonic transitional flow. The breakdown scenarios associated with this mode have been extensively investigated. Representative experimental \citep{kennedy_2022_characterization} and numerical works \citep{sivasubramanian_2015_direct,hader_2019_direct,unnikrishnan_2020_linear} reported that the second-mode fundamental breakdown can be dominant. Apart from the fundamental breakdown, subharmonic breakdown \citep{bountin_2008_evolution,hader_2019_direct,unnikrishnan_2020_linear} and oblique breakdown \citep{zhou_2022_direct,franko_2013_breakdown} of the second mode wave have also been reported as lesser mechanisms. Despite the dominant role of the second mode in the linear stage, the oblique first mode can be, however, characterized by an evidently higher receptivity coefficient within a wide range of forcing frequency for different configurations \citep{zhong_2001_leading,ba_2022_hypersonic,niu_2023_receptivity}. The considerably energetic low-frequency band in wind-tunnel incoming disturbances \citep{schneider_2008_development,duan_2019_characterization,guo_2025_transitional}, the effect of wall thermal conditions \citep{zhu_2022_transitional} and the nonlinear difference interaction between
high-frequency side bands \citep{craig_2019_nonlinear,Zhang_2020_investigation} further elevate the significance of low-frequency instabilities with tens of kilohertz. The transition scenario dominated by the low-frequency first mode has been reported in \cite{mayer_2011_direct}, \cite{franko_2013_breakdown} and \citet{guo_2022_heat} as the oblique breakdown mechanism.\par
The aforementioned breakdown scenarios merely involve nonlinear interactions between the primary wave itself and its tonal components. However, in a more realistic situation where the incoming disturbance exhibits broadband nature, multiple primary waves can co-exist and interact with each other. New transition mechanisms and energy transfer pathways are thus possible. The interactions between the low- and high-frequency waves in high-speed boundary layers were frequently observed in wind-tunnel experiments \citep{bountin_2008_evolution,zhu_2016_transition,craig_2019_nonlinear,Zhang_2022_nonlinear} and investigated by nonlinear stability analysis \citep{chen_2017_interactions}. It has to be noted that in these references, there is still only one dominant primary instability, e.g., the second mode. Given the above factors that augment the low-frequency instabilities, the first mode can be as important as the second mode to trigger the transition. In this case, multiple resonance mechanisms has been identified \citep{guo_2023_interaction}, in particular the combination resonance that generates the detuned mode was detected. The detuning phenomenon during the transition was also reported or remarked by experiments \citep{bountin_2008_evolution,craig_2019_nonlinear}. With the presence of multiple primary instabilities, the conventional secondary stability analysis \citep{herbert_1988_secondary}, such as the bi-global analysis and the Floquet analysis, tends to become invalid. There is a need to construct a framework to describe how primary and  high-order instabilities grow or decay subject to multiple primary instabilities.\par
Apart from the methodology for more complicated transitional flows, some open questions also remain on the physical level. First of all, what are the specific contributions from nonlinear forcings to the response pattern in the transitional flow, and how do they interact with the mean flow to sustain the growth and re-distribution of energy among multiple instability waves? This question is relevant to how spectral broadening is achieved in arbitrary transitional flows. Second, what are the physical natures of the higher-order modes? It has been recognized that first and second modes are of vortical and acoustic nature, respectively, while not for higher-order instabilities. Third, how does the second mode lose or regain its acoustic signature in the complete transition? To this end, the combined effects of the linear and nonlinear mechanisms should be considered. Here, `linear' and `nonlinear' mechanisms refer to the growth contributed by the interaction with the mean flow only and with the other modes, respectively. The mean flow denotes the laminar base flow plus the time- and spatial-averaged distortion. Acting as a quantifier, an energy transfer equation can be derived from the linearized equation governing with the nonlinear term kept. As shown in \citet{ding_2025_mode} and \citet{huang_2025_spatio}, transfer of kinetic energy provides useful insights in incompressible turbulent channel flows. In the field of hypersonic transitional flow, there exist different energy components, including acoustic, vortical and entropic ones. A more complete framework is required, not limited to the kinetic energy.\par
The remainder of the paper is summarized as follows. The problem description, the optimal forcings that produce the two co-existing primary waves, and the triadic resonance state are introduced in $\S$ \ref{sec:BLsimulation}. The analysis techniques, including the input--output formulation and the spectral energy equation are shown in $\S$ \ref{sec:method}. In $\S$ \ref{sec:development}, the direct numerical simulation (DNS) results are shown to describe the development of Fourier modes and spectra broadening into higher-order waves. Contributions from triadic interactions to the forcings as well as the DNS modes are quantified and compared in $\S$ \ref{sec:IO}. In $\S$ \ref{sec:transfer}, the linear and nonlinear energy transfer mechanisms associated with the developments of multiple waves are analyzed and discussed. The main findings and conclusions are summarized in $\S$ \ref{sec:conclusion}.\par

\section{The examined transitional boundary layer}\label{sec:BLsimulation}
\subsection{Flow configuration and direct numerical simulation}\label{sec:dns}
Provided that the tool to describe the energy re-distribution has been available, an example of transitional flow is needed for the application. We consider a flat-plate boundary layer subject to the wind-tunnel-experiment freestream condition of \citet{bountin_2013_stabilization}: Mach number $\mbox{\textit{Ma}}_\infty=6$, static temperature $T_\infty^\ast=43.18\ {\rm{K}}$, and the unit Reynolds number $\Rey_{1\infty}^\ast=1.05\times10^7\ {\rm{m}}^{-1}$ (the superscript '$\ast$' denotes dimensional variable). The flat plate model extends $0.6\ {\rm{m}}$ along the streamwise direction, with a constant room temperature $T_{\mathit{w}}^\ast=293\ {\rm{K}}$ assumed for the wall. A Cartesian coordinate system is constructed, with its origin located at the leading edge of the flat plate and the $x$-, $y$- and $z$-axes along the streamwise, wall-normal, and spanwise directions, respectively. Hereafter, all physical quantities are  nondimensionalized by the corresponding freestream values, except for the pressure, which is normalized by $\rho_{\infty}^{\ast}{u_{\infty}^{\ast}}^2$. The reference length scale is $L_{\mathit{r}}^{\ast}=1\ {\rm{mm}}$. \par
The flow is governed by the compressible Navier--Stokes equation, which can be written in the conservative form:\par
\begin{equation}\label{NS}
  \frac{{\partial}\boldsymbol{Q}}{{\partial}t}+\frac{{\partial}\boldsymbol{F}}{{\partial}x}+\frac{{\partial}\boldsymbol{G}}{{\partial}y}+\frac{{\partial}\boldsymbol{H}}{{\partial}z}=\mathsfbi{B}\boldsymbol{f},
\end{equation}
where $\boldsymbol{Q}=[\rho,{\rho}u,{\rho}v,{\rho}w,{\rho}e]^{\rm{T}}$ is the vector of conservative variables (the superscript `T' represent the scalar transpose), and $\boldsymbol{F}$, $\boldsymbol{G}$, and $\boldsymbol{H}$ are the vectors of the inviscid and viscous fluxes. The symbols $\rho$ and $e$ represent density and total energy per unit mass, respectively, and $u,v,w$ are the respective velocities in $x,y,z$ directions. The detailed expression of the governing equation can be found in \citet{anderson_2002_computational}. In equation (\ref{NS}), the external forcing term $\boldsymbol{f}$ is introduced to initiate the transition, which is constrained to certain spatial extent by the operator $\mathsfbi{B}$. In the present paper, the forcing is applied to the entire wall-normal profile at only $x_0=40$, with its specific form determined by a linear resolvent analysis in $\S$ \ref{sec:IOforcing}.\par
A perfect gas model for air is applied with a constant specific heat ratio $\gamma=1.4$ and Prandtl number $\Pran=0.72$. The range of the computational domain is $[-10, 600]\times[0,160]\times[0,7.5\pi]$ in the streamwise, wall-normal and spanwise directions, respectively. Here, the domain upstream of $x = 0$ is enforced with a slip-wall (inviscid
symmetry) boundary condition to mitigate potential
numerical instabilities. The spanwise width is exactly three times the wavelength of the optimal oblique wave. A computational grid with the resolution $3060\times260\times60$ is used for the spatial discretization. It corresponds to the dimensionless grid spacings ${{\Delta}}{x}^+\approx3.13$, ${{\Delta}}{z}^+\approx5.90$, and ${{\Delta}}{y}^+_{\mathit{min}}\approx0.30$, evaluated in the fully developed turbulent region. As shown in \citet{guo_2023_interaction}, the computational grid used in the current research provides at least 27 grid cells per wavelength of the second mode in the $x$ direction, and exactly 20 cells per wavelength of the first mode in the $z$ direction in the linear and early nonlinear stages.\par
The high-fidelity DNS is then performed using the multi-block parallel finite difference solver OpenCFD, which has been successfully applied to the simulations of high-speed transitional \citep{li_2008_direct,zheng_2019_image} and turbulent \citep{zhang_2014_generalized,she_2018_prediction,xu_2021_effect} flows. A seventh-order weighted essentially non-oscillatory (WENO) scheme \citep{balsara_2000_monotonicity} is used for the reconstruction of the inviscid flux, and a sixth-order central difference scheme is employed for the viscous flux. The time marching is implemented by a third-order Runge--Kutta method. \par

\subsection{Optimal upstream forcings and resonant state}\label{sec:IOforcing}
Prior to the DNS, a resolvent analysis of the 2-D laminar base flow is conducted to determine the most amplified disturbances. The Reynolds decomposition of the vector of primitive variables, $\boldsymbol{q}=[\rho,u,v,w,T]^{\rm{T}}$, is given by\par
\begin{equation}\label{q}
  \boldsymbol{q}(x,y,z,t)=\bar{\boldsymbol{q}}(x,y)+\boldsymbol{q}^\prime(x,y,z,t),
\end{equation}
where the overbar denotes base-flow quantities. Substituting (\ref{q}) into (\ref{NS}) and subtracting the base-flow equation yield the linearized governing equations:\par
\begin{equation}\label{LNS}
  \frac{{\partial}\boldsymbol{q}^\prime}{{\partial}t}=\mathsfbi{L}\boldsymbol{q}^\prime+\boldsymbol{N}^\prime+\mathsfbi{B}\boldsymbol{f}^\prime,
\end{equation}
where $\mathsfbi{L}$ is the linear operator containing the spatial discretization and the base state, and $\boldsymbol{N}^\prime$ is the nonlinear term including all higher-order productions. The nonlinear term can be neglected by taking the small-perturbation assumption, or alternatively, may be lumped into the unknown forcing term as a broadband input to analyze the input--output characteristics. In the resolvent analysis of linear primary instabilities, the nonlinear term $\boldsymbol{N}^\prime$ is dropped so that the linear response to solely the upstream forcing $\mathsfbi{B}\boldsymbol{f}^\prime$ can be quantified. The normal-mode ansatz is introduced for all fluctuating flow quantities and forcing terms, \par
\begin{subequations}\label{ansatz}
\begin{align}
  \boldsymbol{q}^\prime(x,y,z,t)&=\hat{\boldsymbol{q}}(x,y){\rm{exp}}(-{\rm{i}}\omega{t}+{\rm{i}}\beta{z})+\text{c.c.}, \\
  \boldsymbol{f}^\prime(x,y,z,t)&=\hat{\boldsymbol{f}}(x,y){\rm{exp}}(-{\rm{i}}\omega{t}+{\rm{i}}\beta{z})+\text{c.c.},
\end{align}
\end{subequations}
where $\omega$ is the angular frequency, $\beta$ is the spanwise wavenumber and c.c. denotes complex conjugate. By substituting (\ref{ansatz}) into (\ref{LNS}), a linear input--output formulation that describes the response of the system to an external forcing can be obtained as\par
\begin{subequations}\label{IO}
\begin{align}
  \hat{\boldsymbol{q}}&=\mathsfbi{R}\mathsfbi{B}\hat{\boldsymbol{f}}, \\
  \mathsfbi{R}&=\left(-{\rm{i}}\omega-\mathsfbi{L}\right)^{-1}.
\end{align}
\end{subequations}
Here, $\mathsfbi{R}$ is the resolvent operator.\par
The goal of the linear resolvent analysis is to seek the forcing term $\hat{\boldsymbol{f}}$ that maximizes energy amplification, which is measured by the optimal gain:\par
\begin{equation}\label{gain}
  \sigma^2(\omega,\beta)=\underset{\hat{\boldsymbol{f}}}{\rm{max}}\left(\frac{||\hat{\boldsymbol{q}}||_\mathcal{W}}{||\mathsfbi{B}\hat{\boldsymbol{f}}||_\mathcal{W}}\right),
\end{equation}
where $||\cdot||_{\mathcal{W}}$ takes the norm of Chu's energy \citep{chu_1965_energy,bugeat_2019_3d}. As shown by \citet{sipp_2013_characterization} and \citet{garnaud_2013_preferred}, this optimization problem can be transformed into an eigenvalue problem with respect to $\sigma^2$, and the optimal forcing shape $\hat{\boldsymbol{f}}$ is the eigenfunction. The optimal response mode $\hat{\boldsymbol{q}}$ can then be recovered from equation (\ref{IO}). More details regarding the resolvent analysis solver can be found in \citet{hao_2023_response} and \citet{guo_2023_interaction}.\par
\begin{figure}
  \centering
  \includegraphics[width=0.45\textwidth]{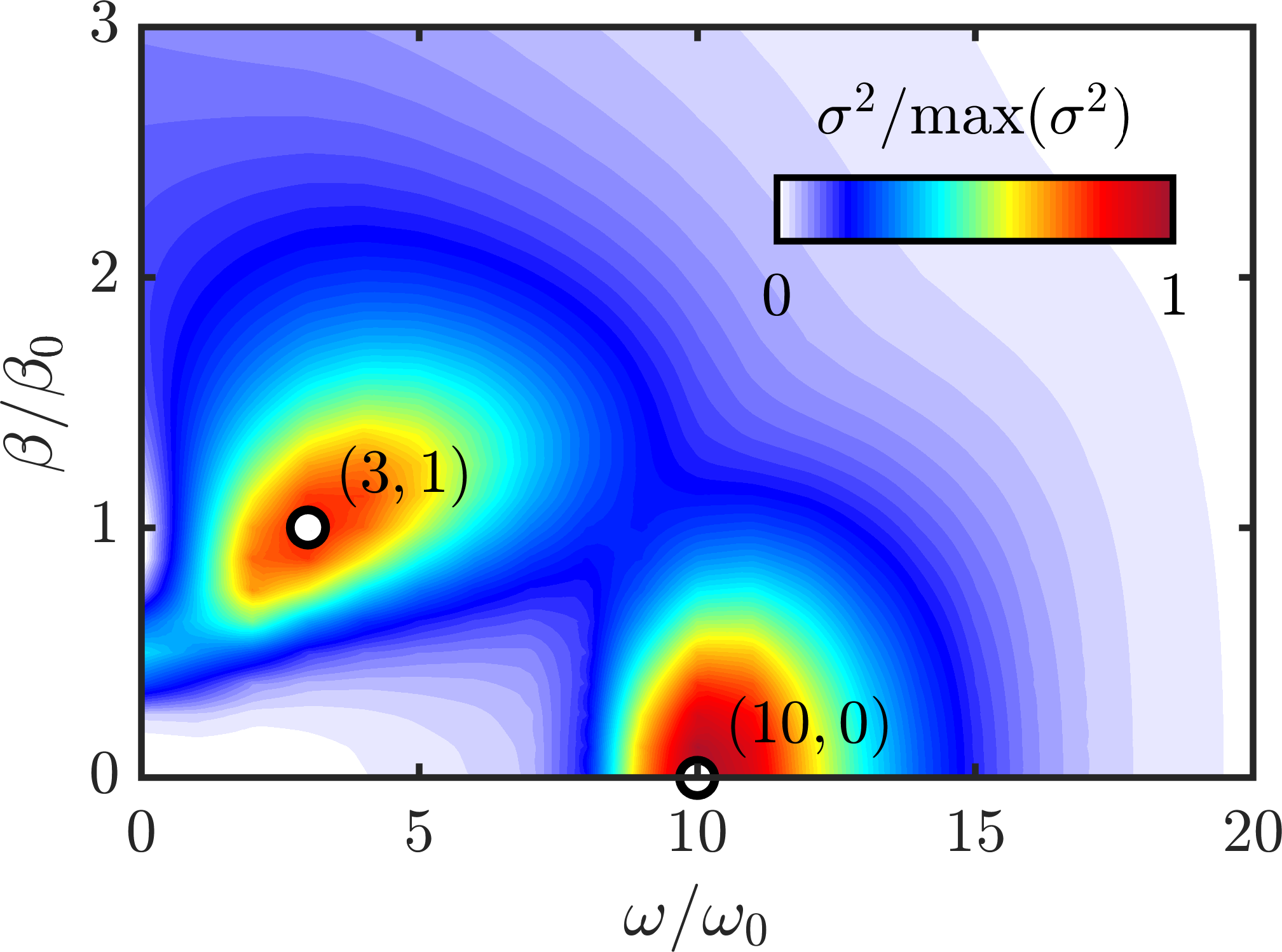}
  \caption{Contours of optimal gain (normalized by the maximum) in the parameter space of the angular frequency and the spanwise wavenumber.}
  \label{fig:resolvent}
\end{figure}

Figure \ref{fig:resolvent} shows the optimal gain as a function of the angular frequency $\omega$ and the spanwise wavenumber $\beta$, normalized by the global maximum. Two prominent local peaks at $(\omega,\beta)=(0.3,0.8)$ and $(1.0,0.0)$ are reported in the contour of optimal gain. These two peaks correspond to the first and second modes in the linear stability analysis (LSA) performed by \citet{guo_2023_interaction}. The notation `mode $(m,n)$' refers to the disturbance wave (Fourier mode) with $(\omega/\omega_0,\beta/\beta_0)=(m,n)$, where $\omega_0=0.1$ and $\beta_0=0.8$ are the fundamental values for normalization. Hence, the two peaks are modes $(3,1)$ and $(10,0)$. Notably, the dimensional frequency of the optimal planar wave $(10,0)$ is $125.8\ {\rm{kHz}}$, which has been shown to peak in the energy spectrum of the experimental boundary layer \citep{bountin_2013_stabilization}.\par
As a result of the resolvent analysis, the forcing terms of a pair of the oblique wave $(3,\pm1)$ and the planar wave $(10,0)$ are imposed at $x_0=40$ in the DNS, since they represent the most linearly amplified waves. The amplitude of the forcing is linearly rescaled in the DNS, such that the maximum $(\rho u)^{\prime}$ is 0.002 for both $(10,0)$ and $(3,\pm1)$ at the nearby location $x=45$ (see more details in \cite{guo_2023_interaction}). This setting allows for the strong interaction between two primary waves of equal significance. The oblique wave (first mode) and planar wave (second mode) are therefore considered as the two primary waves to initiate the transition, which will be referred to as `P1' and `P2' in the following texts for brevity. The number of primary waves is not further increased, since the nonlinear interaction between two representative waves is of sufficient interest. Regarding the onset of transition, \cite{jahanbakhshi_2019_nonlinearly} sought the `most dangerous state' to trigger the flat-plate boundary layer transition nonlinearly. The objective function of optimization was the streamwise integrated squared $C_f$, where $C_f$ denotes the skin friction coefficient. The combination of primary oblique first and planar second modes turned out to be the most effective in triggering the transition. This research justifies the selection of the present primary waves on the nonlinear dynamics level.\par
\begin{figure}
  \centering
  \includegraphics[width=0.6\textwidth]{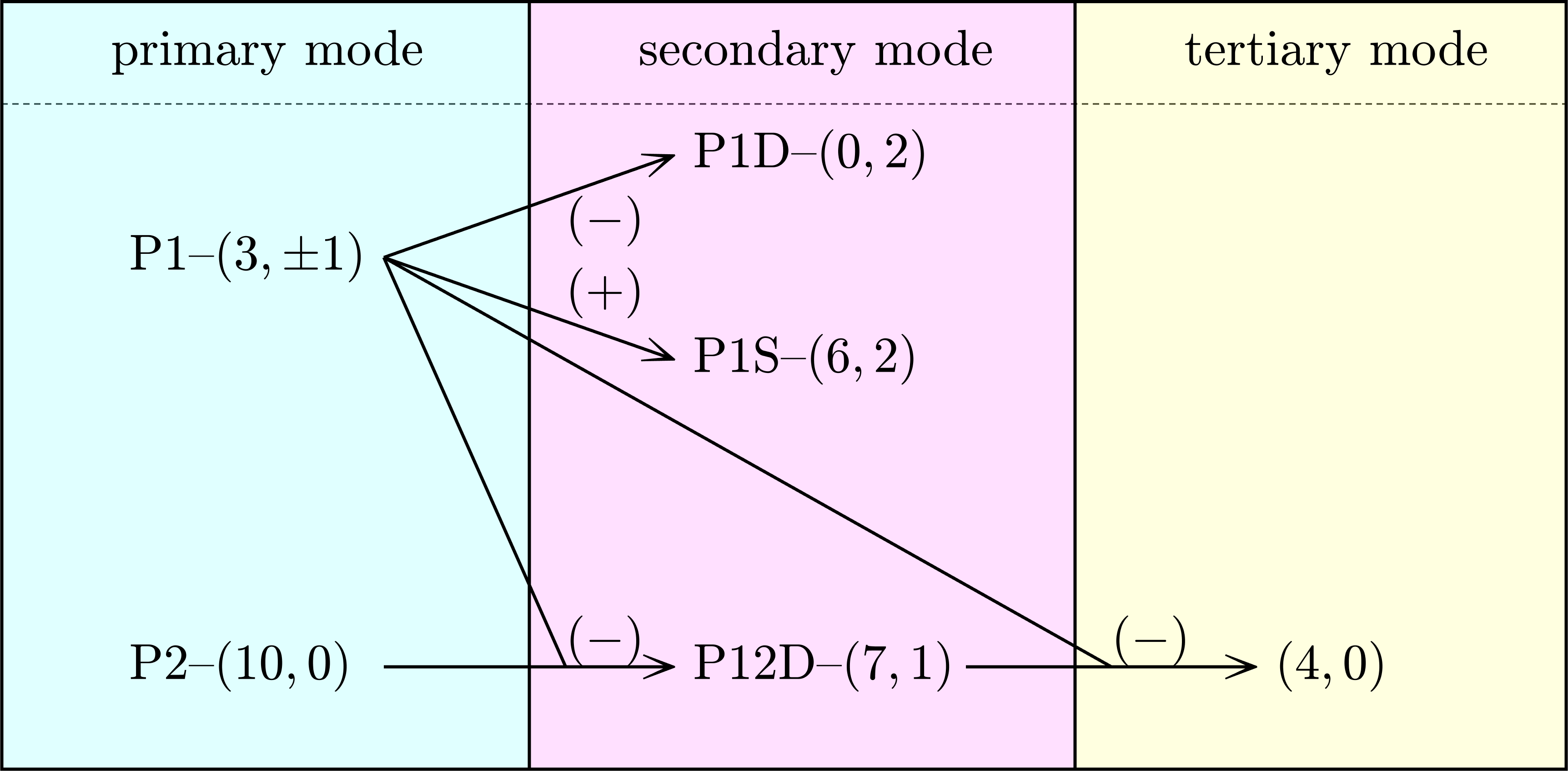}
  \caption{Triadic resonance states involving the primary modes and several representative higher-order modes. Here (+) denotes the sum interaction and (-) denotes the difference interaction.}
  \label{fig:resonance}
\end{figure}
The quadratic nonlinear interaction between Fourier modes is always in the form of sum or difference between the frequency/wavenumber, in that the multiplied sub-terms in the nonlinear term yield the sum or difference in the exponential function. In particular, the difference interaction of two waves can be regarded as the sum interaction between one wave and the conjugate counterpart of the other wave in the Fourier space. Hence, in this work, all possible quadratic interactions including with  conjugate waves are considered, and they are unifiedly expressed in their sum-interaction forms. For example, the difference interaction $(10,0)-(3,1)\rightarrow(7,-1)$ and their conjugate counterparts in the quadrant space are rewritten as $(-3,1)+(10,0)\rightarrow(7,1)$.\par Among the waves generated by nonlinear interactions, those by interactions between the two primary waves $(3,\pm1)$ and $(10,0)$, or the self-interactions of either primary wave, are named the secondary waves. Representative secondary waves include the difference mode `P12D' produced by the difference interaction between two primary waves $(-3,1)+(10,0)\rightarrow(7,1)$, the stationary streak mode `P1D' by the difference interaction between the pair of primary oblique wave $(3,1)+(-3,1)\rightarrow(0,2)$, and the harmonic mode `P1S' by the self-interaction of the primary oblique wave $(3,1)+(3,1)\rightarrow(6,2)$. The tertiary waves are defined as being generated from the interactions involving a primary/secondary and a secondary wave, e.g. from $(-3,1)+(7,-1)\rightarrow(4,0)$. These triadic interactions are depicted in the diagram shown in figure \ref{fig:resonance}. Note that the listed higher-order waves will be identified with high  fractions of energy contributions later in $\S$ \ref{sec:development}. They are associated with rich dynamics and can well represent important physical processes. Therefore, our analysis will mostly focus on the aforementioned ones.\par

\section{Analysis strategy}\label{sec:method}
\subsection{Input--output formulation for nonlinear systems}\label{sec:IO_equation}
In the transitional stage, the nonlinear effect cannot be neglected as has been done in equation (\ref{IO}). The characterization should be distinguished from that in purely linear dynamics. Note that within the linear framework of resolvent analysis, perturbation waves can be regarded as the system response to all external and nonlinear forcing terms. With the known nonlinear forcing terms, the input--output system can be further decomposed, so that how the individual forcing term contributes to the response can be quantified.\par
We seek for the input--output formulation in the Fourier spectral space, which allows for modal orthogonality. To this end, the bi-Fourier decomposition is first applied to the three-dimensional (3-D) DNS data to extract Fourier modes $(m,n)$ defined by $\hat{q}_{\mathit{m,n}}\left(x,y\right)$, which reads \par
\begin{equation}\label{q_hat}
  \boldsymbol{q}^\prime\left(x,y,z,t\right)=\sum\limits_{\mathit{m,n}}\hat{\boldsymbol{q}}_{\mathit{m,n}}\left(x,y\right){\rm{exp}}\left(-{\rm{i}}m\omega_0t+{\rm{i}}n\beta_0z\right).
\end{equation}
Then we revisit the linearized governing equation shown in (\ref{LNS}). For each Fourier mode $\hat{q}_{m,n}\left(x,y\right)$, the full expression of the associated input--output system can be achieved by substituting equation (\ref{q_hat}) into (\ref{LNS}) and retaining all higher-order productions \citep{paredes_2014_advances}. Through harmonic balancing, the general discretized form of each Fourier mode reads \par
\begin{equation}\label{IO_nonlinear}
  \hat{\boldsymbol{q}}_{\mathit{m,n}}=\mathsfbi{R}\left(\underbrace{\mathsfbi{B}\hat{\boldsymbol{f}}}_{\rm{upstream\ forcing}}+\underbrace{\sum\limits_{\mathit{m}_1,\mathit{n}_1}\hat{\boldsymbol{N}}_{\mathit{qq}}(\hat{\boldsymbol{q}}_{\mathit{m}_1,\mathit{n}_1},\hat{\boldsymbol{q}}_{\mathit{m}_2,\mathit{n}_2})}_{\rm{quadratic\ nonlinearity}}+\underbrace{\sum\limits_{\substack{\mathit{m}_1,\mathit{n}_1\\\mathit{m}_2,\mathit{n}_2}}\hat{\boldsymbol{N}}_{\mathit{qqq}}(\hat{\boldsymbol{q}}_{\mathit{m}_1,\mathit{n}_1},\hat{\boldsymbol{q}}_{\mathit{m}_2,\mathit{n}_2},\hat{\boldsymbol{q}}_{\mathit{m}_3,\mathit{n}_3})}_{\rm{cubic\ nonlinearity}}\right),
\end{equation}
where $\hat{\boldsymbol{N}}_{\mathit{qq}}$ and $\hat{\boldsymbol{N}}_{\mathit{qqq}}$ are the terms of quadratic and cubic nonlinearity, respectively. For mode $(m,n)$, the resonant condition $(m_1,n_1)+(m_2,n_2)=(m,n)$ should be fulfilled for any quadratic nonlinear term to form a valid triadic interaction \citep{craik_1971_nonlinear}. Similarly, any valid cubic nonlinear term is subject to the condition $(m_1,n_1)+(m_2,n_2)+(m_3,n_3)=(m,n)$. \par 
Our research interest concentrates on the role of triadic interactions in the formation of total perturbation waves. Therefore, only the responses to quadratic nonlinear forcings will be discussed term by term. By contrast, the cubic nonlinear forcings will be combined with the upstream forcing, regarded as the background forcing term $\hat{\boldsymbol{f}}_0$. In fact, the magnitude of cubic nonlinear terms will be shown to be significantly lower than that of quadratic terms. The responses to several important triadic interactions can well represent important dynamics. Moreover, among all triadic interactions, the one involving the mean-flow distortion (MFD) $\hat{\boldsymbol{q}}_{0,0}$ and the mode $\hat{\boldsymbol{q}}_{\mathit{m,n}}$ itself is also incorporated into the background forcing term, which is expressed by\par
\begin{equation}\label{f_0}
  \hat{\boldsymbol{f}}_{0}=\mathsfbi{B}\hat{\boldsymbol{f}}+\hat{\boldsymbol{N}}_{\mathit{qq}}(\hat{\boldsymbol{q}}_{0,0},\hat{\boldsymbol{q}}_{\mathit{m,n}})+\sum\limits_{\substack{\mathit{m}_1,\mathit{n}_1\\\mathit{m}_2,\mathit{n}_2}}\hat{\boldsymbol{N}}_{\mathit{qqq}}(\hat{\boldsymbol{q}}_{\mathit{m}_1,\mathit{n}_1},\hat{\boldsymbol{q}}_{\mathit{m}_2,\mathit{n}_2},\hat{\boldsymbol{q}}_{\mathit{m}_3,\mathit{n}_3}).
\end{equation}\par
The effect of individual triadic interaction on each Fourier mode can be then isolated and inspected based on\par
\begin{equation}\label{q_decompose}
  \hat{\boldsymbol{q}}_{\mathit{m,n}}=\hat{\boldsymbol{q}}_{\mathit{m,n}}^{(0)}+\sum\limits_{\mathit{k}=1}^{\mathit{k}_{\rm{max}}}\hat{\boldsymbol{q}}_{\mathit{m,n}}^{(\mathit{k})}=\mathsfbi{R}\left(\hat{\boldsymbol{f}}_{0}+\sum\limits_{\mathit{k}=1}^{\mathit{k}_{\rm{max}}}\hat{\boldsymbol{N}}_{\mathit{qq}}^{(\mathit{k})}\right),
\end{equation}
where $\hat{\boldsymbol{q}}_{\mathit{m,n}}^{(\mathit{k})}$ denotes the contribution from the $k$-th triadic interaction $\hat{\boldsymbol{N}}_{\mathit{qq}}^{(\mathit{k})}$ to the response. Acting as one-to-one correspondence, the response of the system to the background forcing $\hat{\boldsymbol{f}}_0$ is described by $\hat{\boldsymbol{q}}_{\mathit{m,n}}^{(0)}$. Within this input--output framework, how different forcing terms amplify the perturbation wave in the transitional flow can be therefore discussed in a quantitative manner.

\subsection{Spectral energy equation}\label{sec:energy_equation}
The amplification or attenuation of waves can arise from the mean flow. However, this effect cannot re-distribute energy among high-order harmonics. Instead, nonlinear forcings are responsible for energy transfers between different wavenumbers and frequencies due to the convolutional nature of nonlinear terms \citep{blanco_2024_linear}.  Different mechanisms are expected to be isolated that drive the energy growth and re-distribution. To this end, a generic energy equation is needed to quantify the energy transfer among the mean flow and waves of primary, secondary and tertiary instabilities.\par
First, different norms used for measuring energies are introduced. The kinetic energy of a given Fourier mode is measured by\par
\begin{equation}\label{ek}
  \hat{e}_{\mathit{k}}=\frac{1}{2}\bar{\rho}\left(\hat{u}^{\rm{H}}\hat{u}+\hat{v}^{\rm{H}}\hat{v}+\hat{w}^{\rm{H}}\hat{w}\right),
\end{equation}
where the superscript '$\rm{H}$' denotes either the complex conjugate of a scalar or the Hermitian transpose of a vector. The internal energy can be calculated by different definitions of the energy norm. Here, a commonly used positive-definite internal energy norm, including the contributions from the potential and thermal energies, is applied. The expression reads\par
\begin{equation}\label{ei}
  \hat{e}_{\mathit{i}}=\frac{1}{2}\left(\frac{\bar{T}\hat{\rho}^{\rm{H}}\hat{\rho}}{{\gamma}\mbox{\textit{Ma}}_{\infty}^2\bar{\rho}}+\frac{\bar{\rho}\hat{T}^{\rm{H}}\hat{T}}{{\gamma}({\gamma}-1)\mbox{\textit{Ma}}_{\infty}^2\bar{T}}\right).
\end{equation}
Note that the sum of $\hat{e}_{\mathit{k}}$ and $\hat{e}_{\mathit{i}}$ yields the Chu's energy density $\hat{e}_{\mathit{c}}$ \citep{chu_1965_energy}. In addition, as the Mack second mode (10,0) is of acoustic nature, it is also important to address the quantification of  acoustic energy \citep{george_2011_chu}. The corresponding norm is defined by\par
\begin{equation}\label{ea}
  \hat{e}_{\mathit{a}}=\frac{1}{2}\frac{\mbox{\textit{Ma}}_{\infty}^2\hat{p}^{\rm{H}}\hat{p}}{\bar{\rho}\bar{T}}.
\end{equation}\par
The rate of local energy transfer is observed with the corresponding material derivative ${\rm{D}}\hat{e}/{\rm{D}}t$. Beforehand, the material derivative of $\hat{q}_s^{\rm{H}}\hat{{q}_{\mathit{s}}}$ should be calculated, which can be split into\par
\begin{equation}\label{DqqDt1}
  \frac{{\rm{D}}(\hat{q}^{\rm{H}}_{\mathit{s}}\hat{q}_{\mathit{s}})}{{\rm{D}}t}=\hat{q}^{\rm{H}}_{\mathit{s}}\frac{{\rm{D}}\hat{q}_{\mathit{s}}}{{\rm{D}}t}+\frac{{\rm{D}}\hat{q}^{\rm{H}}_{\mathit{s}}}{{\rm{D}}t}\hat{q}_{\mathit{s}}.
\end{equation}
Here, a specific subscript $\mathit{s}=1,2,...,5$ introduces the indexes the variable(s) may need for each energy norm. The two terms on the right-hand side (RHS) are the complex conjugate of each other, and thus\par
\begin{equation}\label{DqqDt2}
  \text{RHS of equation (3.8)}=2\times\mathcal{R}\left(\hat{q}^{\rm{H}}_{\mathit{s}}\frac{{\rm{D}}\hat{q}_{\mathit{s}}}{{\rm{D}}t}\right).
\end{equation}\par
Up to now, for each Fourier mode $(m,n)$, the material derivative of kinetic, acoustic or internal energy can be computed by first utilizing the linearized compressible continuity, momentum and energy equations. The material derivative ${\rm{D}}\hat{q}_{\mathit{s}}/{{\rm{D}}t}$ is then substituted into equation  (\ref{DqqDt2}). Similarly to equation (\ref{IO_nonlinear}), all nonlinear forcing terms are retained to account for the energy transfer among different spatio-temporal scales. Moreover, on the left-hand side, appropriate rearrangement should be applied to these equations to construct the material derivative from the time derivative. After the rearrangement, the general form is presented as\par
\begin{equation}\label{DqDt}
  \underbrace{\frac{{\rm{D}}\hat{\boldsymbol{q}}_{\mathit{m,n}}}{{\rm{D}}t}}_{\rm{rate\ of\ change\ in\ fluctuations}}=\underbrace{\mathsfbi{L}\hat{\boldsymbol{q}}_{\mathit{m,n}}}_{\substack{{\rm{amplification/attenuation}}\\{\rm{by\ mean\ flow}}}}+\underbrace{\sum\hat{\boldsymbol{N}}_{\mathit{qq}}+\sum\hat{\boldsymbol{N}}_{\mathit{qqq}}}_{\rm{nonlinear\ transfer\ among\ scales}}.
\end{equation}
Here, $\mathsfbi{L}\hat{\boldsymbol{q}}_{\mathit{m,n}}$ contains the interaction of mode $(m,n)$ and the mean flow. The nonlinear interaction between the mean-flow distortion $\hat{\boldsymbol{q}}_{0,0}$ and mode $(m,n)$ itself is absorbed into the $\mathsfbi{L}\hat{\boldsymbol{q}}_{\mathit{m,n}}$, because they are similar terms acting on mode $(m,n)$ only. In contrast to the nonlinear terms, the $\mathsfbi{L}\hat{\boldsymbol{q}}_{\mathit{m,n}}$ is simply called the linear term later, which characterizes the effect of the mean flow on the development of perturbation waves. While the laminar base flow is commonly employed in the traditional stability analysis, a mean-flow analysis is better for a transitional case. The quadratic and cubic nonlinear terms, $\hat{\boldsymbol{N}}_{\mathit{qq}}$ and $\hat{\boldsymbol{N}}_{\mathit{qqq}}$, are responsible for the energy transfer among different spatio/temporal scales. Note that the MFD-plus-self interaction term is hereafter excluded from $\hat{\boldsymbol{N}}_{\mathit{qq}}$. Following $\S$ \ref{sec:IO_equation}, the quadratic nonlinearity will be discussed term-by-term, while the cubic nonlinearity will be incorporated into the background term.\par
Once equation (\ref{DqDt}) is prepared, the energy transfer can be then obtained using equations (\ref{ek}), (\ref{ei}) and (\ref{DqqDt2}). For the acoustic energy, the pressure is not directly solved from the compressible linearized equation, but computed by rearranging the linearized state equation:\par
\begin{equation}\label{Lstate}
  \frac{{\rm{D}}\hat{p}}{{\rm{D}}t}=\left(\frac{{\rm{D}}(\bar{\rho}\hat{T})}{{\rm{D}}t}+\frac{{\rm{D}}(\bar{T}\hat{\rho})}{{\rm{D}}t}+\frac{{\rm{D}}  (\hat{\rho}\hat{T})}{{\rm{D}}t}\right)/{{\gamma}\mbox{\textit{Ma}}_{\infty}^2}.
\end{equation}
On the RHS of (\ref{Lstate}),  the first two terms can be handled via equation (\ref{DqDt}). The third term is nonlinear, which should fulfill the triadic resonant condition shown for $\hat{\boldsymbol{N}}_{\mathit{qq}}$ in equations (\ref{IO_nonlinear}) and (\ref{DqDt}).\par
As such, the local energy transfer can be described by the Lagrangian scalar equation:\par
\begin{equation}\label{eq_energy}
\frac{{\rm{D}}\hat{e}_{\mathit{def}}}{{\rm{D}}t}=\frac{{\rm{D}}\hat{e}_{\mathit{def,L}}}{{\rm{D}}t}+\sum\frac{{\rm{D}}\hat{e}_{\mathit{def,N}}}{{\rm{D}}t}+\sum\frac{{\rm{D}}\hat{e}_{\mathit{def,NN}}}{{\rm{D}}t}, 
\end{equation}
where the definition of energy norm `${\mathit{def}}$' can be the preceding $k,i,a$ or $c$, and the subscript `$(m,n)$' is omitted for brevity. The RHS terms represent the linear, quadratic nonlinear and cubic nonlinear transfers, successively, resulting from the corresponding terms in (\ref{DqDt}). The contribution of each triad can be retrieved from the quadratic nonlinear transfer ${\rm{D}}\hat{e}_{\mathit{def,N}}/{\rm{D}}t$. To quantify the streamwise development of energy transfer,  the wall-normal integrated transfer term $\mathcal{T}$  is also introduced as $\mathcal{T}=\int{\rm{D}}\hat{e}_{\mathit{def}}/{\rm{D}}t{\rm{d}}y$.\par

\section{Development of disturbances}\label{sec:development}
\begin{figure}
  \centering
  \includegraphics[width=1.0\textwidth]{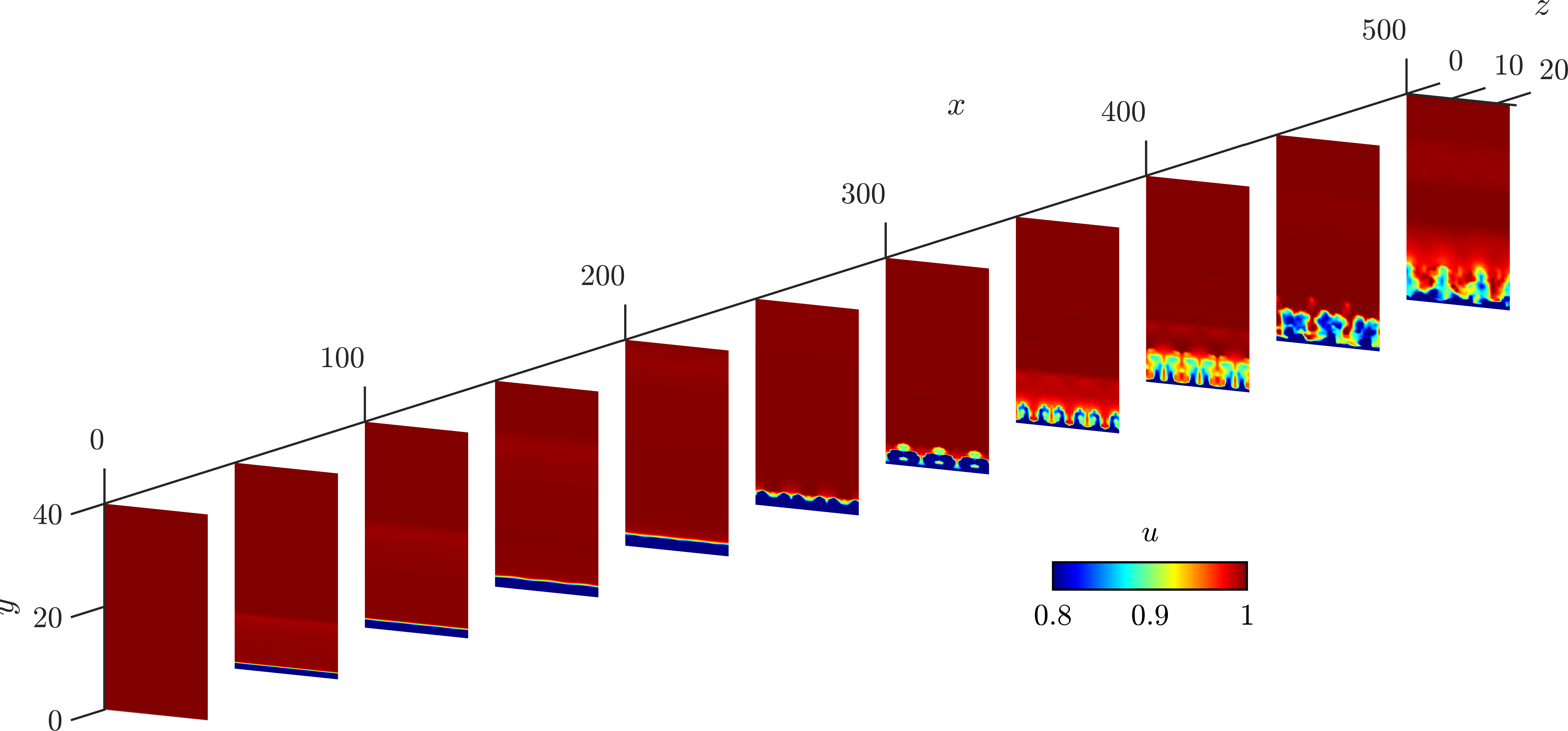}
  \caption{Three-dimensional visualization of the flow field based on the DNS results.}
  \label{fig:dns}
\end{figure}

The considered flow problem describes a transition process initiated by two primary instabilities. Figure \ref{fig:dns} presents the 3-D view of instantaneous contours of the streamwise velocity, extracted from equally spaced $z-y$ slices. In the early stage, three-dimensionality is absent in the boundary-layer flow. Starting from $x=250$, the signature of oblique waves becomes evident. As reported in \citet{guo_2023_interaction} and \citet{chen_2025_role}, the transition onset and end locations are around $x=220$ and $x=570$, respectively, based on the minimum of the spanwise- and time-averaged $C_{\mathit{f}}$ and its intersection with the turbulent correlation. As a result, figure \ref{fig:dns} displays a long-distance transitional process, mimicking a low-amplitude occasion. This gives rise to multiple spatio/temporal scales occurring in the transitional stage, likely due to high-order instabilities. This section intends to provide an inspection of the development of instability waves during the transition process. Meanwhile, the component-wise evolution will be discussed, e.g. acoustic and non-acoustic associated components. \par
\subsection{Saturation of primary waves and spectral broadening}\label{sec:saturation}
\begin{figure}
  \centering   
  \begin{subfigure}[t]{0.49\textwidth}\flushleft
  \subcaption{The primary first mode P1--$(3,1)$}
  \includegraphics[width=1.0\textwidth]{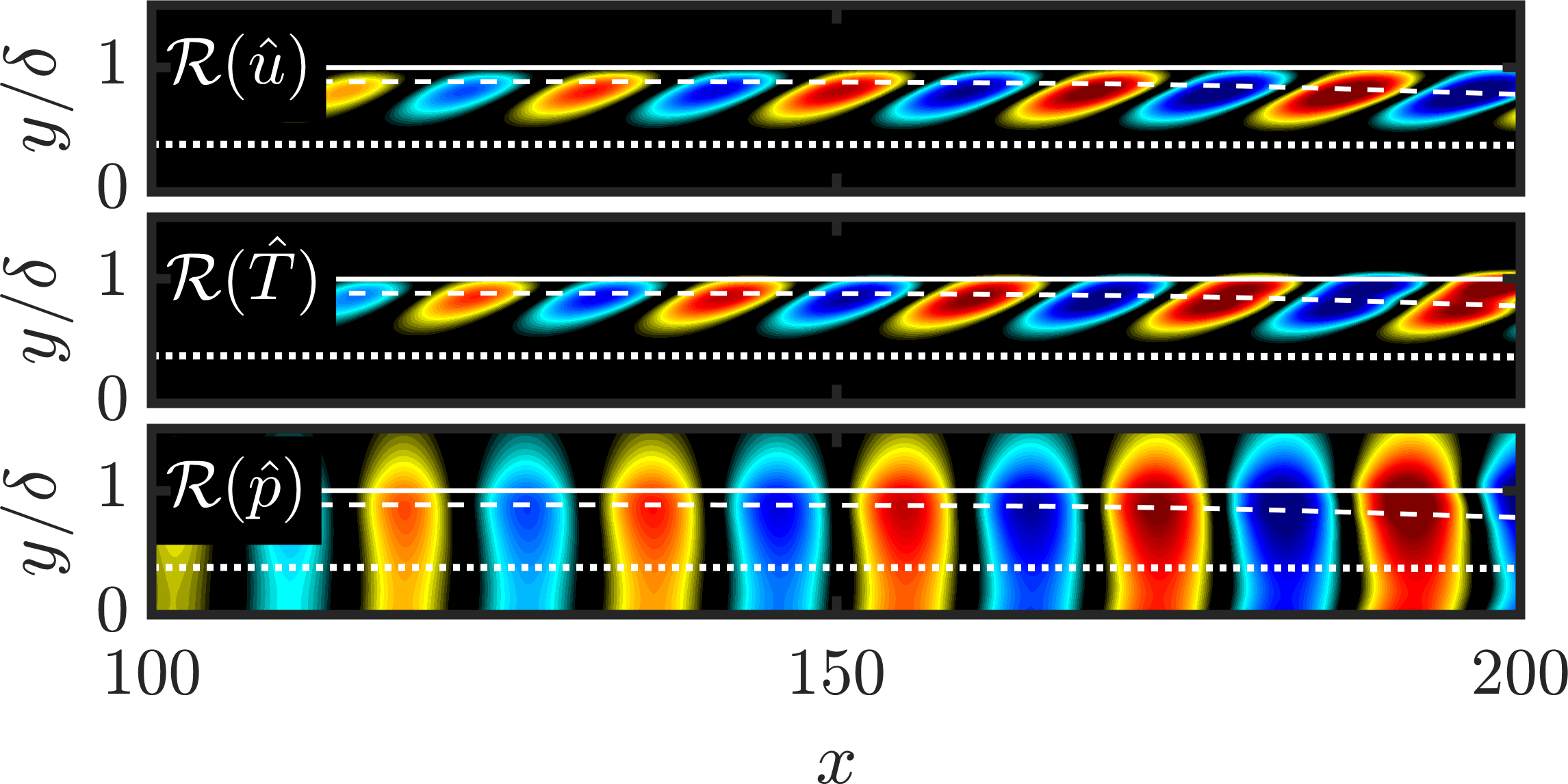}
  \end{subfigure}\vspace{-0cm}
  \begin{subfigure}[t]{0.49\textwidth}\flushleft
  \subcaption{The primary second mode P2--$(10,0)$}
  \includegraphics[width=1.0\textwidth]{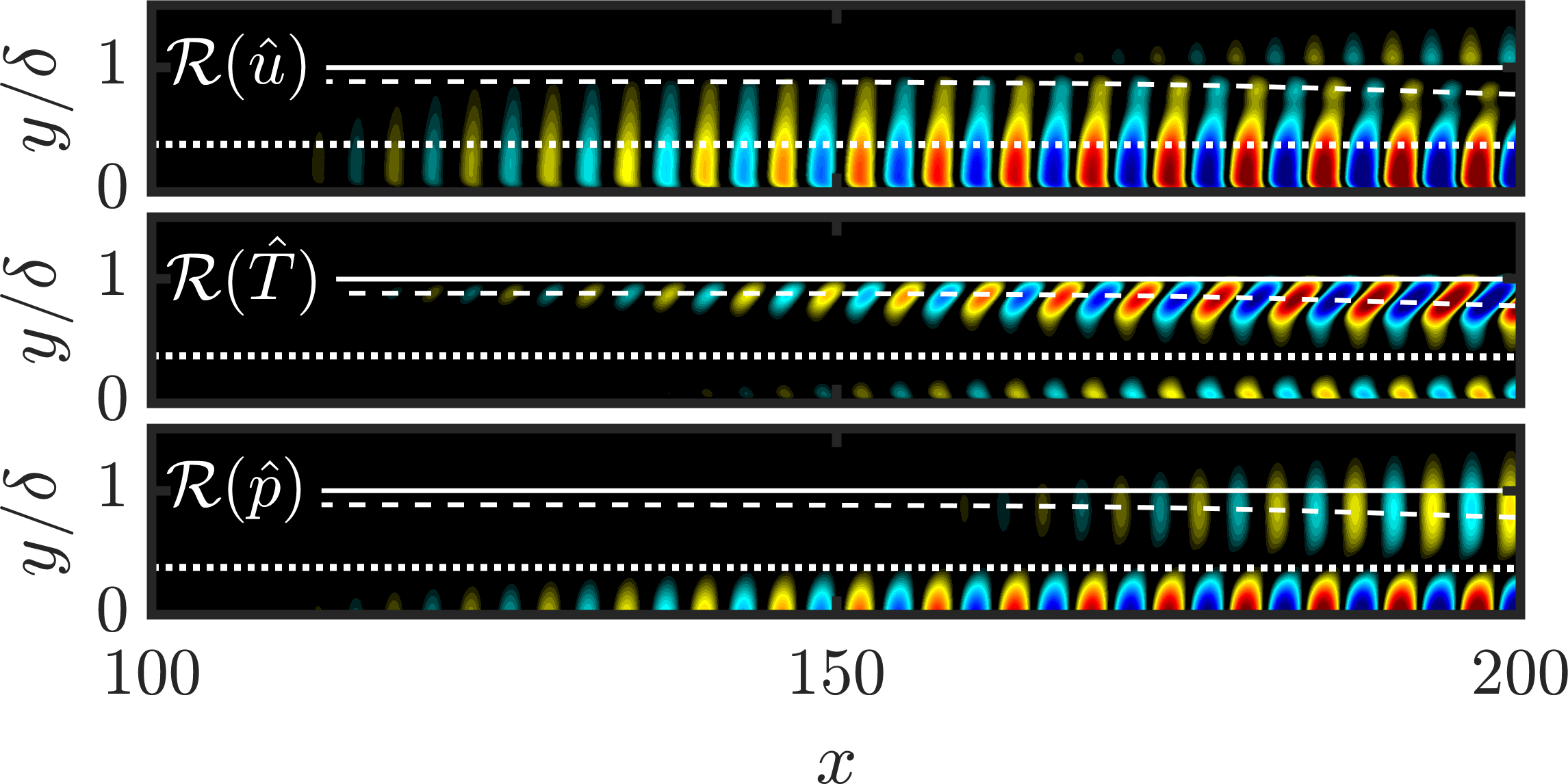}
  \end{subfigure}\vspace{-0cm}
  \caption{Spatial distributions of primary waves visualized by the real parts of the streamwise velocity, temperature and pressure components. The respective solid, dashed and dotted lines in the plots represents the boundary layer edge, the generalized inflection point and the sonic line. Contour bar ({\color[rgb]{0.8 0 0}$\blacksquare$}{\color[rgb]{0 0 0}$\blacksquare$}{\color[rgb]{0 0 0.8}$\blacksquare$}) ranges between $\pm0.8\times{\rm{max}}(|\hat{q}|)$ of each field.}
  \label{fig:mode_primary}
\end{figure}

The development of the two primary waves is investigated first. Figure \ref{fig:mode_primary} shows the distributions of the two primary modes in the linear growth stage, based on the visualization of their $\hat{u}$, $\hat{T}$, and $\hat{p}$ components. Meanwhile, to better relate the waves to the mean flow profile, the boundary layer edge, the generalized inflection point (GIP) and the sonic line are shown in the figure for reference. The $y$-coordinate is normalized at each streamwise location by the local thickness $\delta$ of the mean boundary layer. The GIP of a mean flow profile is defined by $\partial(\Breve{\rho}(\partial\Breve{u}/\partial{y}))/\partial{y}=0$, where $\Breve{(\cdot)}$ represents the mean-flow quantity. The sonic line of the mean flow is defined by the position where the local Mach number $\Breve{\mbox{\textit{Ma}}}=1$. Note that the  acoustic second mode may be more precisely characterized by the `relative sonic line' \citep{fedorov_2011_transition}, while the sonic line is also acceptable. A quantitative comparison between the two lines in hypersonic boundary layers can be found in \citet{roy_2025_disturbance}. The results shown in figure \ref{fig:mode_primary} reveal essential physical properties of the two primary modes similar to previous research: the first mode features amplification of perturbation waves along the GIP, while the second mode manifests as trapped acoustic waves between the wall and the sonic line \citep{mack_1990_inviscid,fedorov_2011_transition,chen_2023_unified}.\par
\begin{figure}
  \centering
  \includegraphics[width=0.625\textwidth]{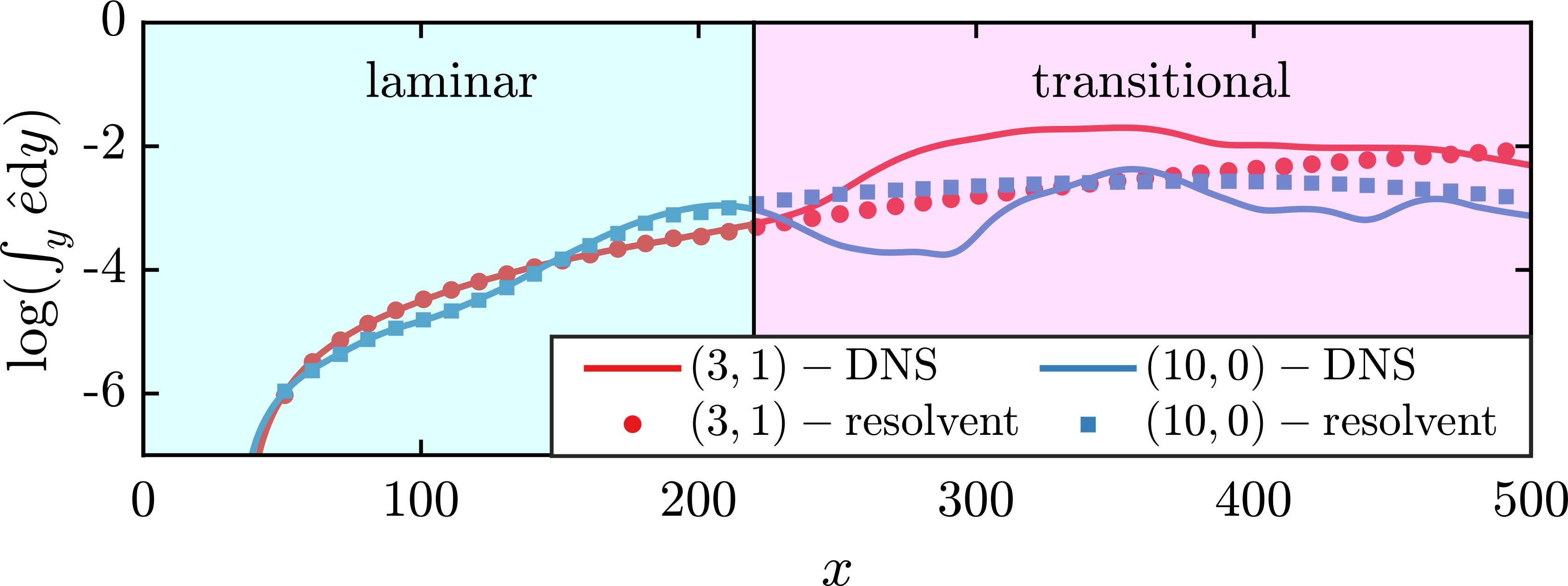}
  \caption{Development of the Chu's energy of the two primary waves.}
  \label{fig:energy}
\end{figure}

As the two primary waves grow continuously, the amplitudes of nonlinear forcing terms can increase to levels comparable to that of the mean-flow effect. Under this circumstance, the development of the primary waves can gradually deviate from the linear resolvent results. The deviation is an important indicator of nonlinear effects and onset of transition. In figure \ref{fig:energy}, the development of the two primary waves, obtained from both the linear resolvent analysis and the DNS, is quantified by the Chu's energy integrated over the wall-normal coordinate. The result is plotted versus the streamwise coordinate. For both primary waves, the DNS results start to show obvious deviation from the linear resolvent results in the vicinity of the transition onset $x=220$, indicating the important role of nonlinearity. In particular, the nonlinear effect drives the amplitude of the second-mode wave to saturate at $x\approx200$, followed by rapid attenuation. On the other hand, the first mode is observed to undergo a secondary growth downstream of the transition onset, with the final saturation location being $x\approx350$. This observation reveals the unique influence from  nonlinear interactions on different primary instabilities.\par
\begin{figure}
  \centering
  \begin{subfigure}[t]{1.0\textwidth}\flushleft
  \subcaption{}
  \includegraphics[width=0.986\textwidth]{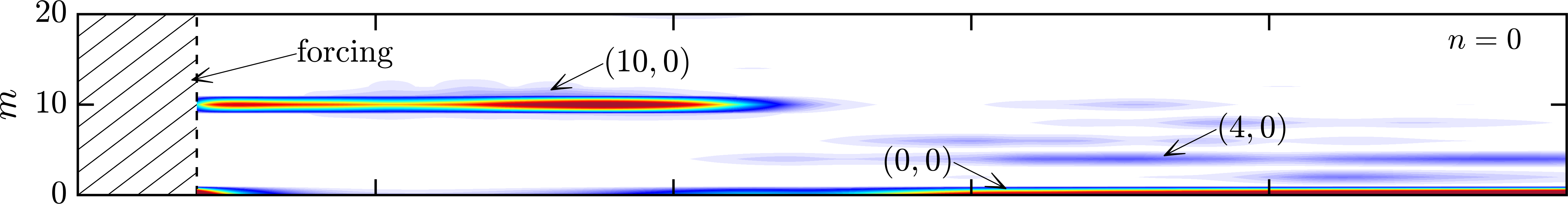}
  \end{subfigure}\vspace{-0cm}
  \begin{subfigure}[t]{1.0\textwidth}\flushleft
  \subcaption{}
  \includegraphics[width=0.986\textwidth]{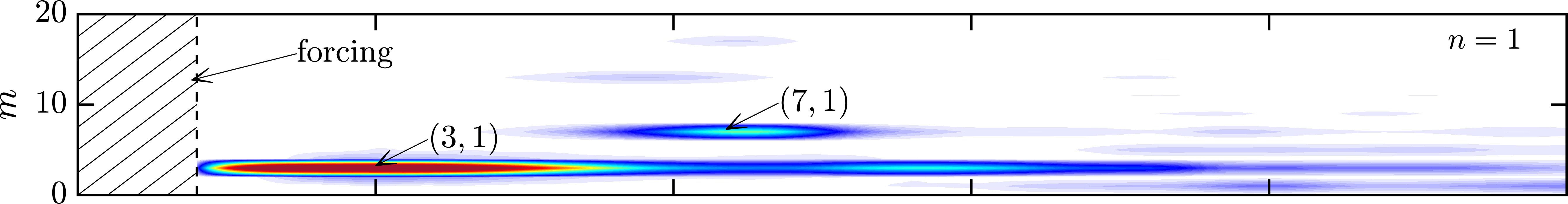}
  \end{subfigure}\vspace{-0cm}
  \begin{subfigure}[t]{1.0\textwidth}\flushleft
  \subcaption{}
  \includegraphics[width=1.0\textwidth]{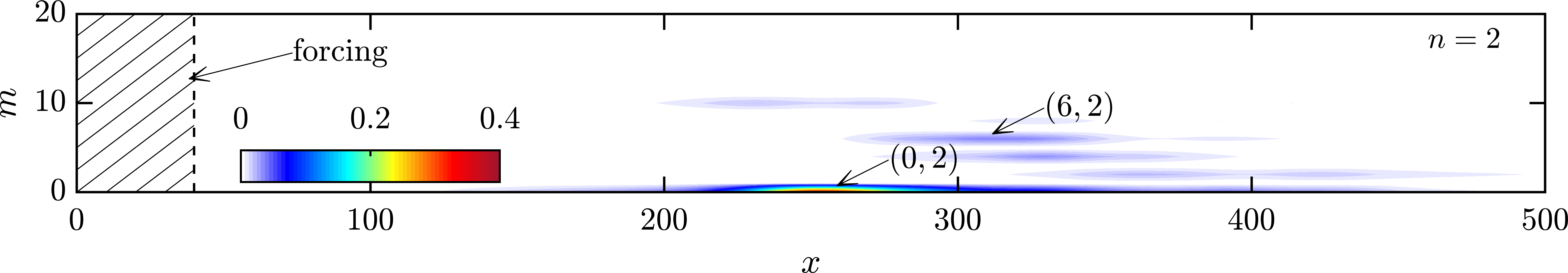}
  \end{subfigure}
  \caption{The percentage of the Chu's energy that each Fourier mode accounts for as a function of streamwise location, $\int_{\mathit{y}}\hat{e}_{\mathit{c}}(m,n){\rm{d}}y/\int_{\mathit{y}}\sum\limits_{\mathit{m,n}}\hat{e}_{\mathit{c}}{\rm{d}}y$, for (\textit{a}) $n=0$,  (\textit{b}) $n=1$ and (\textit{c}) $n=2$.}
  \label{fig:spectra}
\end{figure}

Apart from modulating the primary instabilities, nonlinear interactions also introduce forcing terms into components with different frequencies and wavenumbers. This process sustains the growth of higher-order instability waves, causes the broadening of the energy spectra and facilitates the transition into turbulence. In figure \ref{fig:spectra}, the contribution from  modes $(m,n)$ to the total Chu's energy is visualized. The interaction between the two primary waves gives rise to the difference mode P12D ($(7,1)$ in figure \ref{fig:spectra}(\textit{b})) and the stationary streak P1D ($(0,2)$ in figure \ref{fig:spectra}(\textit{c})), which in turn become significant at $x>200$. Meanwhile, the increased perturbation level leads to growing distortion of the mean flow from the base solution (see $(0,0)$ in figure \ref{fig:spectra}(\textit{a})). Further downstream of $x=300$, the impact of nonlinearity is significantly strengthened, with energy transferred to multiple spatio and temporal scales. In particular, the harmonic mode P1S ($(6,2)$ in figure \ref{fig:spectra}(\textit{c})) and the tertiary mode ($(4,0)$ in figure \ref{fig:spectra}(\textit{a})) can be identified with high energy fractions. The mechanisms associated with the observations in this section will be further investigated.\par

\subsection{Nature of higher-order instabilities: re-distributions of energy components}\label{sec:redistribution}
As aforementioned, the two primary waves are known to be of different physical natures. In short, the first mode is driven by the interplay between the viscous effect and the generalized inflection point. It can be regarded as the supersonic counterpart of the T--S wave \citep{smith_1989_first}. The second mode, on the other hand, originates from the synchronization of the phase speeds of the fast and slow modes. Accompanying the extensive nonlinear interactions, the generated higher-order modes may inherit energy components from the primary waves and re-distribute different components. Even for the primary waves, the nonlinear interactions can possibly alter their physical properties, deviated from the linear results. In order to reveal the physical nature of higher-order modes, the re-distribution of different energy components is discussed.\par
\begin{figure}
  \centering
  \begin{subfigure}[t]{0.49\textwidth}\flushleft
  \subcaption{The streak mode P1D--$(0,2)$}
  \includegraphics[width=1.0\textwidth]{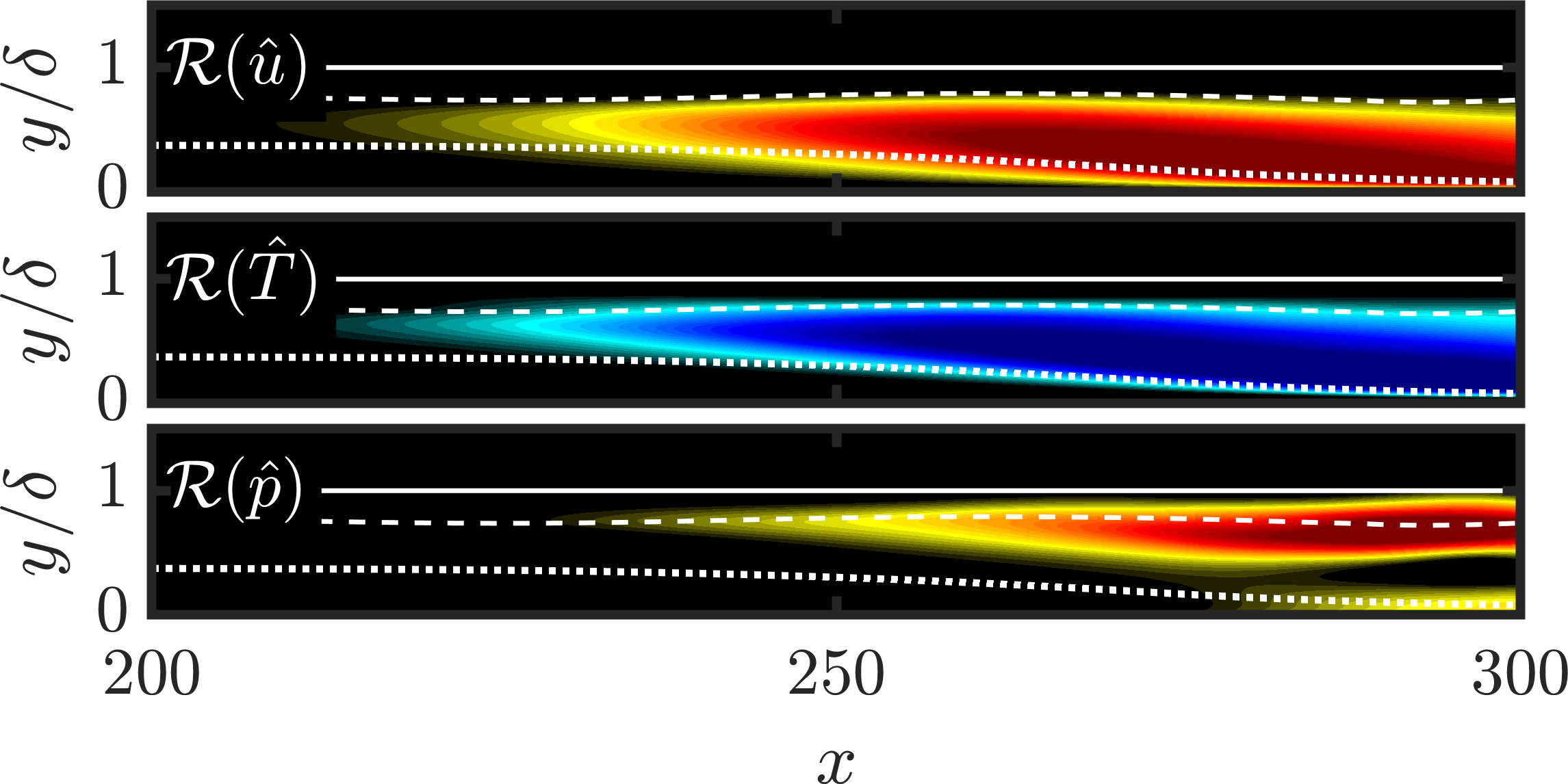}
  \end{subfigure}\vspace{-0cm}
  \begin{subfigure}[t]{0.49\textwidth}\flushleft
  \subcaption{The difference mode P12D--$(7,1)$}
  \includegraphics[width=1.0\textwidth]{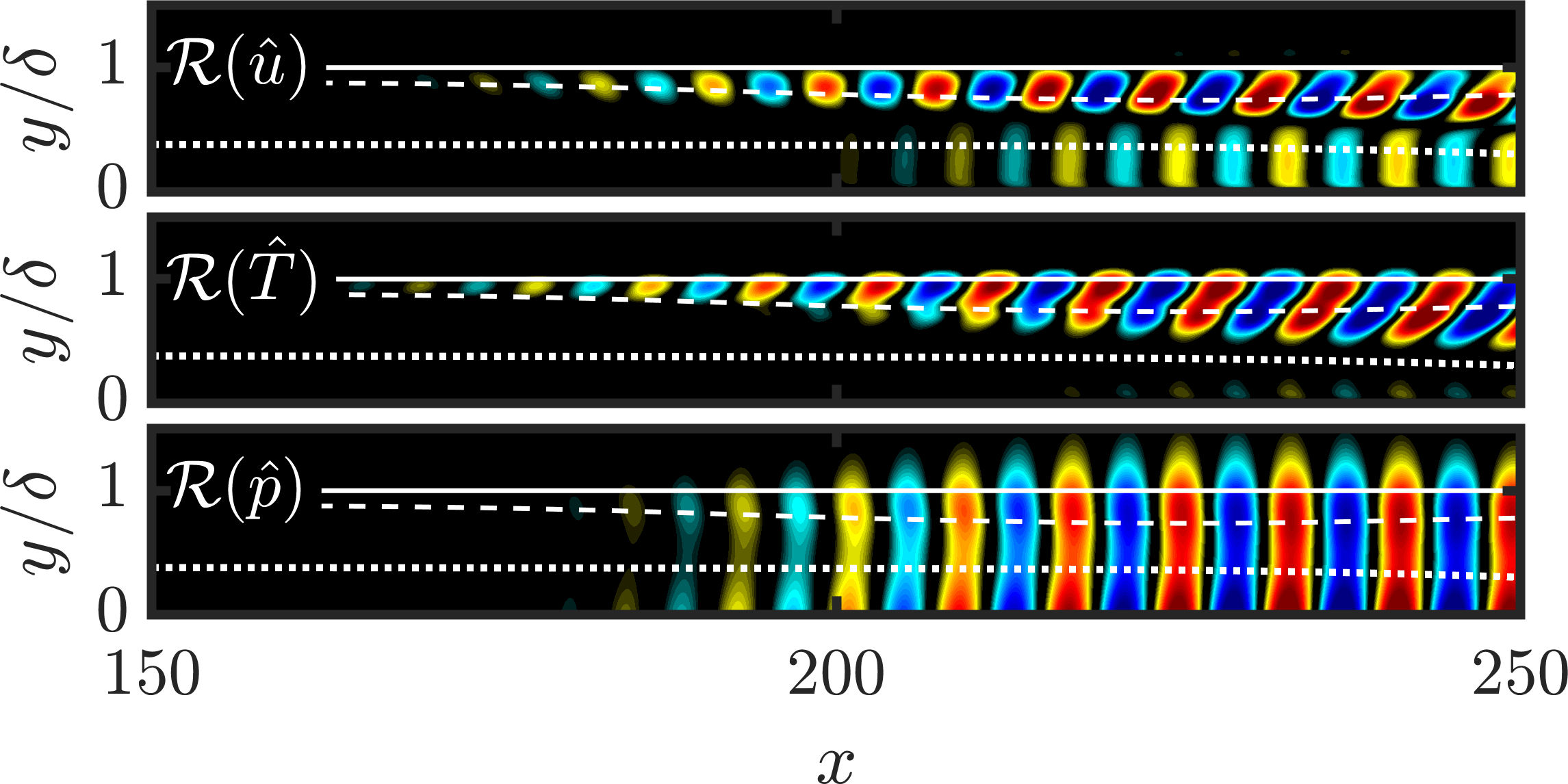}
  \end{subfigure}\vspace{-0cm}
  \begin{subfigure}[t]{0.49\textwidth}\flushleft
  \subcaption{The harmonic mode P1S--$(6,2)$}
  \includegraphics[width=1.0\textwidth]{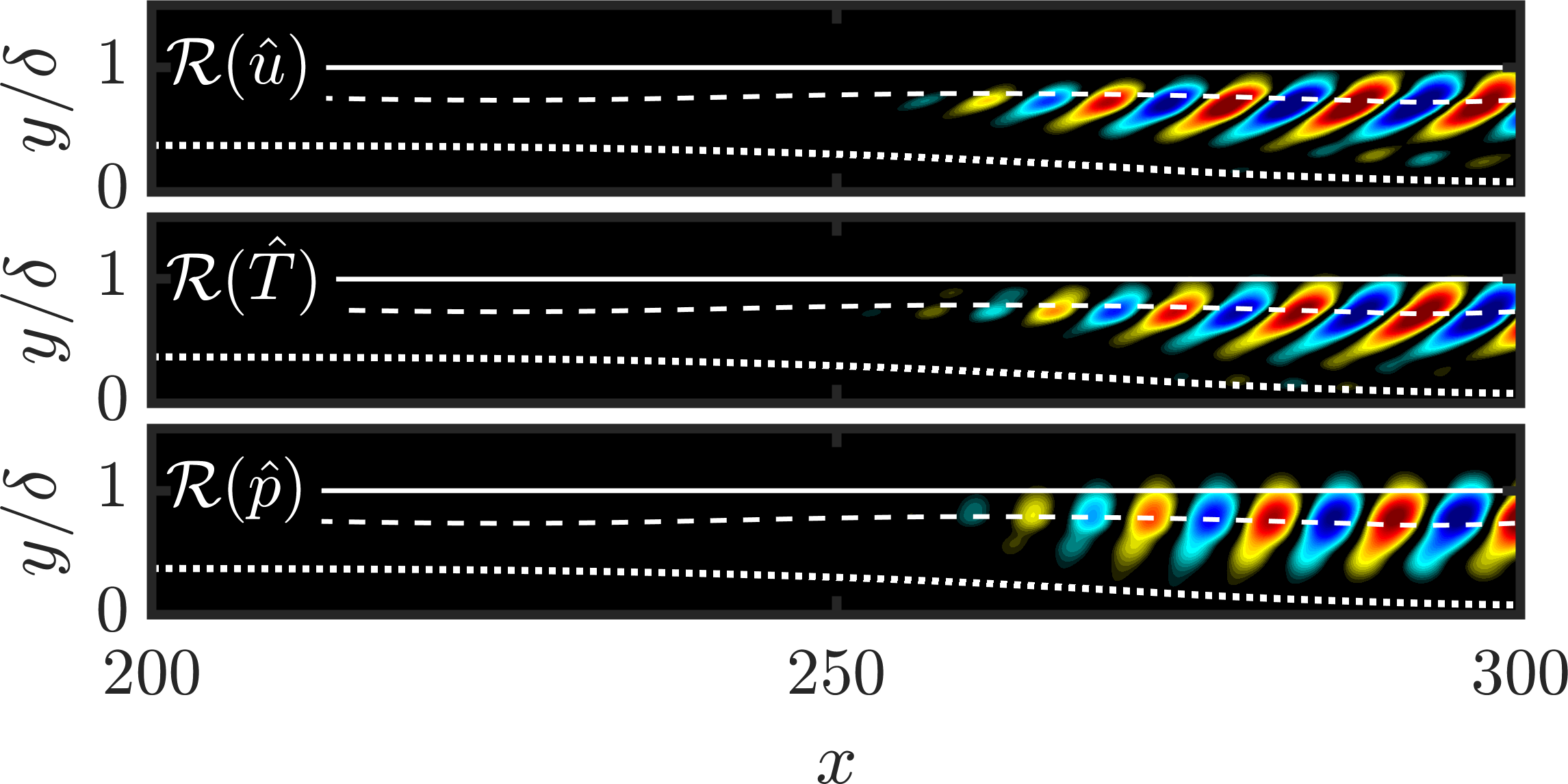}
  \end{subfigure}\vspace{-0cm}
  \begin{subfigure}[t]{0.49\textwidth}\flushleft
  \subcaption{The tertiary mode $(4,0)$}
  \includegraphics[width=1.0\textwidth]{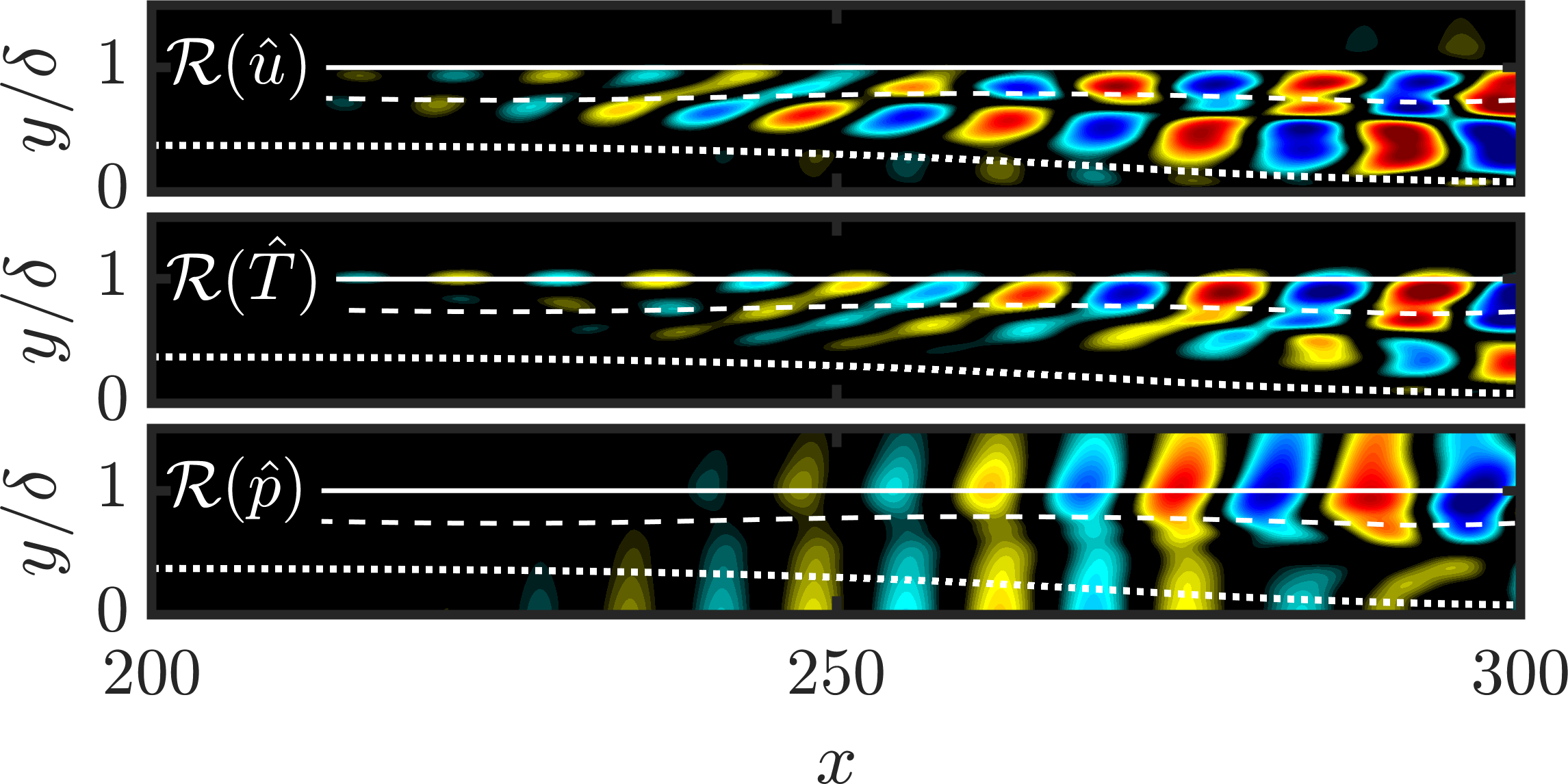}
  \end{subfigure}\vspace{-0cm}
  \caption{Spatial distributions of higher-order Fourier modes visualized by the real parts of the streamwise velocity, temperature and pressure components. The solid, dashed and dotted lines in the plots correspond to figure \ref{fig:mode_primary}. Contour bar ({\color[rgb]{0.8 0 0}$\blacksquare$}{\color[rgb]{0 0 0}$\blacksquare$}{\color[rgb]{0 0 0.8}$\blacksquare$}) ranges between $\pm0.8\times{\rm{max}}(|\hat{q}|)$ of each field.}
  \label{fig:mode}
\end{figure}
\begin{figure}
  \centering
  \begin{subfigure}[t]{1.0\textwidth}\flushleft
  \subcaption{}
  \includegraphics[width=1.0\textwidth]{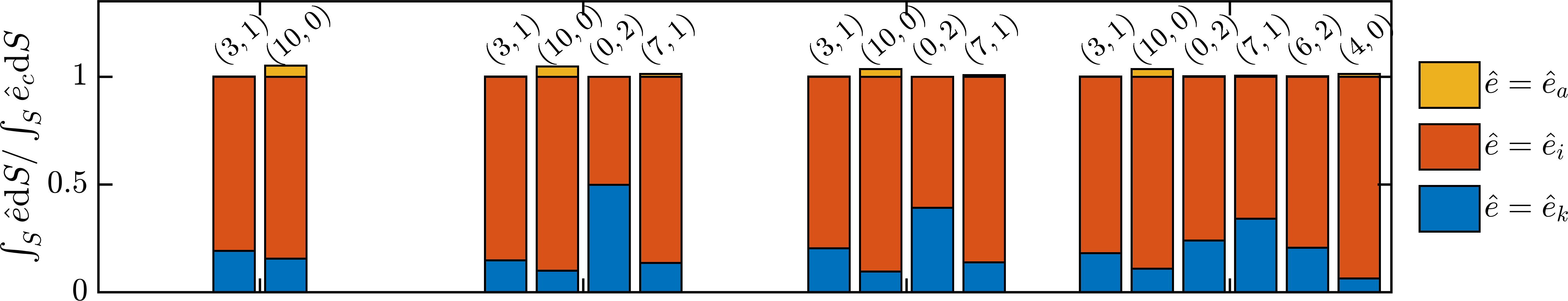}
  \end{subfigure}\vspace{-0cm}
  \begin{subfigure}[t]{1.0\textwidth}\flushleft
  \subcaption{}
  \includegraphics[width=1.0\textwidth]{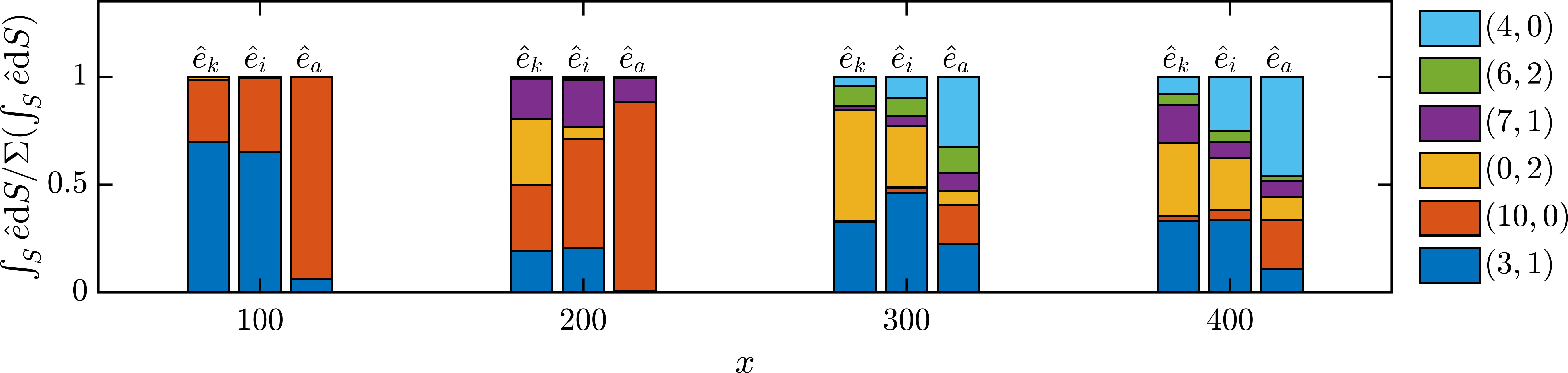}
  \end{subfigure}\vspace{-0cm}
  \caption{Distributions of energy components at different streamwise locations: (\textit{a}) The percentage of the kinetic, internal and acoustic energies to the Chu's energy for selected Fourier modes, and (\textit{b}) The percentage of energy each Fourier mode accounts for to the accumulated kinetic, internal and acoustic energy of selected Fourier modes.}
  \label{fig:bar}
\end{figure}

Apart from the two primary waves, higher-order waves identified with high energy fractions in figure \ref{fig:spectra} are considered, including the streak P1D--$(0,2)$, the difference mode P12D--$(7,1)$, the harmonic mode P1S--$(6,2)$, and the tertiary mode $(4,0)$. In figure \ref{fig:mode}, the distributions of the higher-order modes are first checked based on the visualization of their $\hat{u}$, $\hat{T}$, and $\hat{p}$ components, following the similar manner as in figure \ref{fig:mode_primary}. Note that for the secondary and tertiary modes, their modal shape and active region in boundary layers are scarcely discussed in literature. Furthermore, in figure \ref{fig:bar}, the distributions of different energy components at different streamwise locations are quantified. Both the contributions of different components to each mode and the contribution from each mode to the accumulated energy component are depicted. Herein the integral at the streamwise location $x=x_{\mathit{probe}}$ is computed within the nearby region $[x_{\mathit{probe}}-25,x_{\mathit{probe}}+25]$.\par
First, we focus on the waves P1--$(3, 1)$, P1D--$(0, 2)$ and P1S--$(6, 2)$. They can be justified as waves with tiny acoustic signature according to the contribution from the acoustic energy in figure \ref{fig:bar}(\textit{a}). For the first mode wave P1 and its harmonic mode P1S, the amplification of perturbations along the GIP can be seen from figure \ref{fig:mode_primary}(\textit{a}) and \ref{fig:mode}(\textit{c}). Figure \ref{fig:bar}(\textit{a}) also reveals that the kinetic and internal contributions to the total energy of P1 and P1S are on comparable levels. This observation indicates that these two waves are of similar property. It also implies that the harmonic triad of the primary first mode is an important source to sustain the development of P1S. The streak mode P1D, which is generated as a consequence of the oblique interaction of the first mode, is identified with large contributions from the kinetic part at $x=200$ (figure \ref{fig:bar}(\textit{a})), with the spatial structure locating mostly within the boundary layer (figure \ref{fig:mode}(\textit{a})). This result corresponds well with existing knowledge that the streak structure is characterized by dominant streamwise velocity component \citep{boronin_2013_non,pickering_2020_lift,nekkanti_2023_large}. As the streak develops downstream, the fraction of its kinetic part gradually decreases due to nonlinearity and reaches a level similar to other modes at $x=400$.\par
According to figure \ref{fig:bar}(\textit{b}), the main acoustic energy contribution in $x<200$ can be identified from the second mode P2. Both trapped acoustic waves below the sonic line and thermodynamic amplification along the GIP can be observed for this mode in figure \ref{fig:mode_primary}(\textit{b}). Interestingly, the difference mode P12D--$(7, 1)$ contains a visible acoustic energy component at $x=200$ in figure \ref{fig:bar}(\textit{a}). The mode shape shown in figure \ref{fig:mode}(\textit{b}) exhibits a double-peak structure of the pressure component of P12D along the vertical direction. One peak is below the sonic line, resembling the trapped acoustic wave of the primary second mode P2. Downstream of $x=200$, the acoustic signature still preserves for P2, while gradually decreases for P12D. However, due to the saturation and decay of P2 downstream of the transition onset, its contribution to the accumulated acoustic energy can be observed to decrease rapidly (figure \ref{fig:bar}(\textit{b})). Further downstream, the tertiary wave is generated through high-order interactions, and its mode shape in figure \ref{fig:mode}(\textit{d}) displays complex patterns. The acoustic signature of this wave becomes pronounced at $x=400$, evidenced by the contribution to the accumulated acoustic energy in figure \ref{fig:bar}(\textit{b}).\par
The above observations signify that, the physical properties of primary waves may be inherited by higher-order disturbances. For instance,  the acoustic feature is found to be transferred via P2$\rightarrow$P12D$\rightarrow$tertiary mode, and the path P1$\rightarrow$P1S is associated with the corresponding amplification along the GIP. However, higher-order perturbation waves do not necessarily resemble the features of primary instabilities, represented by the streak mode P1D which exhibits unique feature. The remainder of this paper aims to quantify the sources of these energy re-distributions.\par

\section{Input--output analysis}\label{sec:IO}
$\S$ \ref{sec:development} focuses on the `outcome', while the remaining discussions concentrate on the `reason'. The input--output formulation in $\S$ \ref{sec:IO_equation} enables further decomposition of the system based on the calculated nonlinear forcing terms which evolve from the optimal linear modes. In this section, how different triadic interactions contribute to the total forcing of a certain Fourier mode and in turn determine the response pattern will be examined.\par
\subsection{Triadic contributions to the forcing}\label{sec:forcing}
\begin{figure}
  \centering
  \begin{subfigure}[t]{0.49\textwidth}\flushleft
  \subcaption{}
  \includegraphics[width=0.97\textwidth]{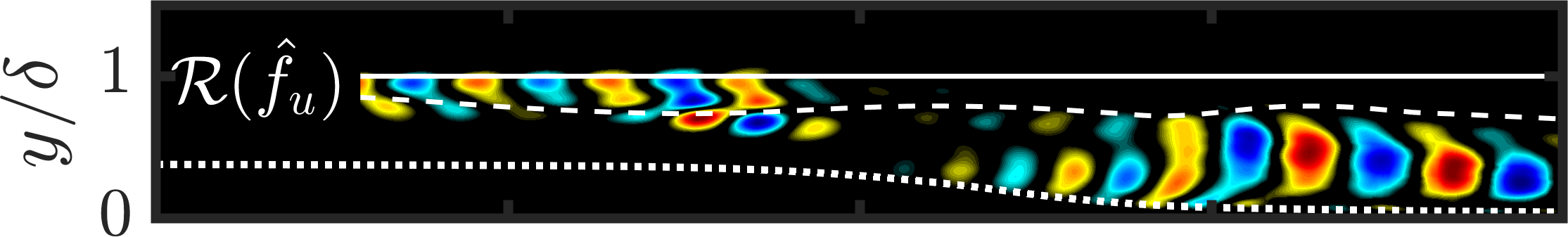}
  \end{subfigure}\vspace{-0cm}
  \begin{subfigure}[t]{0.49\textwidth}\flushleft
  \subcaption{}
  \includegraphics[width=0.97\textwidth]{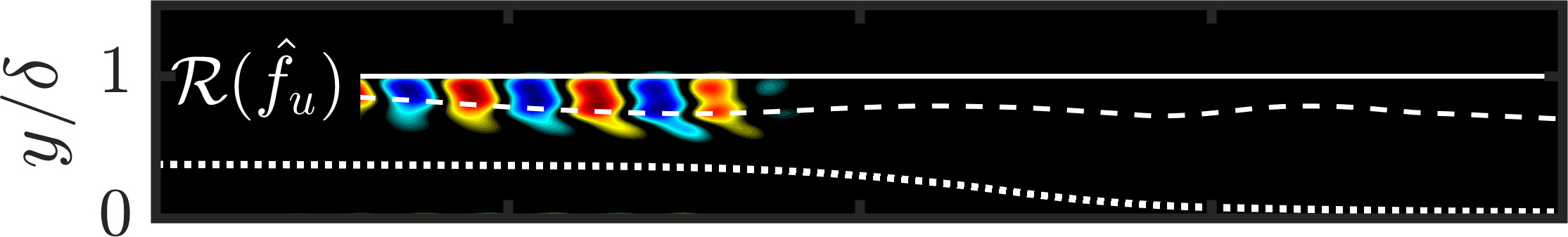}
  \end{subfigure}\vspace{-0cm}
  \begin{subfigure}[t]{0.49\textwidth}\flushleft
  \subcaption{}
  \includegraphics[width=1.0\textwidth]{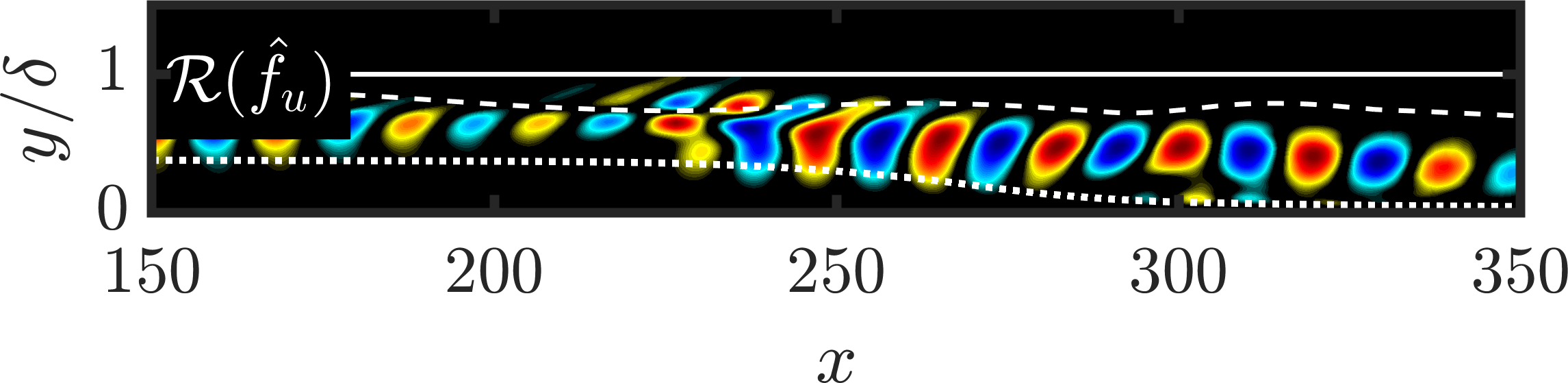}
  \end{subfigure}\vspace{-0cm}
  \begin{subfigure}[t]{0.49\textwidth}\flushleft
  \subcaption{}
  \includegraphics[width=1.0\textwidth]{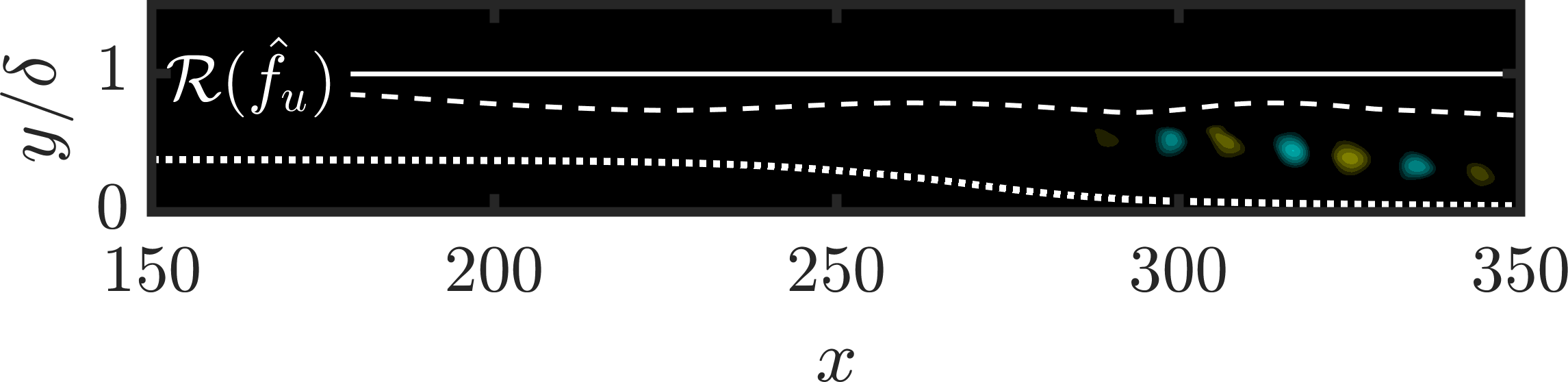}
  \end{subfigure}\vspace{-0cm}
  \caption{Nonlinear forcing terms of P1--$(3,1)$, visualized by the real part of the normalized $\hat{u}$-component: (\textit{a}) $\hat{\boldsymbol{f}}_0$, (\textit{b}) $(-7,1)+(10,0)$, (\textit{c}) $(0,2)+(3,-1)$ and (\textit{d}) $(-3,1)+(6,0)$. The normalization takes the maximum value among the four forcings at each streamwise location. The white solid, dashed and dotted lines in the plots correspond with figure \ref{fig:mode_primary}. Contour bar ({\color[rgb]{0.8 0 0}$\blacksquare$}{\color[rgb]{0 0 0}$\blacksquare$}{\color[rgb]{0 0 0.8}$\blacksquare$}) ranges between $\pm1$.}
  \label{fig:forcing_contour}
\end{figure}

Figure \ref{fig:forcing_contour} shows the contours of representative forcing terms on the $\hat{u}$-equation of the first mode wave P1. These forcings includes the background forcing $\hat{\boldsymbol{f}}_0$ and the triadic counterparts from the interactions $(-7,1)+(10,0)$, $(0,2)+(3,-1)$ and $(-3,1)+(6,0)$. These are chosen based on their dominant role at certain streamwise locations, which will be shown later in this section. \par
Despite the relatively low initial amplitudes, multiple nonlinear interactions are identified to give rise to forcings at comparable levels. This phenomenon can potentially lead to the rapid broadening of spectra. Upstream of the transition onset $x=220$, the triadic contribution from $(-7,1)+(10,0)$ is the most pronounced. Downstream of the onset, the forcing amplitude of the triad $(0,2)+(3,-1)$ increases rapidly and becomes dominant at $x\approx250$, manifesting the feed-back from the secondary streak mode to the primary first mode. Due to the significant mean flow distortion, the $\hat{\boldsymbol{f}}_0$ also plays an important role in the moderate and late transitional stage.\par
\begin{figure}
  \centering
  \begin{subfigure}[t]{1.0\textwidth}\flushleft
  \subcaption{}
  \includegraphics[width=0.97\textwidth]{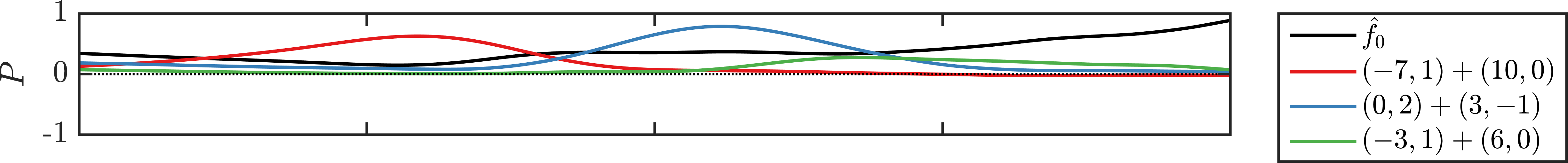}
  \end{subfigure}\vspace{-0.2cm}
  \begin{subfigure}[t]{1.0\textwidth}\flushleft
  \subcaption{}
  \includegraphics[width=0.97\textwidth]{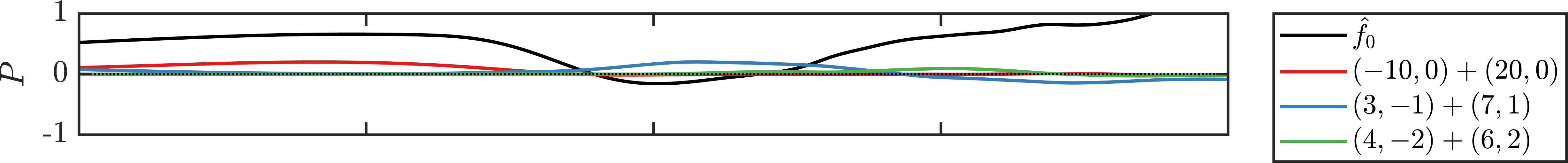}
  \end{subfigure}\vspace{-0.2cm}
  \begin{subfigure}[t]{1.0\textwidth}\flushleft
  \subcaption{}
  \includegraphics[width=0.97\textwidth]{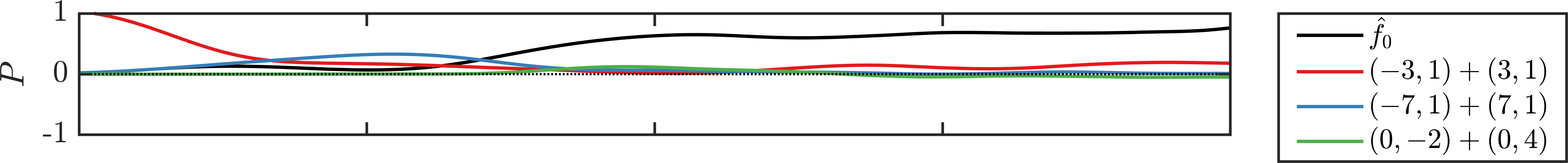}
  \end{subfigure}\vspace{-0.2cm}
  \begin{subfigure}[t]{1.0\textwidth}\flushleft
  \subcaption{}
  \includegraphics[width=0.97\textwidth]{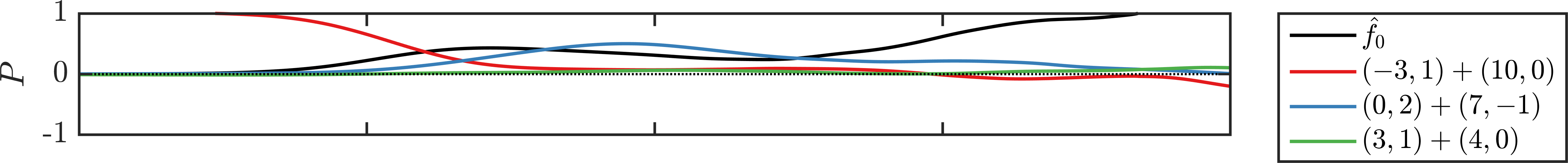}
  \end{subfigure}\vspace{-0.2cm}
  \begin{subfigure}[t]{1.0\textwidth}\flushleft
  \subcaption{}
  \includegraphics[width=0.97\textwidth]{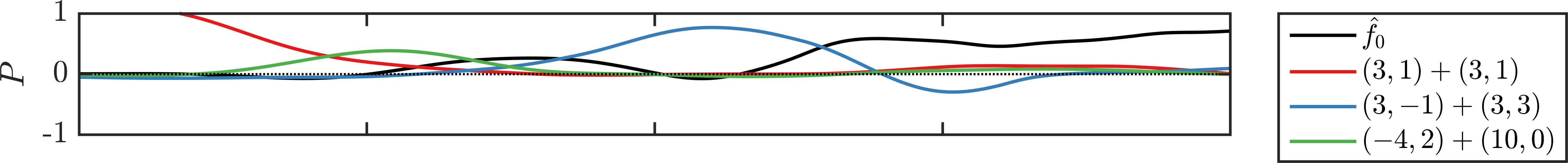}
  \end{subfigure}\vspace{-0.2cm}
  \begin{subfigure}[t]{1.0\textwidth}\flushleft
  \subcaption{}
  \includegraphics[width=0.97\textwidth]{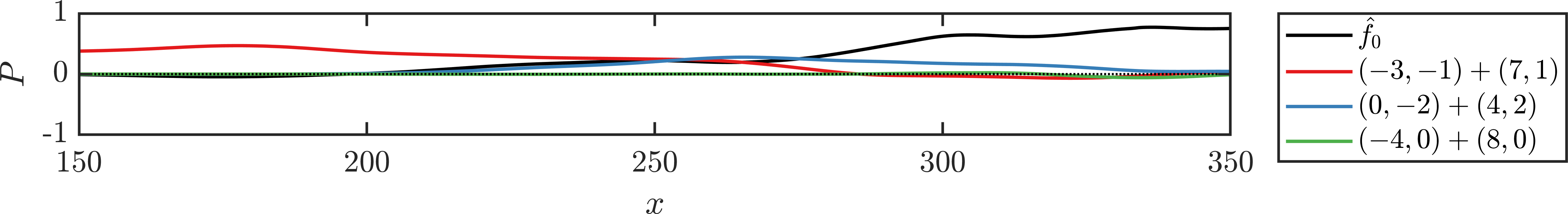}
  \end{subfigure}\vspace{-0.2cm}
  \caption{The projection coefficients of different forcing terms onto the total forcing of: (\textit{a}) The primary first mode P1--$(3,1)$, (\textit{b}) The primary second mode P2--$(10,0)$, (\textit{c}) The streak mode P1D--$(0,2)$, (\textit{d}) The difference mode P12D--$(7,1)$, (\textit{e}) The harmonic mode P1S--$(6,2)$ and (\textit{f}) The tertiary mode $(4,0)$.}
  \label{fig:projection_forcing}
\end{figure}

Note that the results in figure \ref{fig:forcing_contour} are only based on the linearized $u$-equation, which may not be able to fully involve the characteristics of different forcing terms. To address this issue, the vertical profile of each forcing is projected onto the profile of the total forcing:\par
\begin{equation}\label{P}
  P=\mathcal{R}\left(\frac{\left<\hat{\boldsymbol{N}}_{\mathit{qq}}^{(k)},\hat{\boldsymbol{f}}\right>_{\mathit{L}_2}}{\left<\hat{\boldsymbol{f}},\hat{\boldsymbol{f}}\right>_{\mathit{L}_2}}\right).
\end{equation}
Here, $\left<\cdot\right>_{\mathit{L}_2}$ denotes the inner product \citep{von_2024_role}. By taking the real part of the inner product, the phase relation between the two complex vectors is also at play. The projection coefficient of a selected triadic forcing onto the total forcing is calculated locally. Thus, the coefficient is depicted as a function of the streamwise location in figure \ref{fig:projection_forcing}.\par
The quantification of the triadic contribution associated with P1 in figure \ref{fig:projection_forcing}(\textit{a}) accords well with the discussion regarding figure \ref{fig:forcing_contour}. For the second mode P2 upstream of the transition onset in figure \ref{fig:projection_forcing}(\textit{b}), the MFD-associated $\hat{\boldsymbol{f}}_0$ and the harmonic feedback forcing from $(-10,0)+(20,0)$ are the first and second leading factors, respectively. The forcing from the triad $(3,-1)+(7,1)$ then increases downstream of the onset. In the moderate and late transitional stages, the MFD-associated triadic interaction dominates due to the large-amplitude distortion of the mean flow, resembling what has been observed for P1.\par
For forcings of higher-order waves, a common feature upstream of the transition onset is the existence of a leading, specific nonlinear triad, as shown in figure \ref{fig:projection_forcing}(\textit{a-f}). These dominant triadic forcings may either be responsible for the direct growth of higher-order waves or serve as the initial seed to be convected by the base flow. Further downstream, the forcings on different waves can be characterized by complex combinations of triadic interactions, given the fact of the spectra broadening in the transitional flow. In moderate and late transitional stages, the prominent contribution from the background forcing $\hat{\boldsymbol{f}}_0$ is the shared observation for all higher-order waves.\par

\subsection{Responses to triadic interactions}\label{sec:response}
\begin{figure}
  \centering
  \begin{subfigure}[t]{0.49\textwidth}\flushleft
  \subcaption{}
  \includegraphics[width=0.97\textwidth]{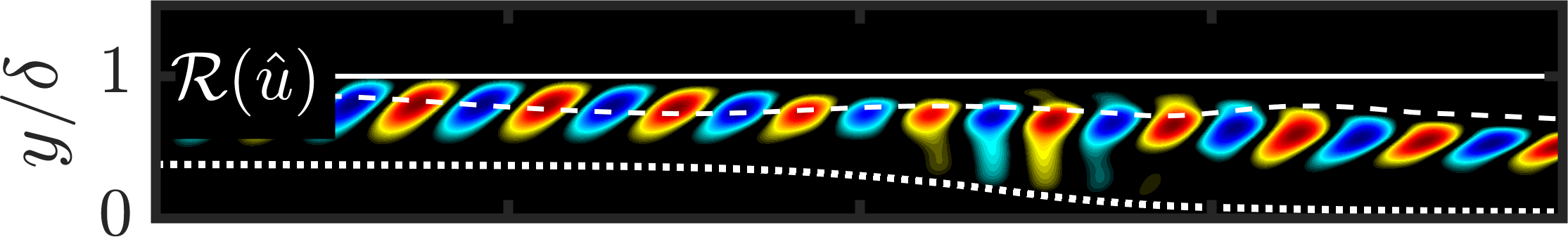}
  \end{subfigure}\vspace{-0cm}
  \begin{subfigure}[t]{0.49\textwidth}\flushleft
  \subcaption{}
  \includegraphics[width=0.97\textwidth]{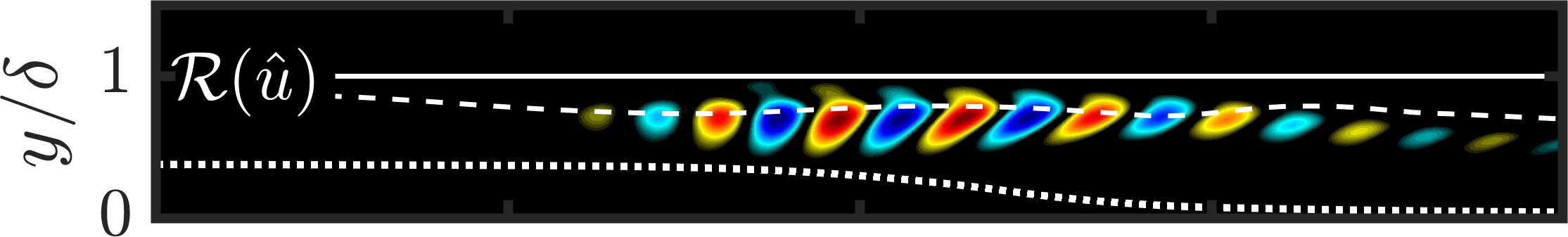}
  \end{subfigure}\vspace{-0cm}
  \begin{subfigure}[t]{0.49\textwidth}\flushleft
  \subcaption{}
  \includegraphics[width=1.0\textwidth]{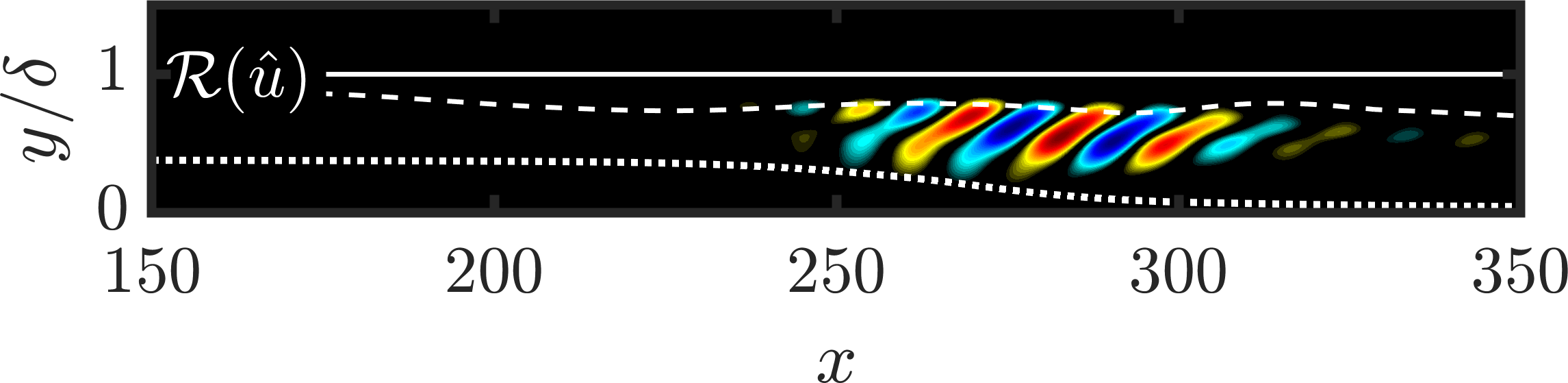}
  \end{subfigure}\vspace{-0cm}
  \begin{subfigure}[t]{0.49\textwidth}\flushleft
  \subcaption{}
  \includegraphics[width=1.0\textwidth]{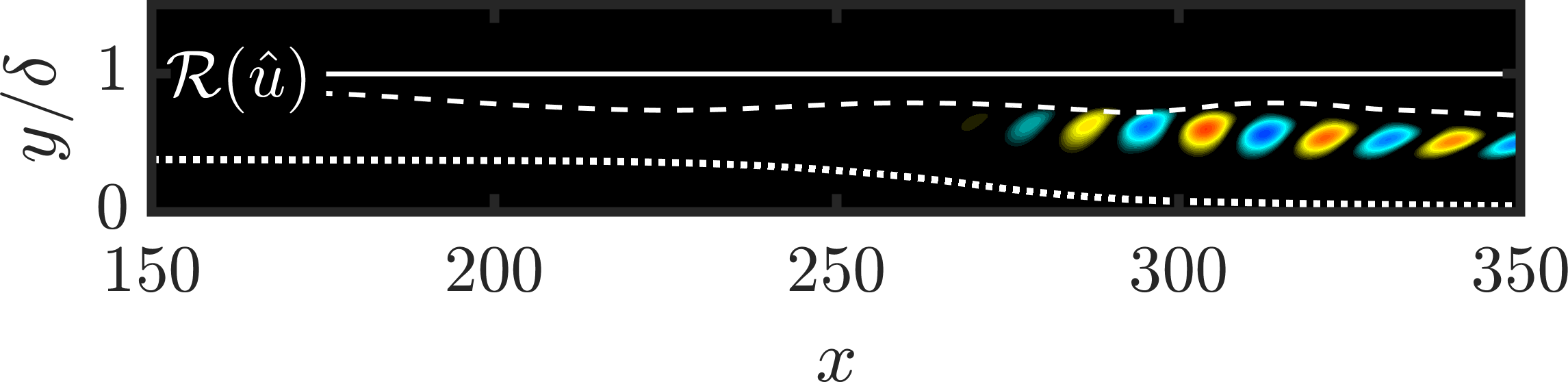}
  \end{subfigure}\vspace{-0cm}
  \caption{Response to different nonlinear forcing terms of P1--$(3,1)$, visualized by the real part of the normalized $\hat{u}$-component: (\textit{a}) $\hat{\boldsymbol{f}}_0$; (\textit{b}) $(-7,1)+(10,0)$; (\textit{c}) $(0,2)+(3,-1)$; (\textit{d}) $(-3,1)+(6,0)$. The normalization takes the maximum value among the four response modes at each streamwise location. The white solid, dashed and dotted lines in the plots correspond to figure \ref{fig:mode_primary}. Contour bar ({\color[rgb]{0.8 0 0}$\blacksquare$}{\color[rgb]{0 0 0}$\blacksquare$}{\color[rgb]{0 0 0.8}$\blacksquare$}) ranges between $\pm1$.}
  \label{fig:response_contour}
\end{figure}

Resembling figure \ref{fig:forcing_contour}, the responses to the respective forcing terms are visualized for the first mode P1 in figure \ref{fig:response_contour}. Due to the external upstream forcing incorporated in $\hat{\boldsymbol{f}}_0$, the response manifests the linear instability feature, and $\hat{\boldsymbol{f}}_0$ dominates the laminar-flow stage. In the transitional stage, the nonlinear forcings are amplified to reshape the response pattern. As shown in figure \ref{fig:response_contour}, the responses associated with (\textit{b}) the difference mode P12D, (\textit{c}) the stationary streak mode P1D and (\textit{d}) the first-mode harmonic successively exhibit significance.\par
\begin{figure}
  \centering
  \begin{subfigure}[t]{1.0\textwidth}\flushleft
  \subcaption{}
  \includegraphics[width=0.97\textwidth]{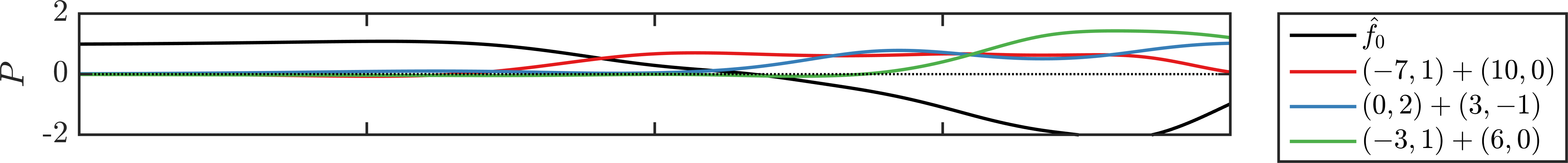}
  \end{subfigure}\vspace{-0.2cm}
  \begin{subfigure}[t]{1.0\textwidth}\flushleft
  \subcaption{}
  \includegraphics[width=0.97\textwidth]{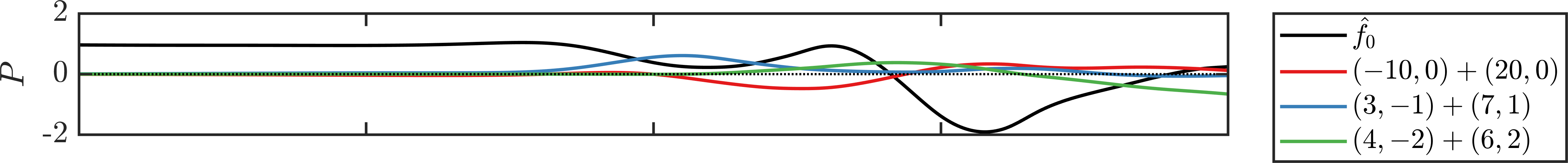}
  \end{subfigure}\vspace{-0.2cm}
  \begin{subfigure}[t]{1.0\textwidth}\flushleft
  \subcaption{}
  \includegraphics[width=0.97\textwidth]{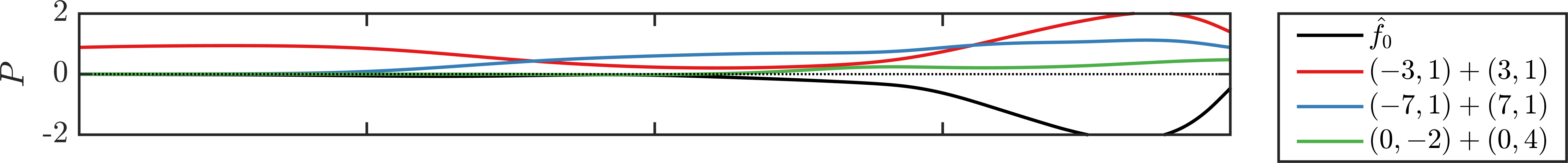}
  \end{subfigure}\vspace{-0.2cm}
  \begin{subfigure}[t]{1.0\textwidth}\flushleft
  \subcaption{}
  \includegraphics[width=0.97\textwidth]{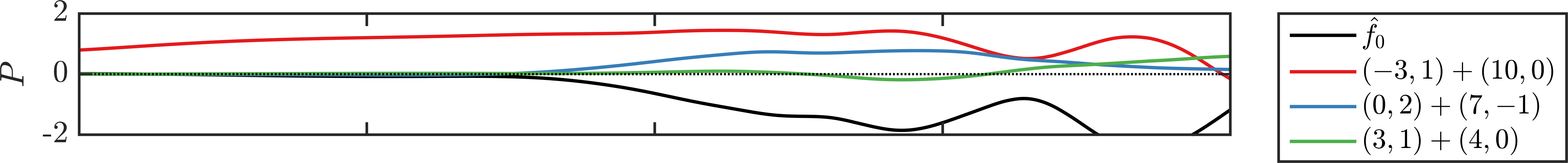}
  \end{subfigure}\vspace{-0.2cm}
  \begin{subfigure}[t]{1.0\textwidth}\flushleft
  \subcaption{}
  \includegraphics[width=0.97\textwidth]{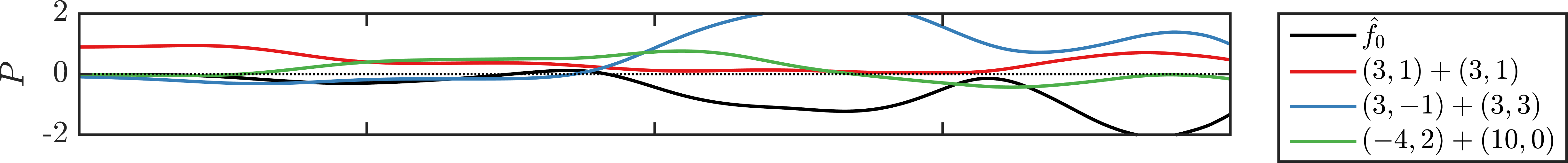}
  \end{subfigure}\vspace{-0.2cm}
  \begin{subfigure}[t]{1.0\textwidth}\flushleft
  \subcaption{}
  \includegraphics[width=0.97\textwidth]{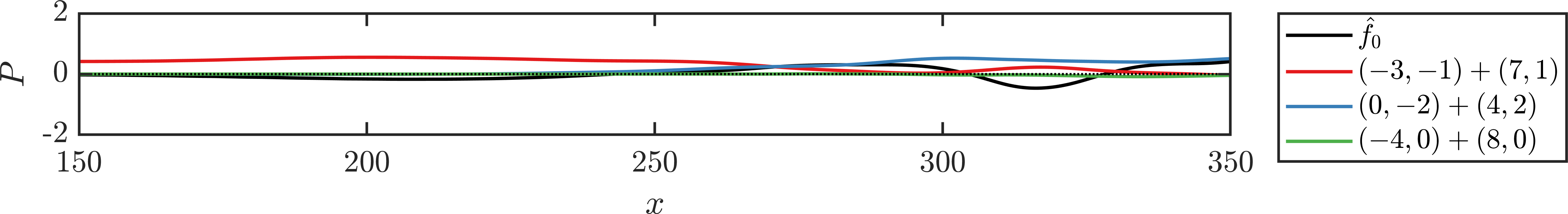}
  \end{subfigure}\vspace{-0.2cm}
  \caption{The projection coefficients of different response mode onto the associated DNS mode of: (\textit{a}) The primary first mode P1--$(3,1)$, (\textit{b}) The primary second mode P2--$(10,0)$, (\textit{c}) The streak mode P1D--$(0,2)$, (\textit{d}) The difference mode P12D--$(7,1)$, (\textit{e}) The harmonic mode P1S--$(6,2)$ and (\textit{f}) The tertiary mode $(4,0)$.}
  \label{fig:projection_response}
\end{figure}
Further quantification of the contributions from selected triadic interactions is made based on the projection coefficients. Similar to the definition in equation (\ref{P}), the coefficient is computed by projecting the response mode $\hat{\boldsymbol{q}}^{(\mathit{k})}_{\mathit{m,n}}$ onto the actual DNS mode $\hat{\boldsymbol{q}}_{\mathit{m,n}}$, as defined in equation (\ref{q_decompose}). In figure \ref{fig:projection_response}, the projection coefficients are depicted.
For the two primary waves, the response to triadic forcings can be regarded as a less important factor upstream of the transition onset. There, the features of primary waves are mostly determined by the base flow. By contrast, higher-order waves manifest solely in response to the identified dominant forcings in figure \ref{fig:projection_forcing}.\par
In the moderate and late transitional stages, one triadic interaction may be observed to contribute differently to the total forcing and the DNS mode at any fixed streamwise location. This can be found by comparing the results for several representative triads shown in figure \ref{fig:projection_response} and figure \ref{fig:projection_forcing}. For instance, for P1, $(-3,1)+(6,0)$ in the range $x>300$ contributes largely to the actual DNS mode in figure \ref{fig:projection_response}(\textit{a}) while not evidently to the total forcing in figure \ref{fig:projection_forcing}(\textit{a}); for P12D, $(-3,1)+(10,0)$ in $x>220$ proves pronounced to the DNS mode in figure \ref{fig:projection_response}(\textit{d}) while not to the total forcing in figure \ref{fig:projection_forcing}(\textit{d}). This observation reveals that, for any given Fourier mode, the associated forcing terms may not be equally transferred to the response by the resolvent operator. In fact, the response mode $\hat{\boldsymbol{q}}^{(\mathit{k})}_{\mathit{m,n}}$ defined in equation (\ref{q_decompose}) comprises the contributions from both the (linear and nonlinear) forcings and the base-flow effect. The base-flow one arises from the resolvent operator as defined in equation (\ref{IO}\textit{b}). Therefore, the unequal amplifications of forcings via the resolvent operator further highlights the importance of the base-flow modulation effect during the development of perturbation waves.\par
\begin{figure}
  \centering
  \begin{subfigure}[t]{0.80\textwidth}
  \subcaption{}
  \includegraphics[width=0.97\textwidth]{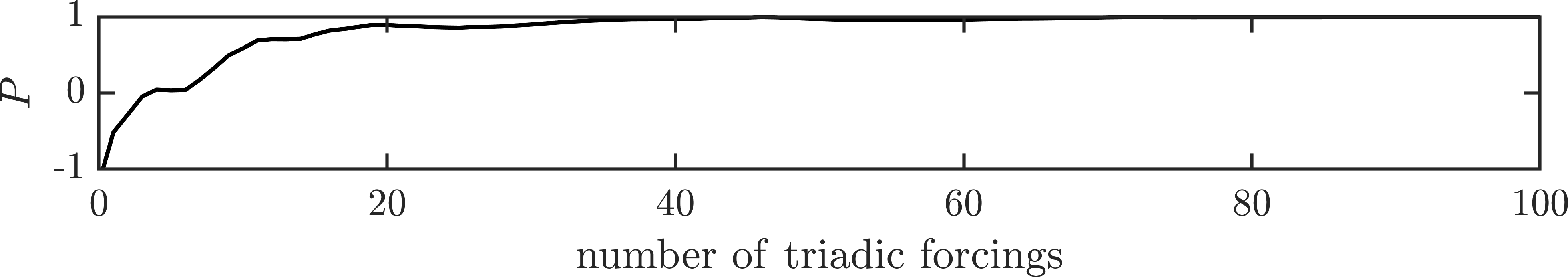}
  \end{subfigure}\vspace{-0cm}
  \begin{subfigure}[t]{0.49\textwidth}\flushleft
  \subcaption{}
  \includegraphics[width=0.97\textwidth]{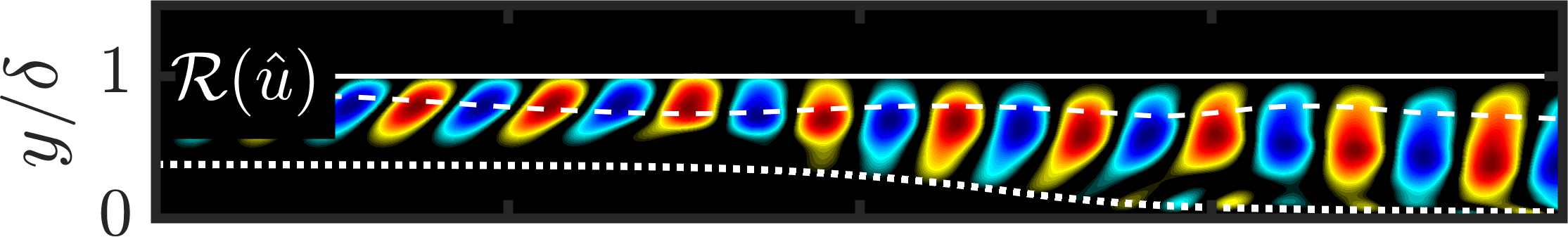}
  \end{subfigure}\vspace{-0cm}
  \begin{subfigure}[t]{0.49\textwidth}\flushleft
  \subcaption{}
  \includegraphics[width=0.97\textwidth]{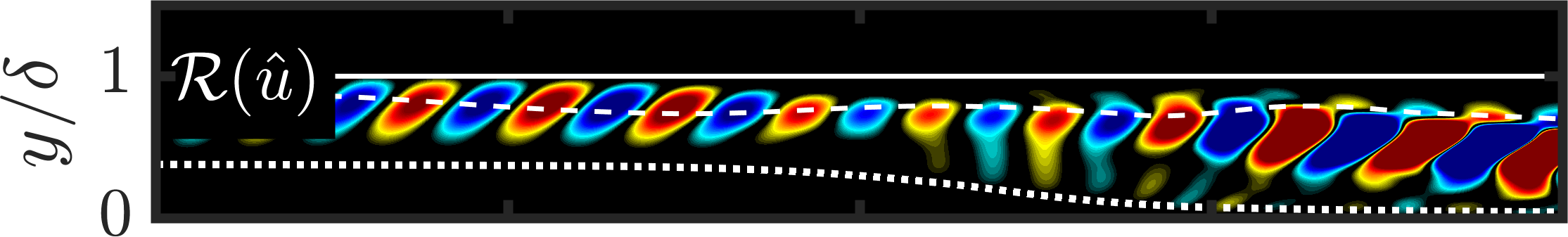}
  \end{subfigure}\vspace{-0cm}
  \begin{subfigure}[t]{0.49\textwidth}\flushleft
  \subcaption{}
  \includegraphics[width=1.0\textwidth]{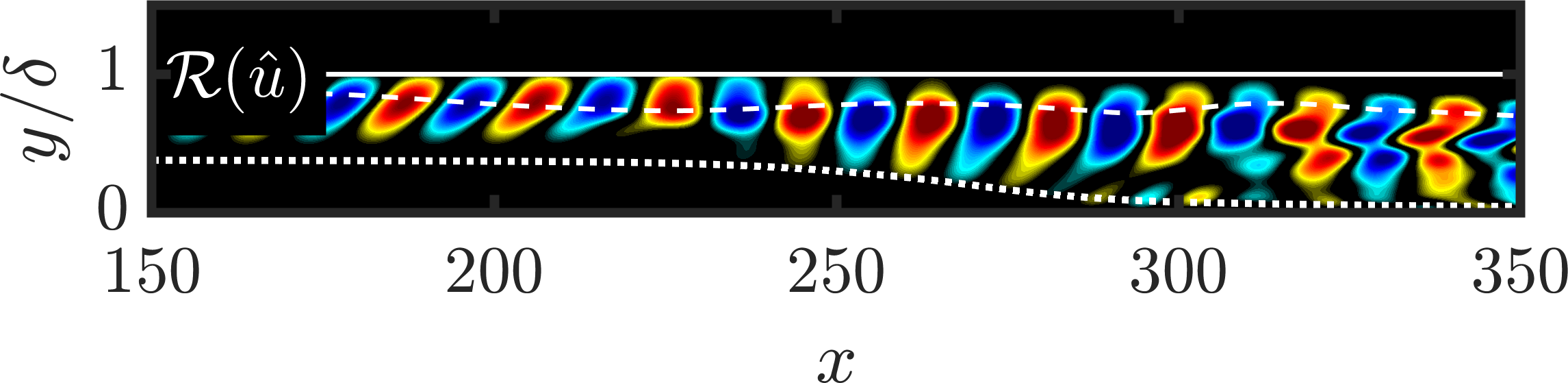}
  \end{subfigure}\vspace{-0cm}
  \begin{subfigure}[t]{0.49\textwidth}\flushleft
  \subcaption{}
  \includegraphics[width=1.0\textwidth]{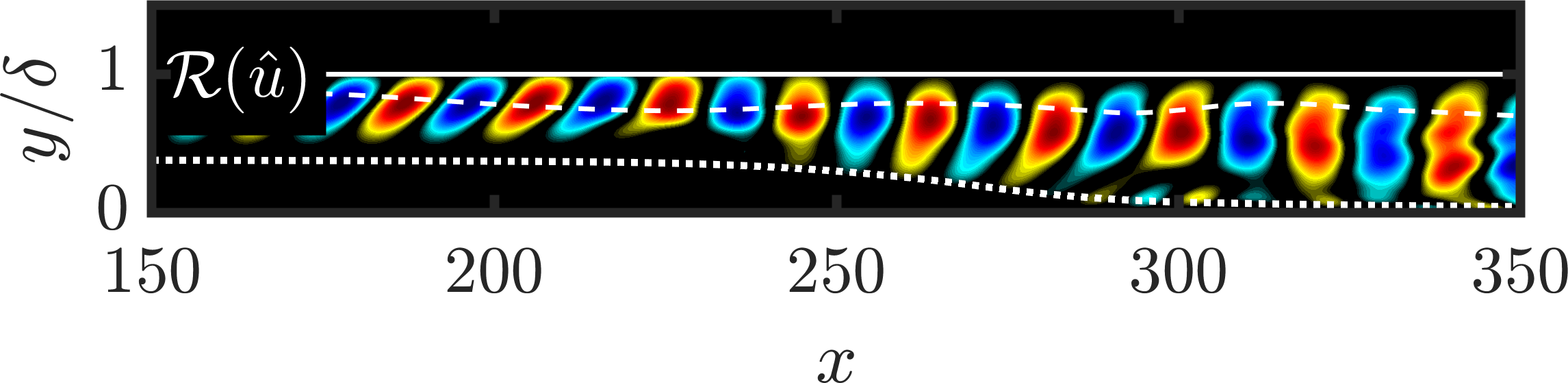}
  \end{subfigure}\vspace{-0cm}
  \caption{Comparison of the normalized $\hat{u}$-components between the accumulated response mode and the DNS mode for P1--$(3,1)$. Panel (\textit{a}) shows the projection coefficient of the accumulated response mode onto the DNS mode as a function of the number of triadic forcings, with 0 triadic forcing being only subject to $\hat{\boldsymbol{f}}_0$. Panels (\textit{b--e}) show the real parts of the normalized $\hat{u}$-components of the DNS mode and the respective accumulated response modes with 0, 20 and 60 leading triadic forcings. The result is normalized by the maximum of the DNS mode at each streamwise location. The white solid, dashed and dotted lines in the plots correspond with figure \ref{fig:mode_primary}. Contour bar ({\color[rgb]{0.8 0 0}$\blacksquare$}{\color[rgb]{0 0 0}$\blacksquare$}{\color[rgb]{0 0 0.8}$\blacksquare$}) ranges between $\pm1$.}
  \label{fig:response_accumulated}
\end{figure}

As the transitional flow develops along the streamwise direction, the number of the involving mode--mode interactions is increased. Hence, as the transition progresses, an increasing number of triadic forcings may be required to recover the main feature of the total response. Here, the primary first mode P1 is examined to find out the minimum number of `skeleton interactions', whose forcings can given sufficient response to reconstruct the transitional flow field. The triadic forcings are first ranked based on their integrated Chu's energy within $150\le x\le350$. Then the response modes are computed by solving equation (\ref{q_decompose}), using the sum of $\hat{\boldsymbol{f}}_0$ and the leading $k$ triadic forcings. The comparison between the accumulated response mode and the DNS mode is made in figure \ref{fig:response_accumulated}. Note that, different from figure \ref{fig:projection_forcing} and figure \ref{fig:projection_response}, the projection coefficient in figure \ref{fig:response_accumulated}(\textit{a}) is computed using the 2-D field re-arranged into the vector form, which quantifies the similarity in the global perspective.\par
Surprisingly, the nonlinear effect on the perturbation evolution can be distinct even upstream of the onset location, justified by the deviation between figure \ref{fig:response_accumulated}(\textit{b}) (DNS mode) and figure \ref{fig:response_accumulated}(\textit{c}) (0 triadic forcing). The deviation starts early from $x\approx200$. Without the restriction from the nonlinear effect, the response to only $\hat{\boldsymbol{f}}_0$ shows a significantly higher amplitude in $x>300$. Note also in figure \ref{fig:energy} that the resolvent forcing $\mathsfbi{B}\hat{\boldsymbol{f}}$ leads to amplitude reduction compared with DNS. Therefore, the resulting magnitude of the response amplitude presents the relation: resolvent forcing $<$ DNS total forcing $<$ $\hat{\boldsymbol{f}}_0$. According to equations (\ref{f_0}) and(\ref{q_decompose}), the total DNS forcing comprises $\hat{\boldsymbol{f}}_0$ plus the triadic forcing, while $\hat{\boldsymbol{f}}_0$ consists of the resolvent forcing $\mathsfbi{B}\hat{\boldsymbol{f}}$ plus the MFD- and cubic-associated forcings. We then conclude from the above relation `DNS total forcing $<$ $\hat{\boldsymbol{f}}_0$' that the summed triadic forcing suppresses the P1 mode in the nonlinear stage $x>300$. We also summarize from `resolvent forcing $<$ DNS total forcing' that the MFD- and cubic-associated forcings support the secondary growth of P1 in this stage. \par
With 20 leading triadic forcings, despite minor discrepancies in the wave amplitude, the accumulated response mode shown in figure \ref{fig:response_accumulated}(\textit{d}) can well represent the perturbation pattern of the DNS result in $x<300$. This observation corresponds well with the quantification result in figure \ref{fig:response_accumulated}(\textit{a}), in which the projection coefficient of $P>0.9$ can be achieved with 20 triadic forcings. Further downstream, triadic forcings being ranked as `less important' can be sufficiently amplified by the base flow. The absence of these lesser forcings leads to the distortion of the total response. Based on the projection coefficient result, these high-rank triadic forcings can have either positive or negative contributions to the accumulated response mode. As a consequence, at least 60 triadic forcings may be required to reconstruct the flow in the moderate and late transitional stages.\par

\section{Growth and re-distribution of spectral energy}\label{sec:transfer}
In this section, the instability evolution is further analyzed from the perspective of energy transfer. Contributions from the mean-flow effect (linear transfer) as well as mode--mode interactions among scales (nonlinear transfer) will be isolated. We expect to elucidate the physical mechanism of energy re-distribution and spectral broadening, initiated by merely two seeded primary waves.\par
\subsection{Contributions from linear and nonlinear mechanisms}\label{sec:budget}
\begin{figure}
  \centering
  \begin{subfigure}[t]{0.40\textwidth}\flushleft
  \subcaption{}
  \includegraphics[width=0.966\textwidth]{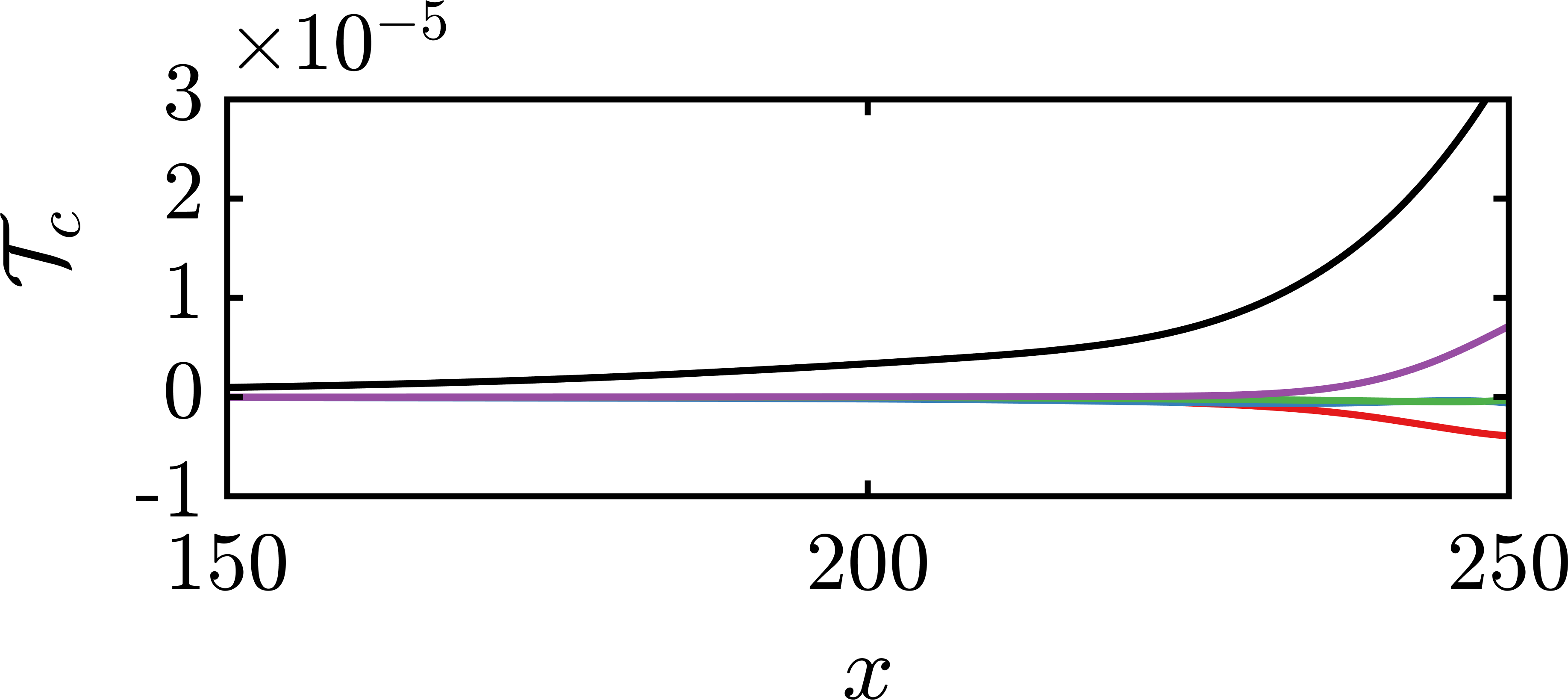}
  \end{subfigure}
  \begin{subfigure}[t]{0.59\textwidth}\flushright
  \subcaption{}
  \includegraphics[width=1\textwidth]{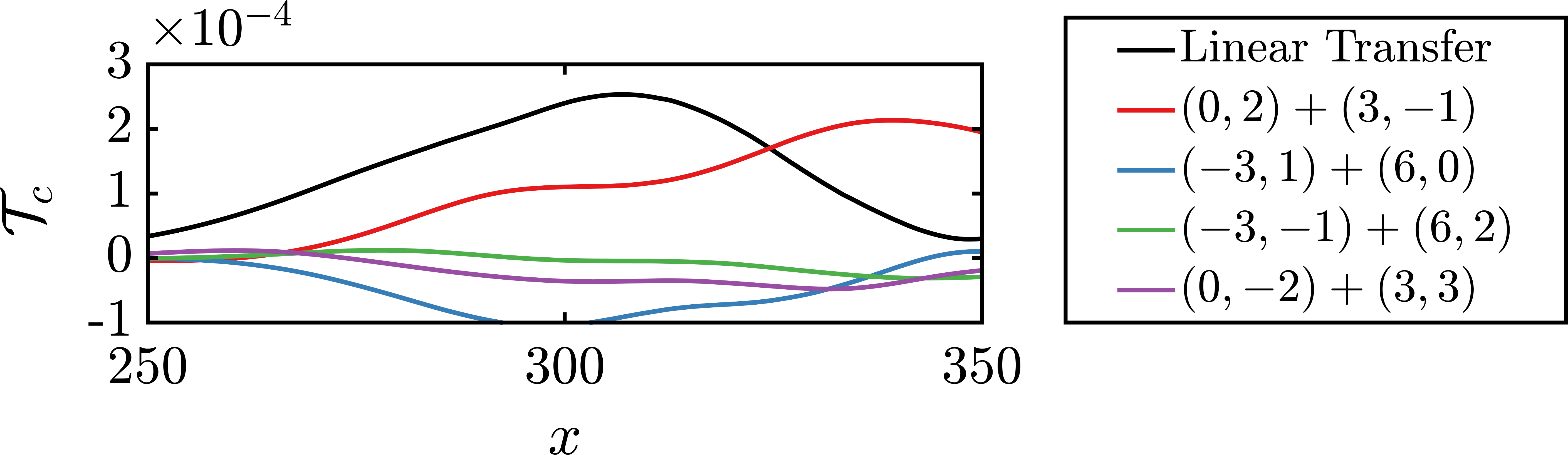}
  \end{subfigure}\vspace{-0.1cm}
  \begin{subfigure}[t]{0.40\textwidth}\flushleft
  \subcaption{}
  \includegraphics[width=0.966\textwidth]{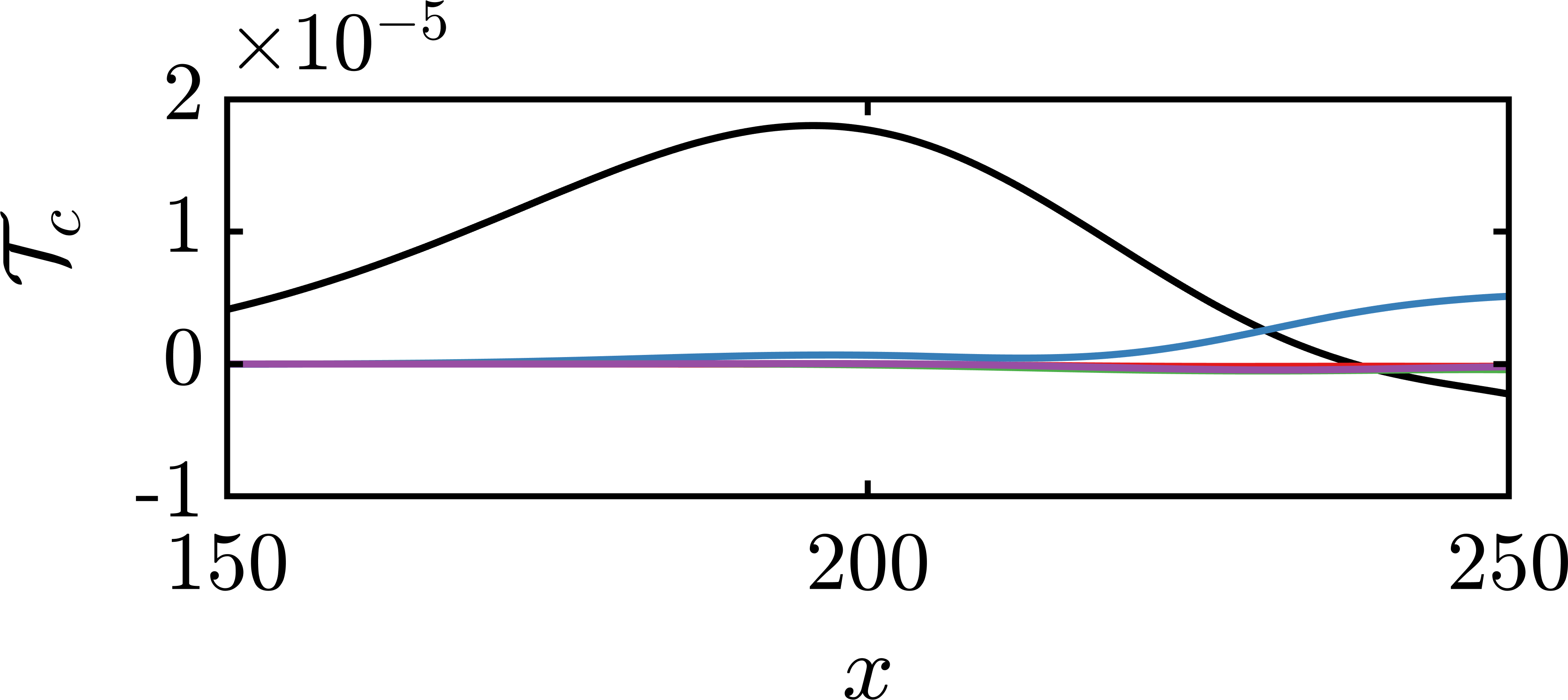}
  \end{subfigure}
  \begin{subfigure}[t]{0.59\textwidth}\flushright
  \subcaption{}
  \includegraphics[width=1\textwidth]{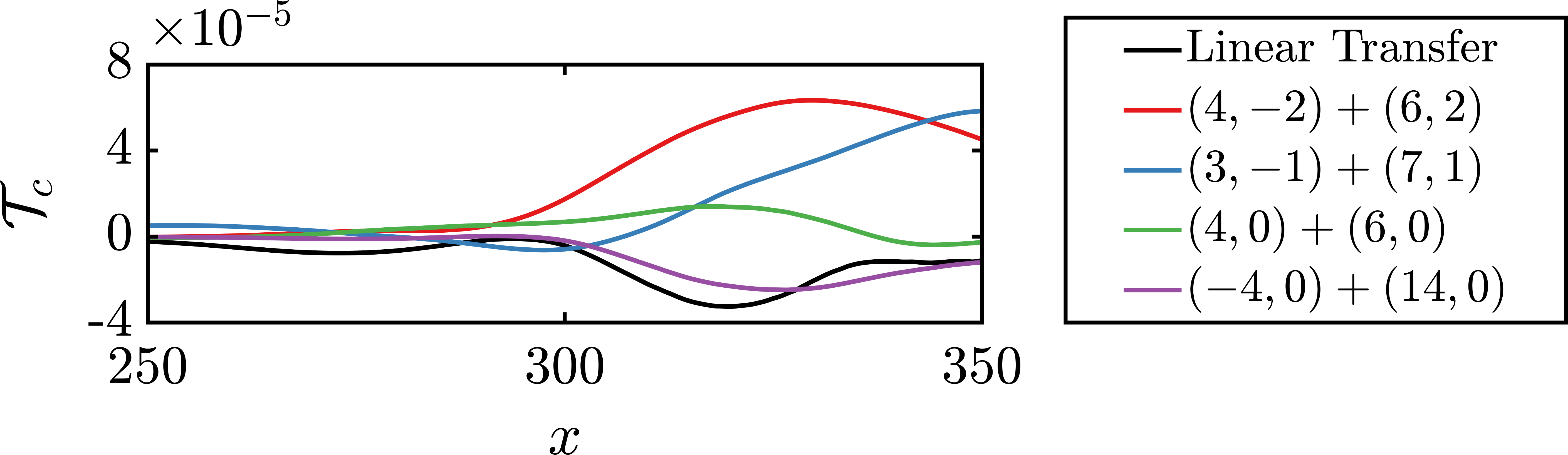}
  \end{subfigure}\vspace{-0.1cm}
  \begin{subfigure}[t]{0.40\textwidth}\flushleft
  \subcaption{}
  \includegraphics[width=0.966\textwidth]{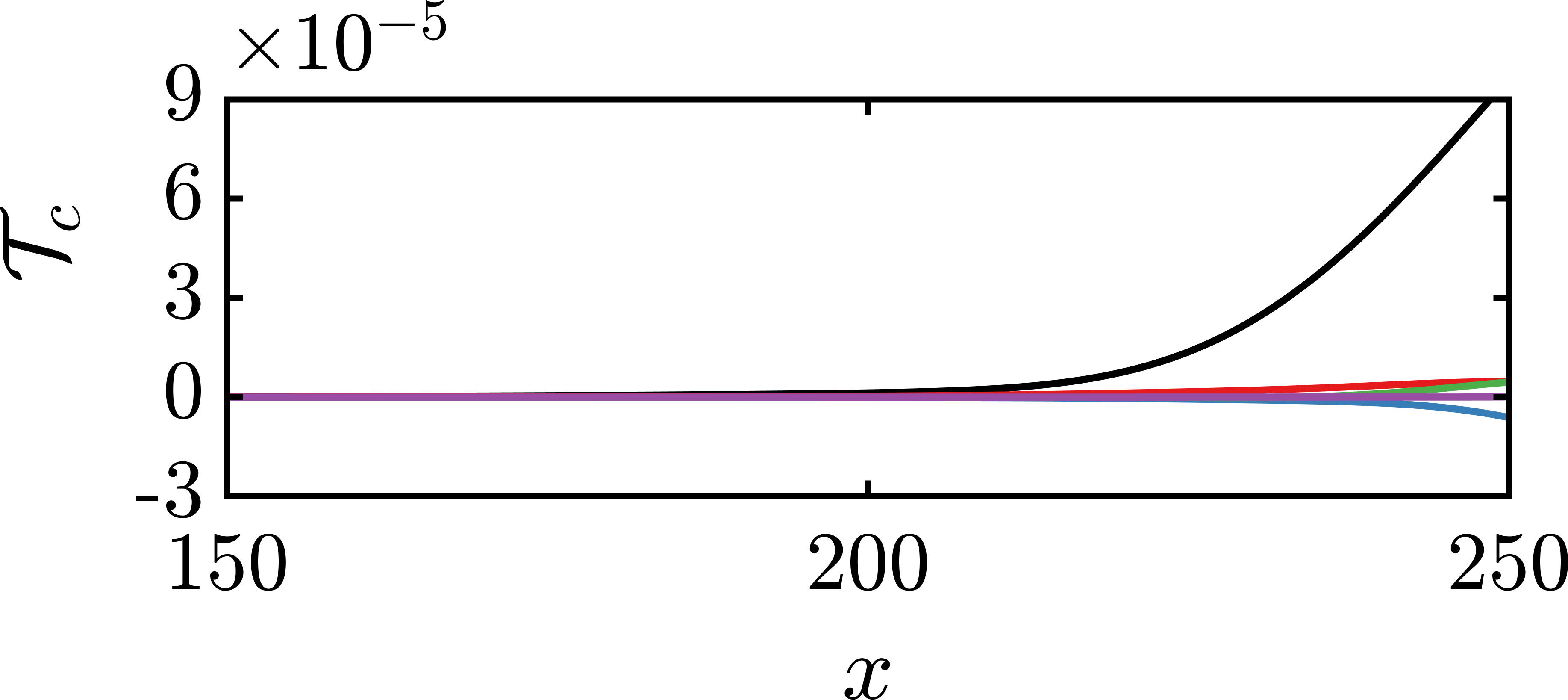}
  \end{subfigure}
  \begin{subfigure}[t]{0.59\textwidth}\flushright
  \subcaption{}
  \includegraphics[width=1\textwidth]{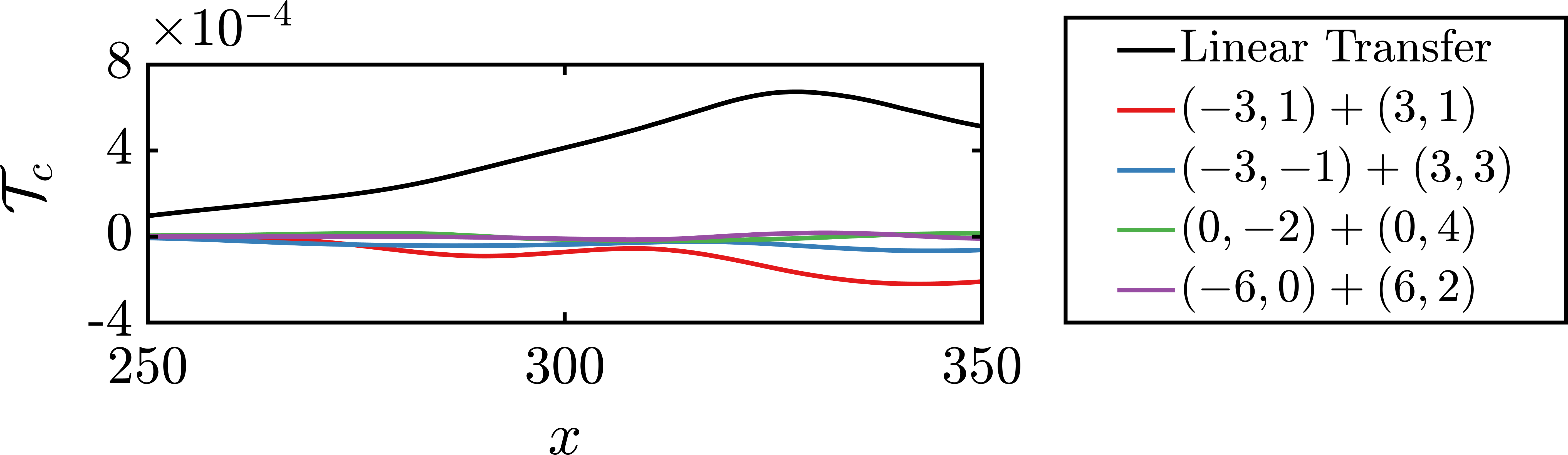}
  \end{subfigure}
  \caption{Integrated linear and nonlinear transfer to the Chu's energy as functions of streamwise location for (\textit{a,b}) the primary first mode P1--$(3,1)$, (\textit{c,d}) the primary second mode P2--$(10,0)$ and (\textit{e,f}) the streak mode P1D--$(0,2)$.}
  \label{fig:transfer}
\end{figure}

\begin{figure}
  \centering
  \begin{subfigure}[t]{0.49\textwidth}\flushleft
  \subcaption{The primary first mode P1--$(3,1)$}
  \includegraphics[width=1.0\textwidth]{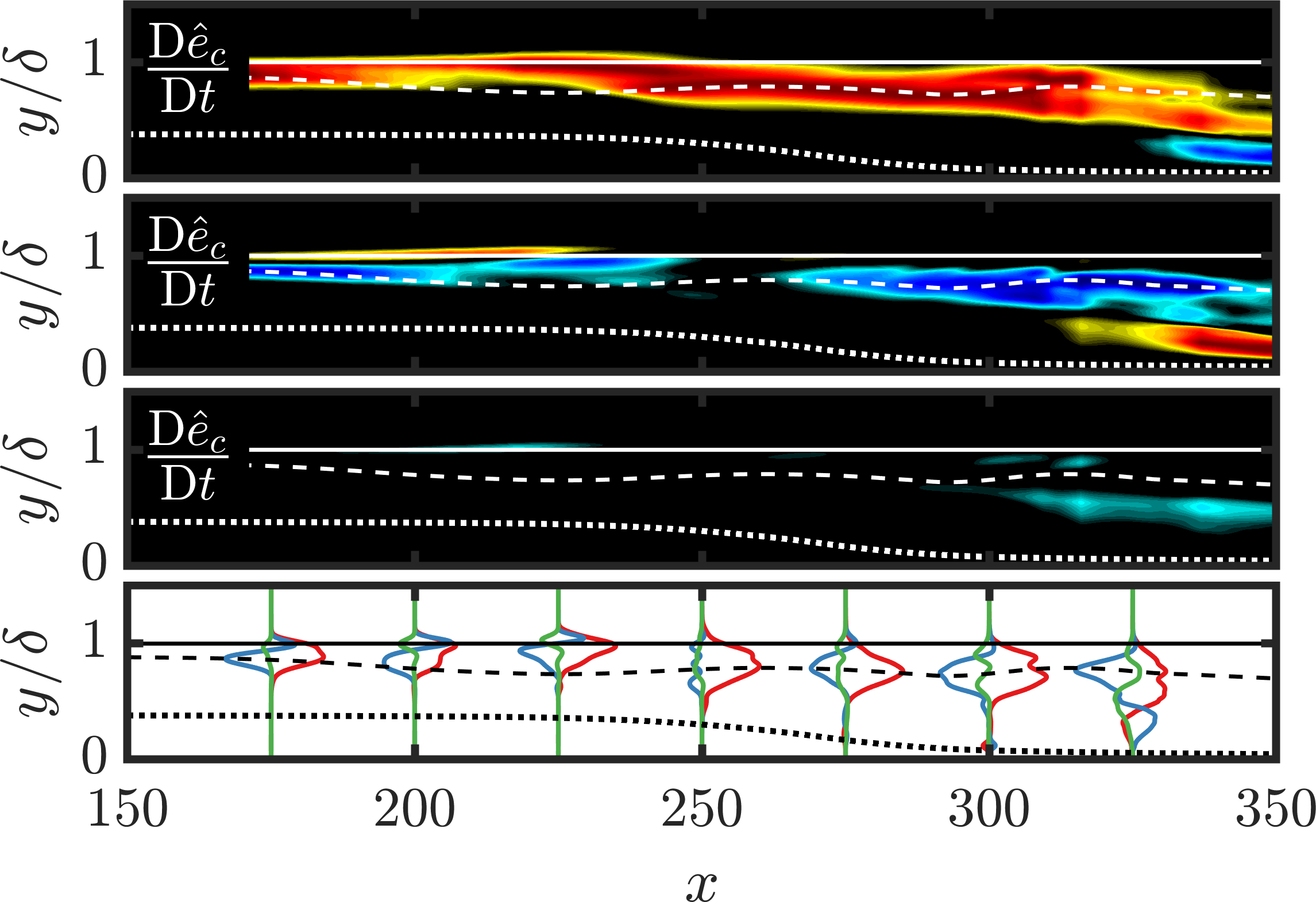}
  \end{subfigure}\vspace{-0cm}
  \begin{subfigure}[t]{0.49\textwidth}\flushleft
  \subcaption{The primary second mode P2--$(10,0)$}
  \includegraphics[width=1.0\textwidth]{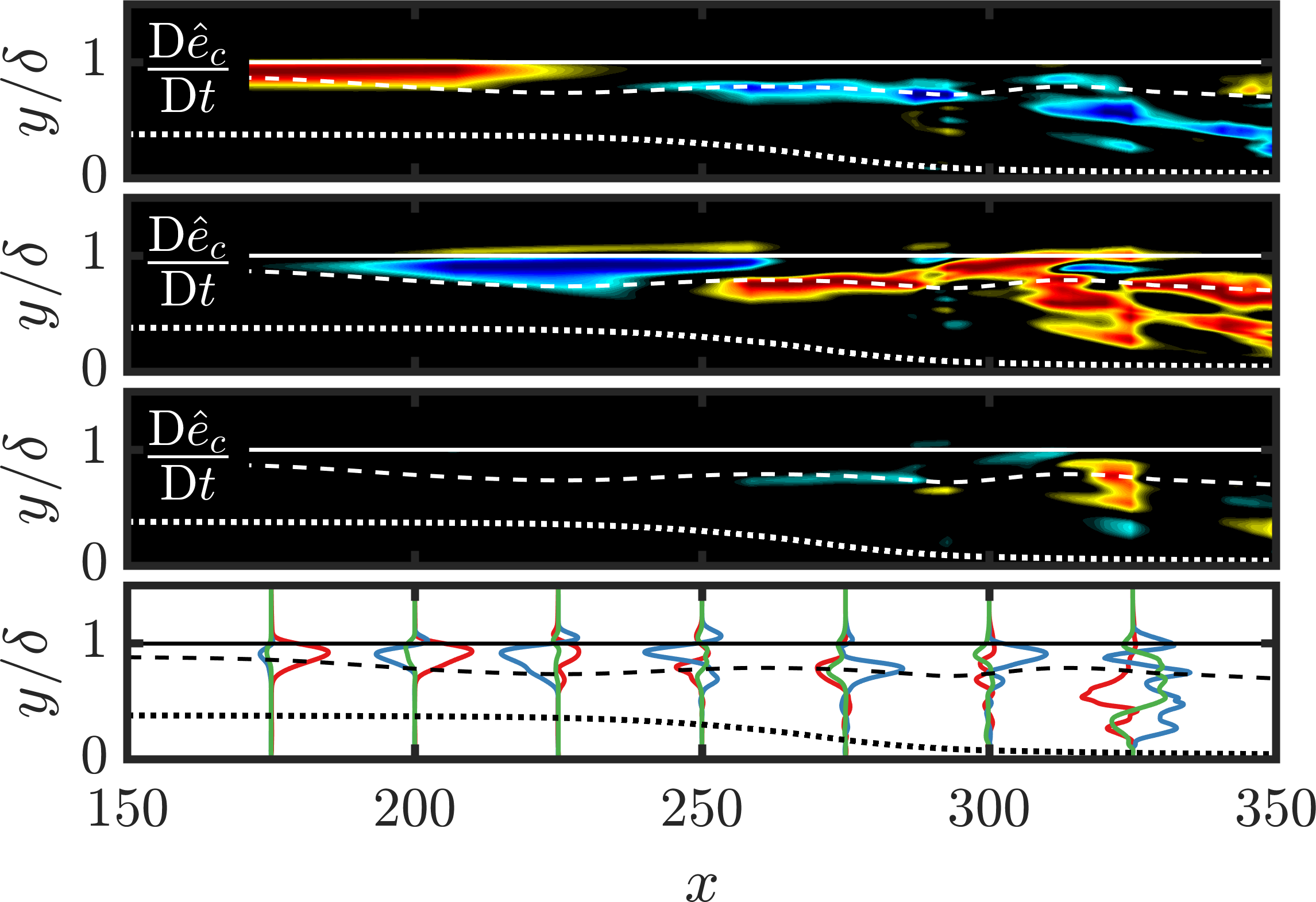}
  \end{subfigure}\vspace{-0cm}
  \begin{subfigure}[t]{0.49\textwidth}\flushleft
  \subcaption{The streak mode P1D--$(0,2)$}
  \includegraphics[width=1.0\textwidth]{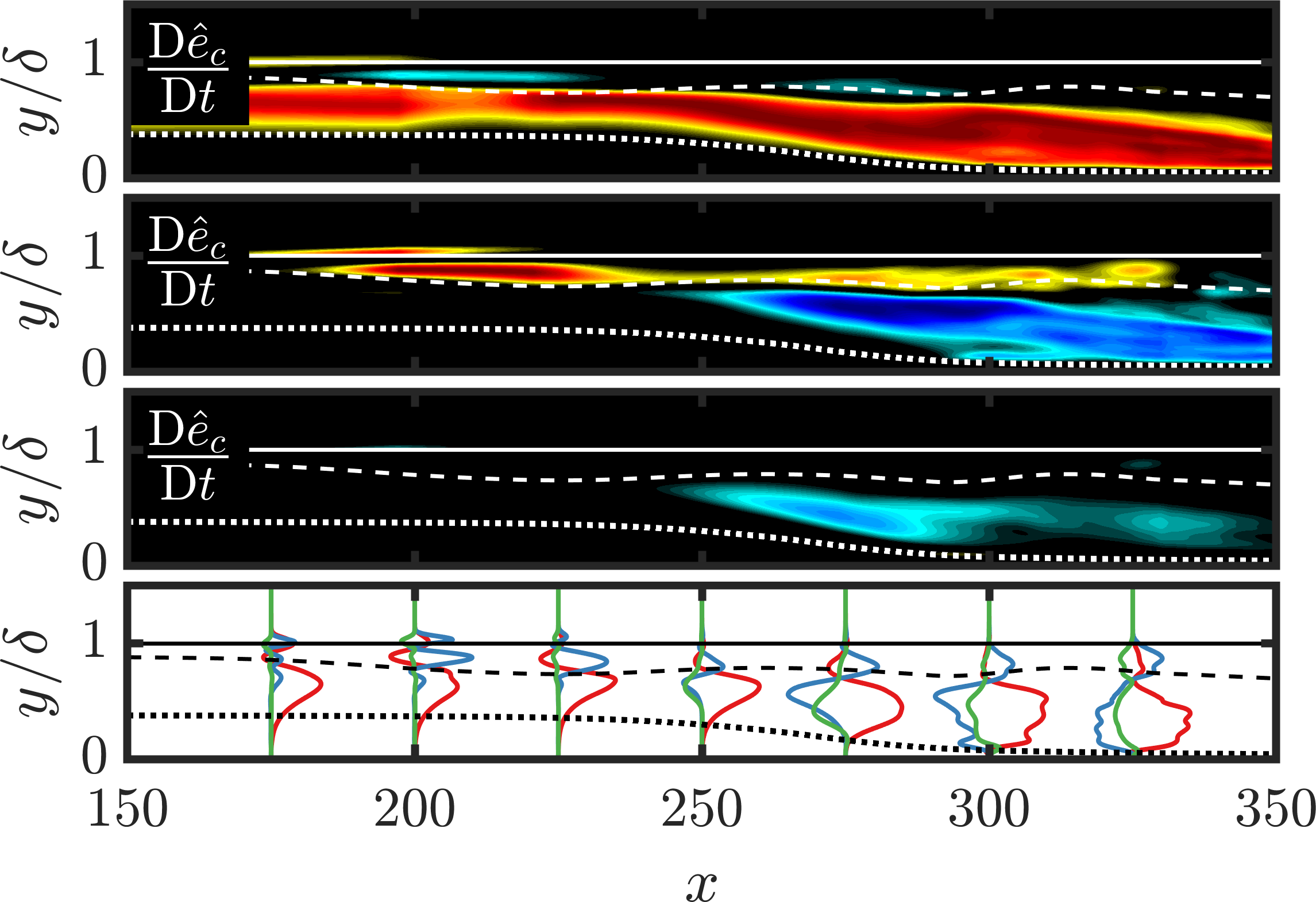}
  \end{subfigure}\vspace{-0cm}
  \caption{The normalized linear, net quadratic nonlinear and net cubic nonlinear transfer to the Chu's energy for different modes, visualized by their contour plots and the comparison between their profiles ({\color[rgb]{0.89 0.10 0.11}---} for the linear transfer, {\color[rgb]{0.22 0.49 0.72}---} for the net quadratic nonlinear transfer, and {\color[rgb]{0.30 0.69 0.29}---} for the net cubic nonlinear transfer) at different streamwise locations. The energy transfers are normalized by the maximum value among the three terms at each streamwise location. The white or black solid, dashed and dotted lines in the plots correspond with figure \ref{fig:mode_primary}. Color bar ({\color[rgb]{0.8 0 0}$\blacksquare$}{\color[rgb]{0 0 0}$\blacksquare$}{\color[rgb]{0 0 0.8}$\blacksquare$}) ranges between $\pm1$.}
  \label{fig:Tc_contour}
\end{figure}

The contributions from both linear and nonlinear energy transfers are first compared. For the two primary waves and the streak, figure \ref{fig:transfer} illustrates the integrated linear and nonlinear transfer to the Chu's energy as a function of the streamwise coordinate. Here, the four leading nonlinear triads within the shown streamwise range are displayed. Note that the importance of nonlinear triads is ranked based on the integrated rate of change in energy $\int_{\mathit{S}}\left|{\rm{D}}\hat{e}_{\mathit{c,n}}/{{\rm{D}}t}\right|{\rm{d}}S,$ where the integral is taken in the domain $150<x<350$. Moreover, the linear and net nonlinear transfer (for the respective quadratic and cubic nonlinear terms) contours and profiles are plotted in figure \ref{fig:Tc_contour}. One can identify spatial extents with intense energy transfer and associated linear/nonlinear mechanisms.\par
In the initial linear growth stage, a competition scenario between the first and second modes is observed based on figure \ref{fig:transfer}(\textit{a,c}). Upstream of the transition onset $x=220$, the primary second mode P2 is identified with the most prominent linear gain, which aligns with the resolvent analysis. The region with the maximum energy gain is found in figure \ref{fig:Tc_contour}(\textit{b}) to be between the boundary layer edge and the GIP. This finding indicates that the outer thermodynamic growth overtakes the near-wall acoustic energy growth. The Chu's energy of the primary first mode P1 is also amplified in this region. Downstream of transition onset, the linear mechanism of P2 starts to decay and draws energy backward as $x$ exceeds 240. In the meantime, the linear energy supply to P1 continues to grow along the GIP (dashed line) and becomes the most significant one. The competition between the two primary waves has also been reported in both simulation and experimental studies \citep{chen_2017_interactions,qiu_2024_boundary}. It is currently captured in a systematic quantification framework, involving different energy norms and detailed nonlinear sub-terms.\par
As the amplitudes of the two primary waves reach certain threshold, the quadratic nonlinear interaction comes into play. Upstream of the transition onset, the magnitude of net quadratic nonlinear transfer is already comparable to that of linear transfer for P1, as evidenced by figure \ref{fig:Tc_contour}(\textit{a}). For the second mode P2, the quadratic nonlinear interactions is a lesser mechanism upstream of its saturation location, as shown in figure \ref{fig:Tc_contour}(\textit{b}). One may also notice in the profile plot that the nonlinear term is at play more upstream for P1 than P2. \par
Downstream of the transition onset, the streak mode P1D signifying the oblique breakdown mechanism \citep{mayer_2011_direct,franko_2013_breakdown}, is notable in figure \ref{fig:transfer}(\textit{a},\textit{e}). According to figure \ref{fig:Tc_contour}(\textit{a,c}), the region where energy is nonlinearly transferred from P1 to P1D is located between the boundary layer edge and the GIP. Then the streak P1D grows continuously via the linear lift-up mechanism \citep{hanifi_1996_transient,brandt_2014_lift,laible_2016_continuously}, and feeds energy back to the first mode P1 starting from $x=260$ in figure \ref{fig:transfer}(\textit{b},\textit{f}). This nonlinear feedback process is detected in figure \ref{fig:Tc_contour}(\textit{c}) to be the most prominent inside the streak structure, leading to the secondary growth of P1 as stated for figure \ref{fig:energy}. This finding in the high-speed boundary layer coincides with the concept `self-sustaining cycle of the streak' in other low-speed shear flows \citep{waleffe_1997_self,dessup_2018_self,nekkanti_2025_bispectral}.\par
Apart from the interactions associated with the streak, the self interaction of P1 around the GIP also generates the harmonic mode P1S. This mode can draw energy from the primary oblique wave P1 (not shown). For the primary second mode P2 near the saturation point, the quadratic nonlinear transfer grows to a level comparable to the linear mechanism. Farther downstream ($300<x<350$), quadratic nonlinear interactions are found to be important mechanisms resulting in the secondary growth of P2, as shown by figure \ref{fig:energy}. Figure \ref{fig:transfer}(\textit{d}) reveals that the interaction between the tertiary and secondary instabilities, $(4,2)+(6,-2)$, becomes the leading one within $300<x<350$, followed subsequently by the triad $(3,1)+(7,-1)$ triggering P12D. As far as we know, the significant role of tertiary instabilities in triggering the secondary growth of Mack second mode in the transitional stage has not been reported before.  \par
Regarding the cubic nonlinear terms, figure \ref{fig:transfer} shows that the associated net energy transfers can be escalated to be significant enough in the moderate transitional stage. However, compared to the linear and net quadratic nonlinear transfer terms, the energy re-distribution due to the cubic nonlinearity is still a lesser mechanism. The quadratic nonlinear transfer will be focused on.\par

\subsection{Mode--mode interactions: dominant nonlinear triads}\label{sec:triads}
\begin{figure}
  \centering
  \begin{subfigure}[t]{0.39\textwidth}\centering \small energy gain
  \end{subfigure}
  \begin{subfigure}[t]{0.39\textwidth}\centering \small energy loss
  \end{subfigure}
  \begin{subfigure}[t]{0.20\textwidth}\centering \ 
  \end{subfigure}\vspace{-0.4cm}
  \begin{subfigure}[t]{0.39\textwidth}\flushleft
  \subcaption{}
  \includegraphics[width=0.967\textwidth]{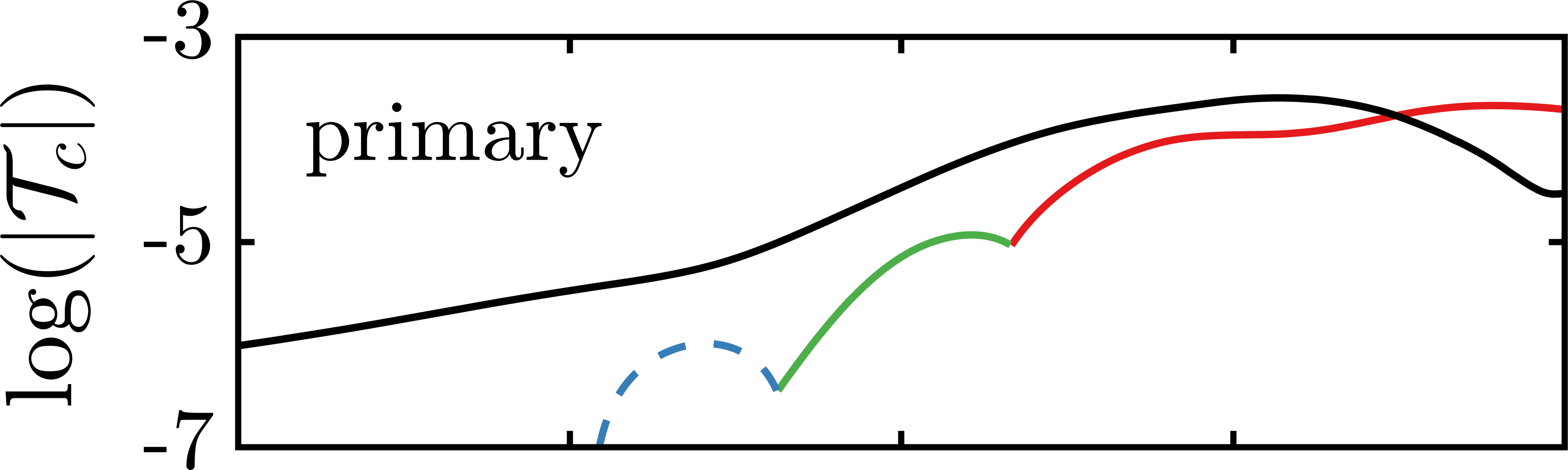}
  \end{subfigure}
  \begin{subfigure}[t]{0.60\textwidth}\flushright
  \subcaption{}
  \includegraphics[width=1\textwidth]{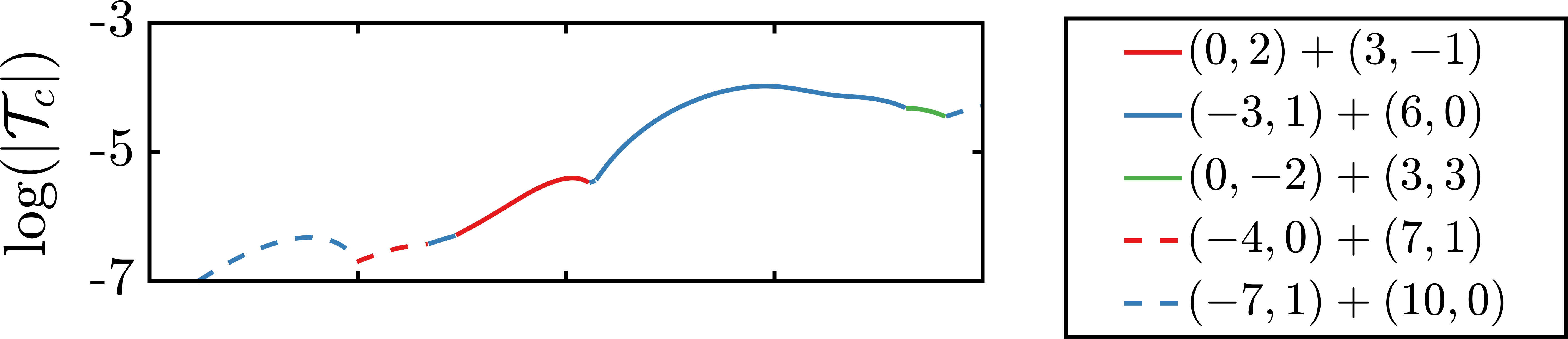}
  \end{subfigure}\vspace{-0.2cm}
  \begin{subfigure}[t]{0.39\textwidth}\flushleft
  \subcaption{}
  \includegraphics[width=0.967\textwidth]{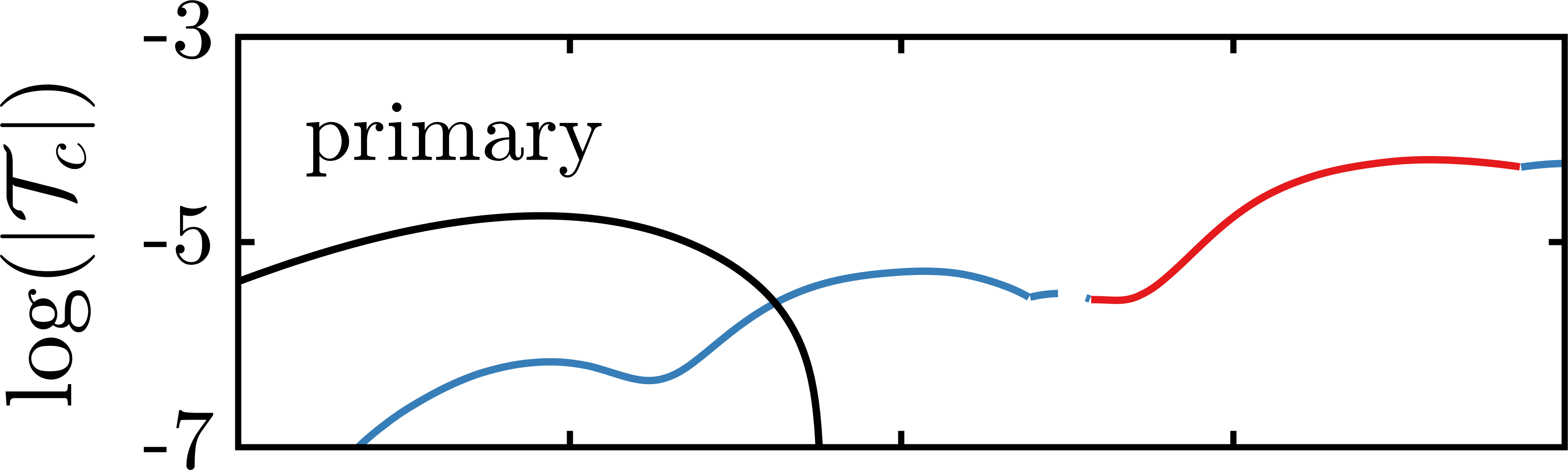}
  \end{subfigure}
  \begin{subfigure}[t]{0.60\textwidth}\flushright
  \subcaption{}
  \includegraphics[width=1\textwidth]{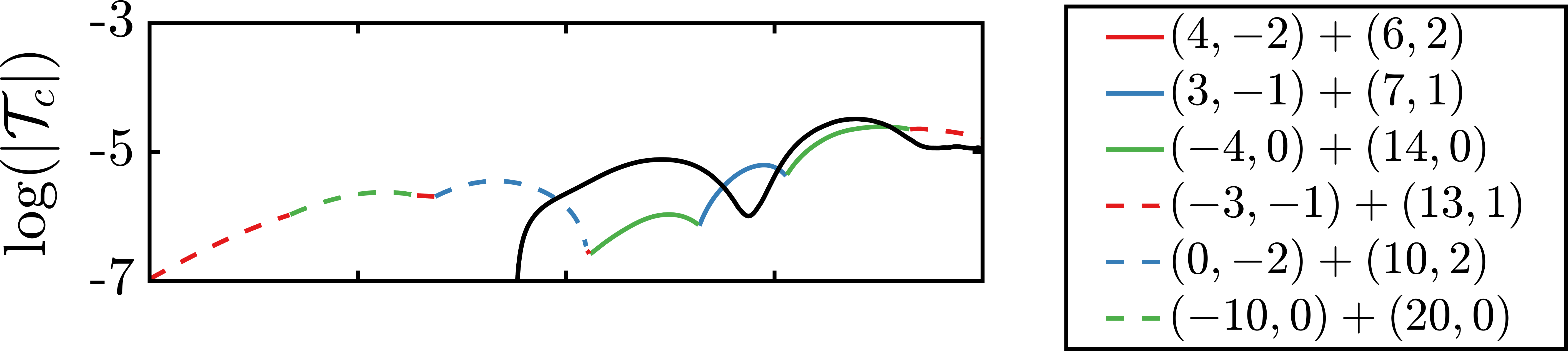}
  \end{subfigure}\vspace{-0.2cm}
  \begin{subfigure}[t]{0.39\textwidth}\flushleft
  \subcaption{}
  \includegraphics[width=0.967\textwidth]{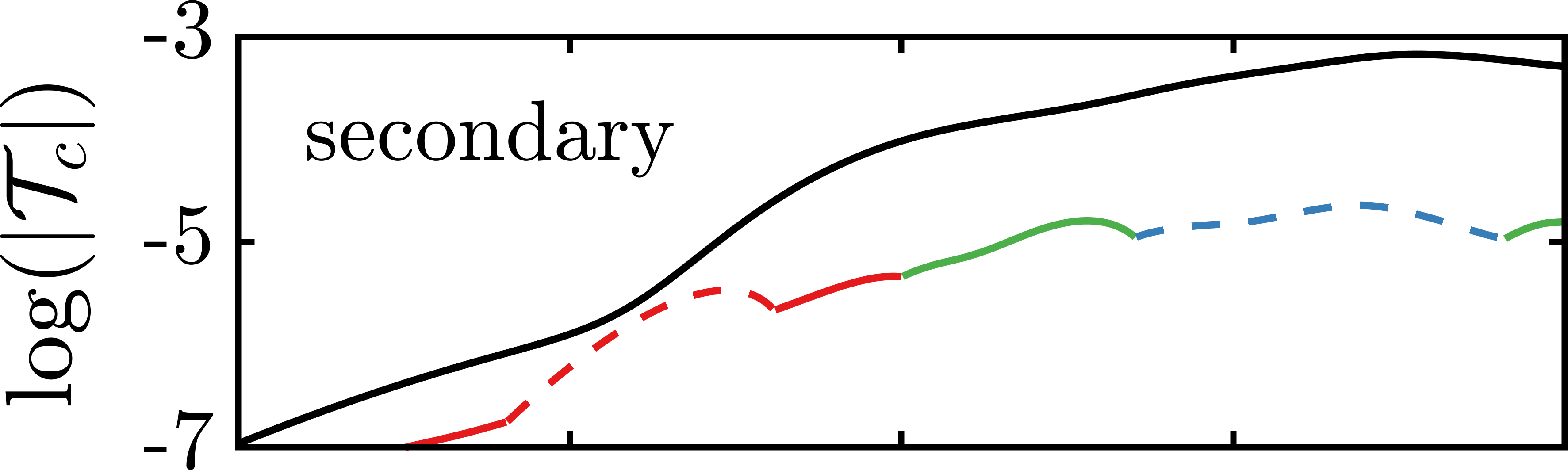}
  \end{subfigure}
  \begin{subfigure}[t]{0.60\textwidth}\flushright
  \subcaption{}
  \includegraphics[width=1\textwidth]{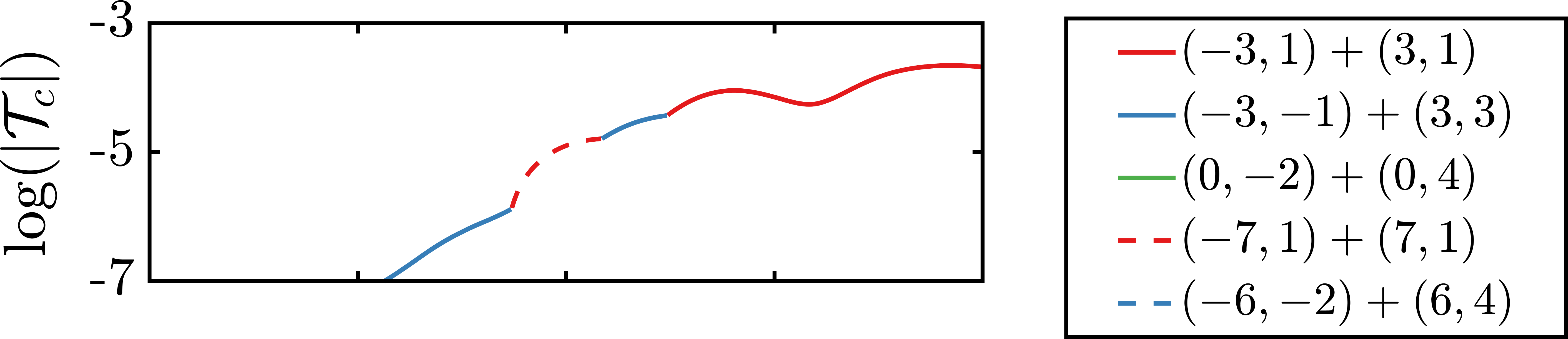}
  \end{subfigure}\vspace{-0.2cm}
  \begin{subfigure}[t]{0.39\textwidth}\flushleft
  \subcaption{}
  \includegraphics[width=0.967\textwidth]{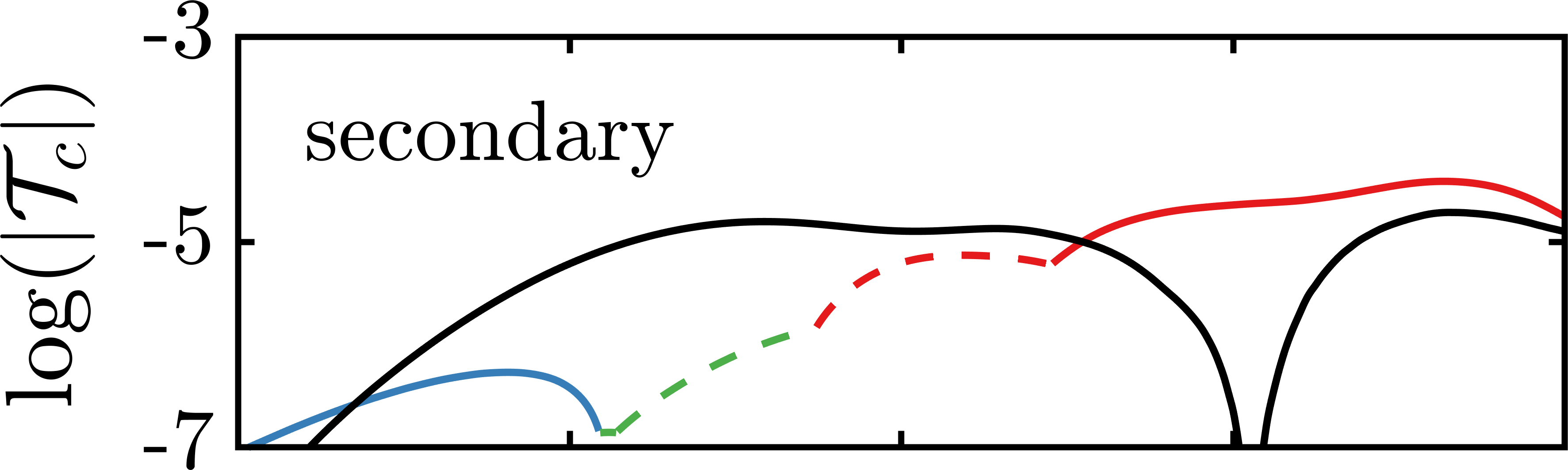}
  \end{subfigure}
  \begin{subfigure}[t]{0.60\textwidth}\flushright
  \subcaption{}
  \includegraphics[width=1\textwidth]{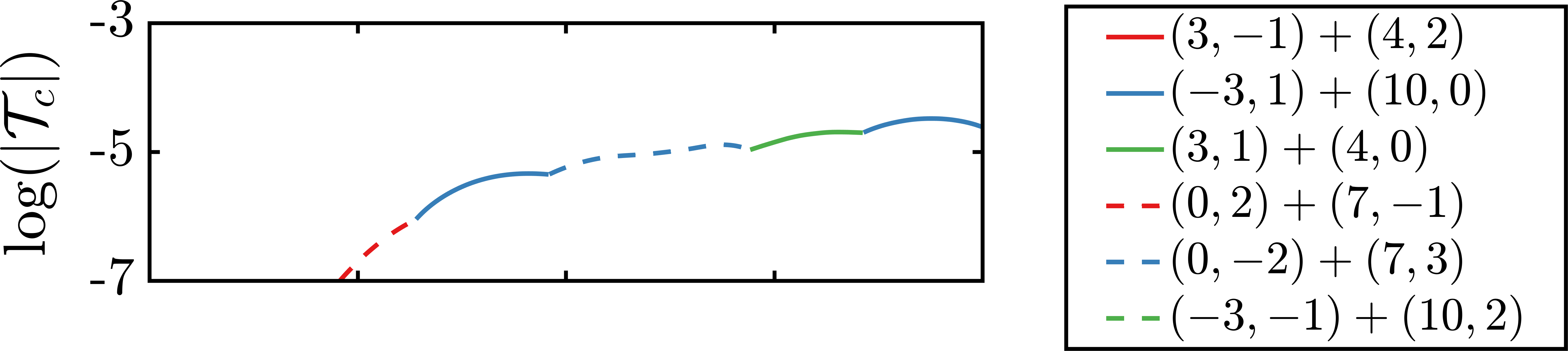}
  \end{subfigure}\vspace{-0.2cm}
  \begin{subfigure}[t]{0.39\textwidth}\flushleft
  \subcaption{}
  \includegraphics[width=0.967\textwidth]{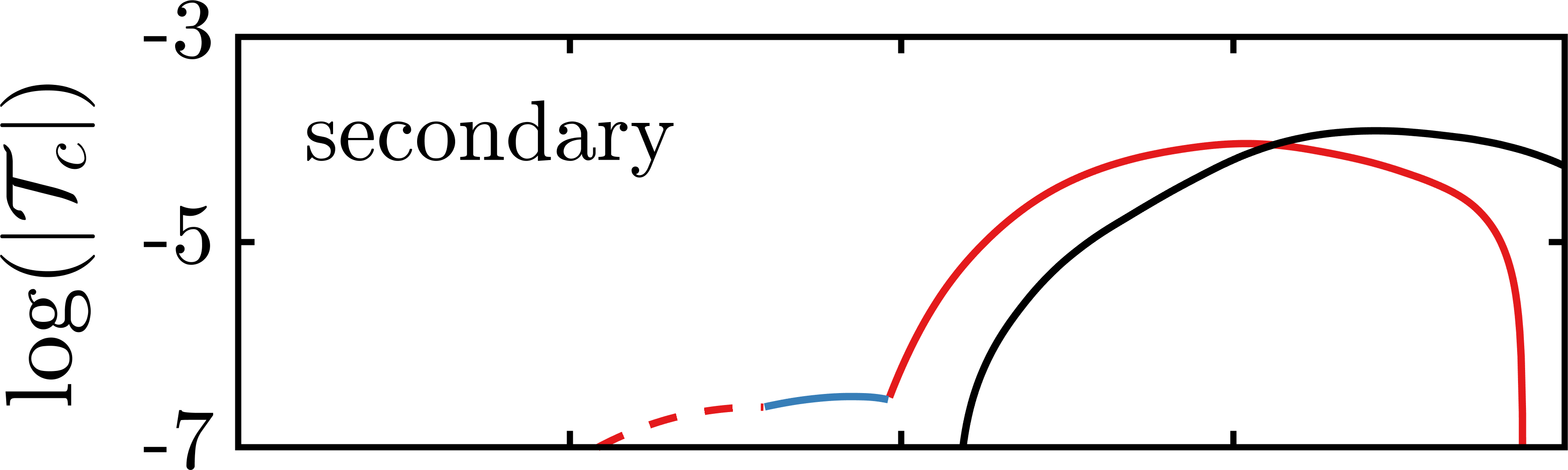}
  \end{subfigure}
  \begin{subfigure}[t]{0.60\textwidth}\flushright
  \subcaption{}
  \includegraphics[width=1\textwidth]{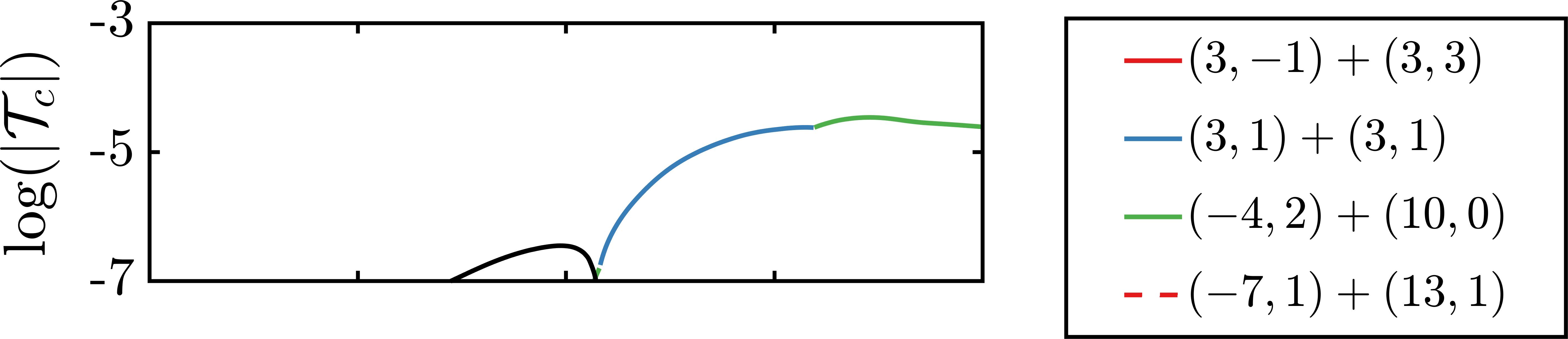}
  \end{subfigure}\vspace{-0.2cm}
  \begin{subfigure}[t]{0.39\textwidth}\flushleft
  \subcaption{}
  \includegraphics[width=1\textwidth]{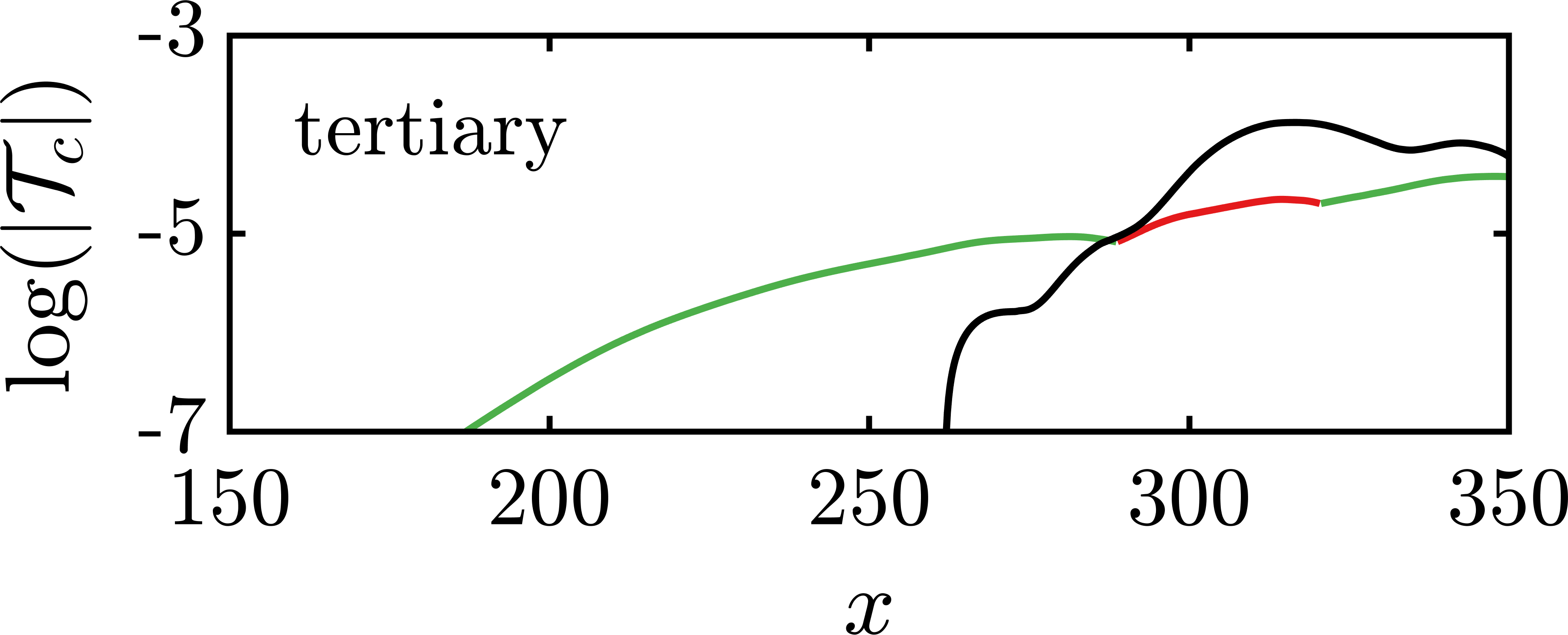}
  \end{subfigure}
  \begin{subfigure}[t]{0.60\textwidth}\flushright
  \subcaption{}
  \includegraphics[width=1\textwidth]{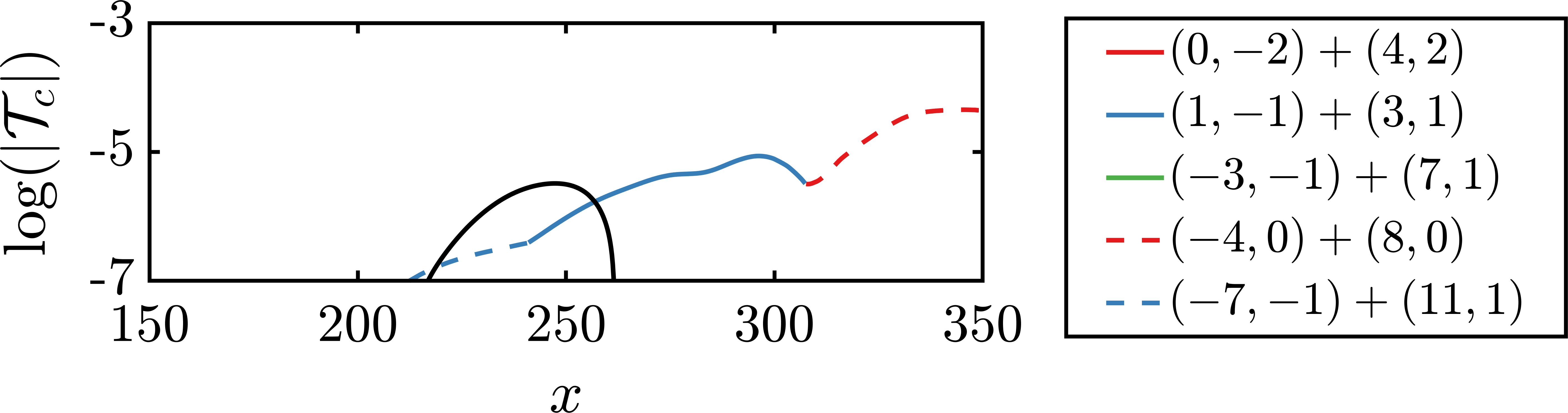}
  \end{subfigure}
  \caption{The local most significant nonlinear triad contributing to the gain (left column) and loss (right column) of the wall-normal integrated Chu's energy for selected Fourier mode at different streamwise locations: (\textit{a,b}) The primary first mode P1--$(3,1)$, (\textit{c,d}) The primary second mode P2--$(10,0)$, (\textit{e,f}) The streak mode P1D--$(0,2)$, (\textit{g,h}) The difference mode P12D--$(7,1)$, (\textit{i,j}) The harmonic mode P1S--$(6,2)$ and (\textit{k,l}) The tertiary mode $(4,0)$. The integrated linear transfer for each mode ({\color[rgb]{0 0 0}---}) is also shown for comparison.}
  \label{fig:triad}
\end{figure}

The co-existence of multiple primary and high-order modes can increase the complexity of the quadratic interaction scenario. The nonlinear triads, important in the global sense in figure \ref{fig:transfer}, may not contribute significantly to the modal growth at each streamwise station. Therefore in this subsection, for each Fourier mode detected with high energy fraction (see figure \ref{fig:spectra}), we extract the associated most significant nonlinear triads locally, since the transition is a streamwise varying process. Here, the locally dominant nonlinear triad is determined based on the maximum wall-normal integrated transfer of Chu's energy,  $\mathcal{T}_c$, as defined in $\S$ \ref{sec:energy_equation}. Collectively, figure \ref{fig:triad} provides the locally dominant triad that contributes to the gain (left panel) and loss (right panel) of the wall-normal integrated Chu’s energy for selected Fourier modes.\par
We first focus on the interaction between the two primary waves, which generates the sum mode $(13,1)$ and the difference mode P12D--$(7,1)$. As illustrated in figure \ref{fig:spectra}, the sum mode is not as evident as the difference mode P12D. Figure \ref{fig:triad}(\textit{d}) demonstrates that the sum mode only plays a leading role in the energy-loss process of the second mode wave P2 through $(-3,-1)+(13,1)\rightarrow(10,0)$. The streamwise range is the linearly unstable stage of the second mode ($x<180$) and where the flow is highly nonlinear ($x>330$).\par
In contrast, the energy transfer among the difference mode and the two primary modes is pronounced in re-distributing their energy. In figure \ref{fig:triad}(\textit{g}), the difference mode P12D is generated upstream of the transition onset ($x<220$) through the triad $(10,0)+(-3,1)\rightarrow(7,1)$. P12D grows due to both the linear and nonlinear energy gain, and the active transfer region is along the GIP (unshown and similar to figure \ref{fig:Tc_contour}). Simultaneously, figure \ref{fig:triad}(\textit{b}) shows that the primary oblique wave P1 loses energy in the triadic interaction $(-7,1)+(10,0)\rightarrow(3,1)$, while figure \ref{fig:triad}(\textit{c}) reveals that the second mode P2 gains energy through the triad $(3,-1)+(7,1)\rightarrow(10,0)$. In other words, in this streamwise range, the primary first mode P1 serves as the energy supplier among the discussed P1, P2 and P12D. Subsequently, as the modes pass $x=220$, the energy transfer direction reverses and P12D takes over the primary mode P1 to become the supplier, evidenced by figure \ref{fig:triad}(\textit{a,c,h}). It is interesting to observe the forward and reverse net energy transfer in this early stage near the transition onset point. Furthermore, downstream of the onset in figure \ref{fig:triad}(\textit{d}), the mean flow can reversely extract energy from the second mode P2. The second mode P2 is observed to extract energy from either P1 or P12D, and this behavior extends until $x\approx280$. Farther downstream in figure \ref{fig:triad}(\textit{d}), the second mode P2 starts to lose energy via the interaction $(3,1)+(7,-1)$ in between $x\approx280$ and $x\approx300$. \par
In addition to the interaction with the two primary waves, the difference mode P12D is also reported in figure \ref{fig:triad}(\textit{g}) to gain energy through $(0,2)+(7,-1)$ and $(3,-1)+(4,2)$ as it develops. Here, a noteworthy finding is the interaction of the difference mode with the streak. In $\S$ \ref{sec:budget}, the energy transfer between P1 and P1D is identified with a globally dominant role in the streaky growth. Nevertheless, the difference mode P12D can occasionally lead the streak generation through the self-interaction $(-7,1)+(7,1)\rightarrow(0,2)$ (figure \ref{fig:triad}(\textit{e})), followed by the reverse energy flow from the streak to P12D via $(0,2)+(7,-1)\rightarrow(7,1)$ (figure \ref{fig:triad}(\textit{g})). Without this local inspection, the leading role of the P12D in the streaky generation might be overlooked. We also report that the spatial region of this P1D-P12D interaction is consistent with the oblique interaction between P1 and P1D. Therefore, provided that multiple primary instabilities exist, the self-sustaining cycle of the streak can be facilitated by both the oblique first mode and the difference mode.\par
Some additional observations are summarized as follows. For the first-mode harmonic P1S $(6,2)$, the primary oblique wave $(3,1)$ plays an important role through first the self-interaction $(3,1)+(3,1)$ and then considerably by $(3,-1)+(3,3)$, as shown by figure \ref{fig:triad}(\textit{i}). The tertiary instability mode $(3,3)$ can be generated via $(3,1)+(0,2)\rightarrow(3,3)$. This result aligns well with the shape resemblance in figures \ref{fig:mode_primary}(\textit{a}) and \ref{fig:mode}(\textit{c}) that both the oblique first mode and its first harmonic are augmented along the GIP. For the tertiary mode $(4,0)$, figure \ref{fig:triad}(\textit{k}) reveals the leading supply triad as $(-3,1)+(7,-1)\rightarrow(4,0)$ throughout a broad streamwise range. As a result, the tertiary mode $(4,0)$ comprises complex physical signature, resembling multiple primary and secondary waves (see figures \ref{fig:mode}(\textit{d}) and \ref{fig:bar}). Usually, the secondary instability is recognized important in spectral broadening in literature. Nonetheless, the present data demonstrates that the triadic interaction with tertiary modes can play a leading role as well even in the early transitional stage.\par
\begin{figure}
  \centering
  \includegraphics[width=0.5\textwidth]{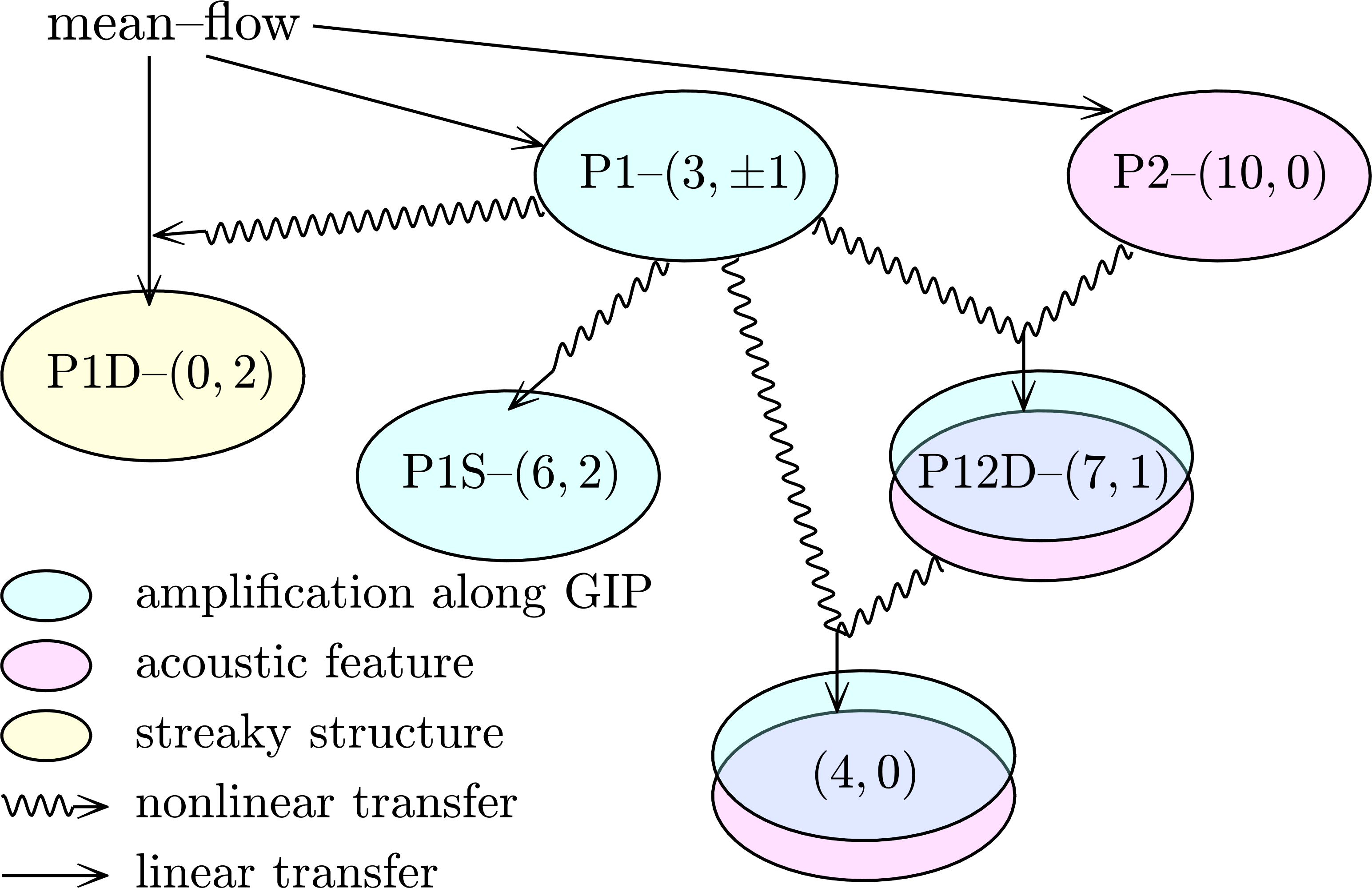}
  \caption{Inheritance of physical properties from primary to higher-order waves during their generation stages.}
  \label{fig:physical_property}
\end{figure}
By quantifying linear and nonlinear energy transfer associated with concerned Fourier modes, the inheritance of physical properties from primary to higher-order disturbances during their initial growth stages, as observed in $\S$ \ref{sec:redistribution}, may be conceivable. The conceptual conclusion is summarized by figure \ref{fig:physical_property}. On one hand, the oscillatory modes are mostly affected by the source they originate from. The oscillatory `child modes', e.g. the difference mode P12D, the harmonic mode P1S and the tertiary mode, share some signatures with either P1 or P2 (parent modes). These signatures are inherited from parent modes to child modes. For example, P1S is amplified along GIP similarly to the vortical-like first mode P1; P12D comprises both the vortical-wave feature with P1 and the acoustic feature with P2. On the other hand, the steady streaky mode exhibits unique physical nature. Despite the fact that the streaky mode P1D is generated by P1, every oscillatory mode can affect P1D through self oblique interaction, e.g. the aforementioned $(-7,1)+(7,1)\rightarrow(0,2)$. These nonlinear forcings, however, are not directly responsible for the growth of P1D, but serve as the initial seed, which is then convected and linearly amplified by the mean flow. This dominant role of linear growth may be used to understand why the property of the `parent mode' fades out in the `child' streaky mode after the initial seed. Thus, the steady streaky mode may not be merely determined by a single mode. Instead, it can be sustainably supported by the mean-flow effect as well as various nonlinear interactions. For now, the scope of the conclusion is limited to the resulting interactions generated from two primary instability waves.\par

\subsection{Attenuation and secondary growth of acoustic signature for second mode}\label{sec:acoustic}
\begin{figure}
  \centering
  \includegraphics[width=1.0\textwidth]{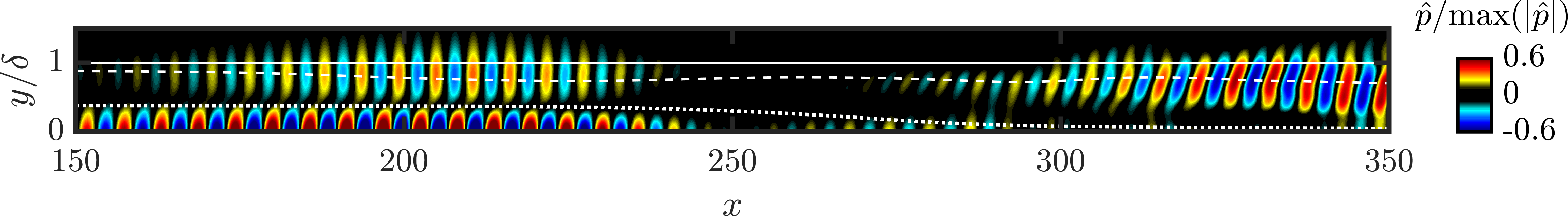}
  \caption{Spatial distribution of the primary second mode visualized by the real part of the pressure component. The white solid, dashed and dotted lines in the plot correspond with figure \ref{fig:mode_primary}.}
  \label{fig:pprime}
\end{figure}

Downstream of the transition onset, the Mack second mode P2 reports a sudden decrease in the Chu's energy (see figure \ref{fig:energy}). Figure \ref{fig:Tc_contour} proves it as a consequence of combined linear and nonlinear mechanisms. The saturation of the second mode can be also detected by wall pressure measurement in experimental studies \citep{zhang_2013_hypersonic}. This region with a low acoustic level is normally called the `quiet zone'. To illustrate this phenomenon, figure \ref{fig:pprime} displays the pressure component of the second mode wave P2. The quiet zone begins at $x\approx240$, where the trapped acoustic wave is observed to lose its signature. In this section, we focus on the mechanism behind by investigating the linear and nonlinear transfer of acoustic energy.\par
\begin{figure}
  \centering
  \begin{subfigure}[t]{0.39\textwidth}\flushleft
  \subcaption{}
  \includegraphics[width=1\textwidth]{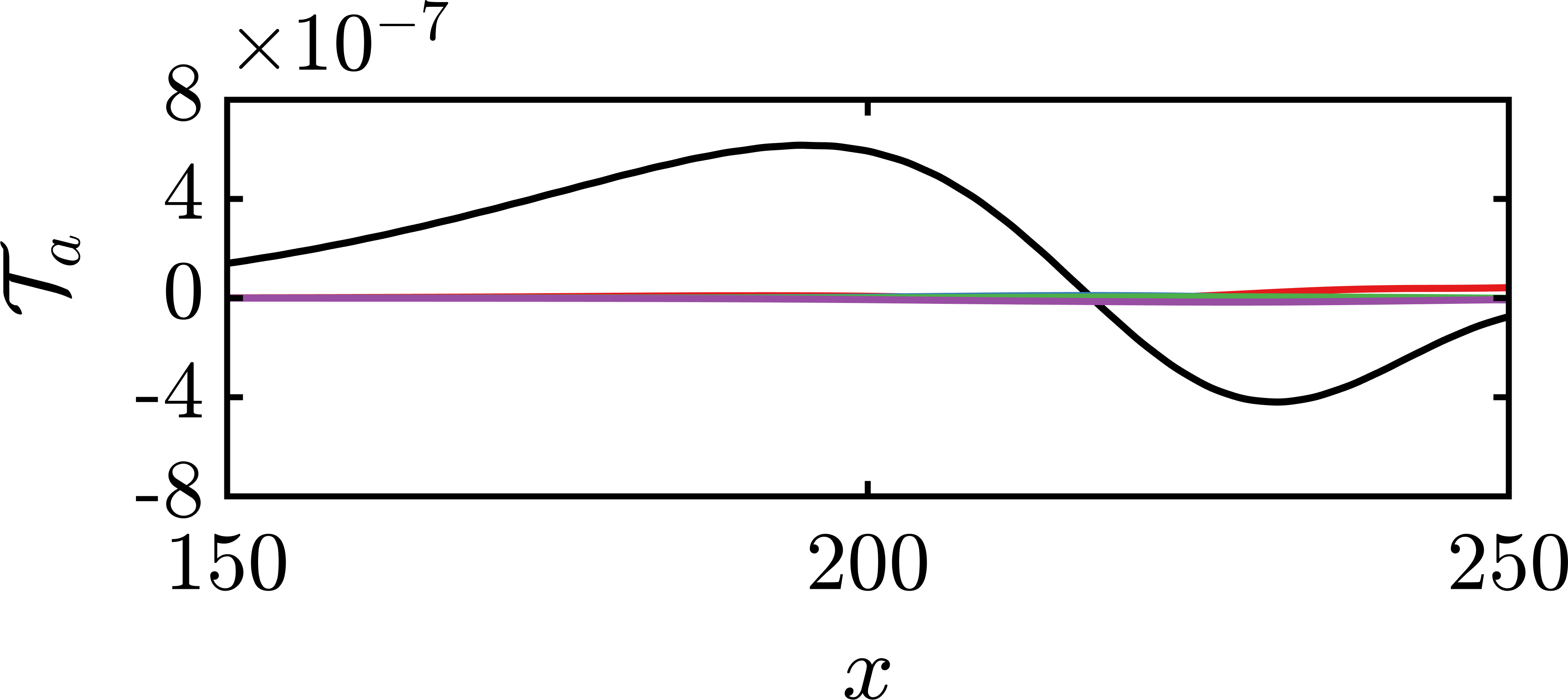}
  \end{subfigure}
  \begin{subfigure}[t]{0.60\textwidth}\flushleft
  \subcaption{}
  \includegraphics[width=1\textwidth]{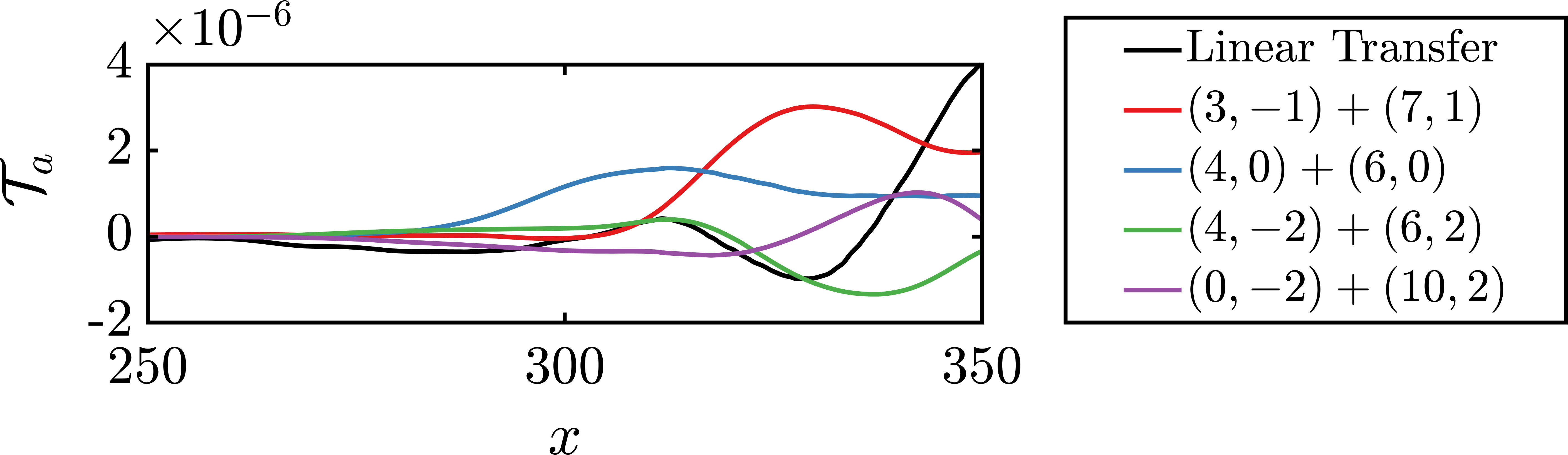}
  \end{subfigure}
  \caption{Integrated linear and nonlinear transfer to the acoustic energy as a function of streamwise location for the primary second mode P2--$(10,0)$.}
  \label{fig:acoustic}
\end{figure}

Figure \ref{fig:acoustic} shows the wall-normal integrated transfer term of the acoustic energy, $\mathcal{T}_a$, for mode P2. The four leading nonlinear triads, determined based on the same rule of figure \ref{fig:transfer}, are considered. Figure \ref{fig:acoustic}(\textit{a}) reveals that the linear transfer of acoustic energy to P2 reaches maximum upstream of $x=200$, followed by a rapid decay. In $220<x<250$ , the linear transfer is negative enough to eliminate the acoustic feature of P2, albeit acoustic energy can be increased via $(3,-1)+(7,1)\rightarrow(10,0)$. Downstream of $x=250$, the linear-transfer term maintains a negative value for a distance, as depicted by figure \ref{fig:acoustic}(\textit{b}). However, multiple nonlinear triads start to grow and contribute acoustic energy to P2. In particular, the triad $(3,-1)+(7,1)\rightarrow(10,0)$ is noticed with a significant growth in acoustic energy transfer. All these intense nonlinear interactions lead to the secondary increase in the acoustic energy of P2. Subsequently, the mean-flow supply of acoustic energy also turns positive owing to the growing nonlinearity. Eventually, the quiet zone is broken with a strong signal along the GIP, as shown by figure \ref{fig:pprime}. The positive nonlinear contribution is ahead of the linear counterpart. We may infer that the nonlinear net transfer is more crucial in this secondary growth. \par
Notably, according to figure \ref{fig:pprime}, the linear acoustic signature below the sonic line vanishes in $x>300$. This observation indicates that the re-emergence of this high-frequency pressure fluctuation is not due to a Mack-mode mechanism. As suggested by figure \ref{fig:Tc_contour}(\textit{b}), the net nonlinear energy transfer to P2 is remarkable along the GIP. Meanwhile, other Fourier modes are pronounced along the GIP. Thus, the secondary growth of the P2 pressure signal is attributable to the nonlinear interaction in the outer layer. This evident signal is away from the wall, thus not able to be captured by experimental PCB sensors. \par
\begin{figure}
  \centering
  \begin{subfigure}[t]{0.49\textwidth}\flushleft
  \subcaption{}
  \includegraphics[width=0.97\textwidth]{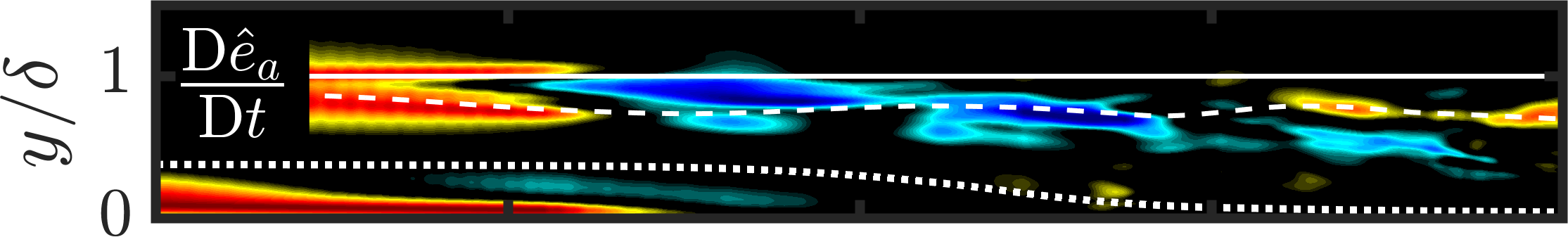}
  \end{subfigure}\vspace{-0cm}
  \begin{subfigure}[t]{0.49\textwidth}\flushleft
  \subcaption{}
  \includegraphics[width=0.97\textwidth]{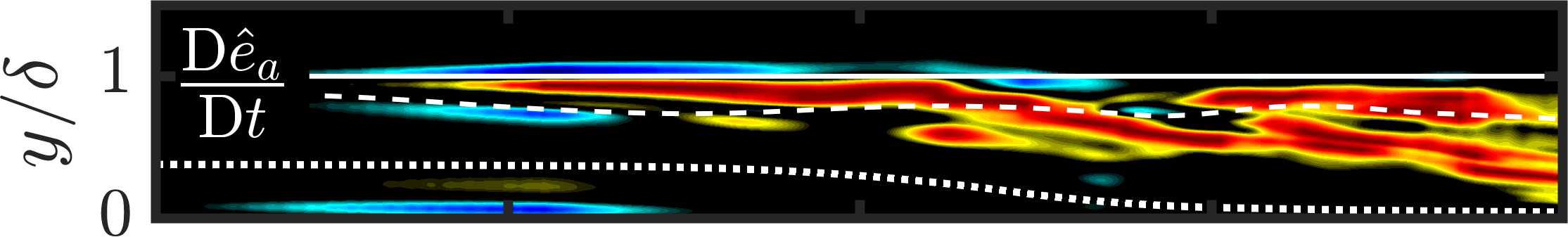}
  \end{subfigure}\vspace{-0cm}
  \begin{subfigure}[t]{0.49\textwidth}\flushleft
  \subcaption{}
  \includegraphics[width=1.0\textwidth]{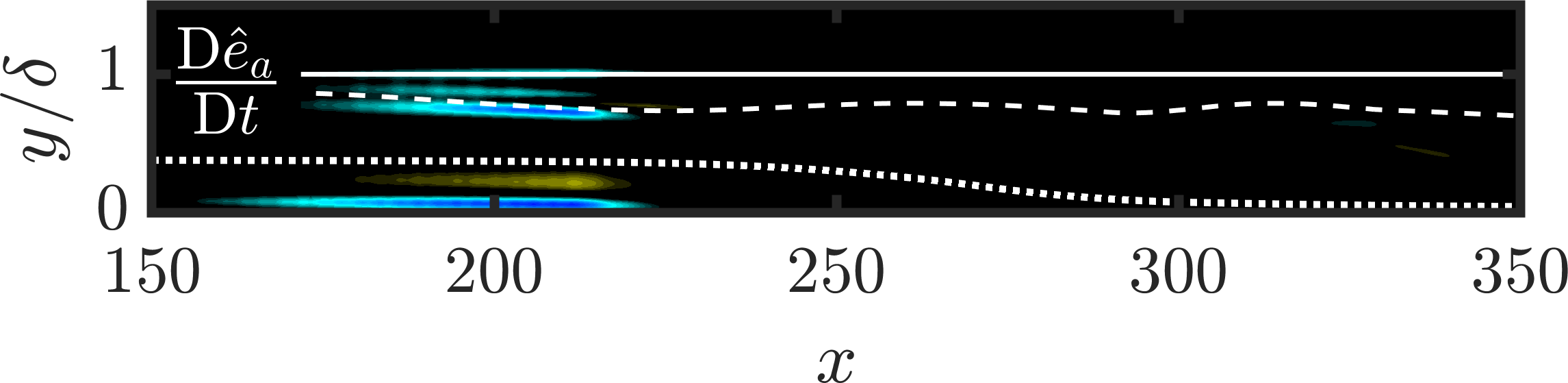}
  \end{subfigure}\vspace{-0cm}
  \begin{subfigure}[t]{0.49\textwidth}\flushleft
  \subcaption{}
  \includegraphics[width=1.0\textwidth]{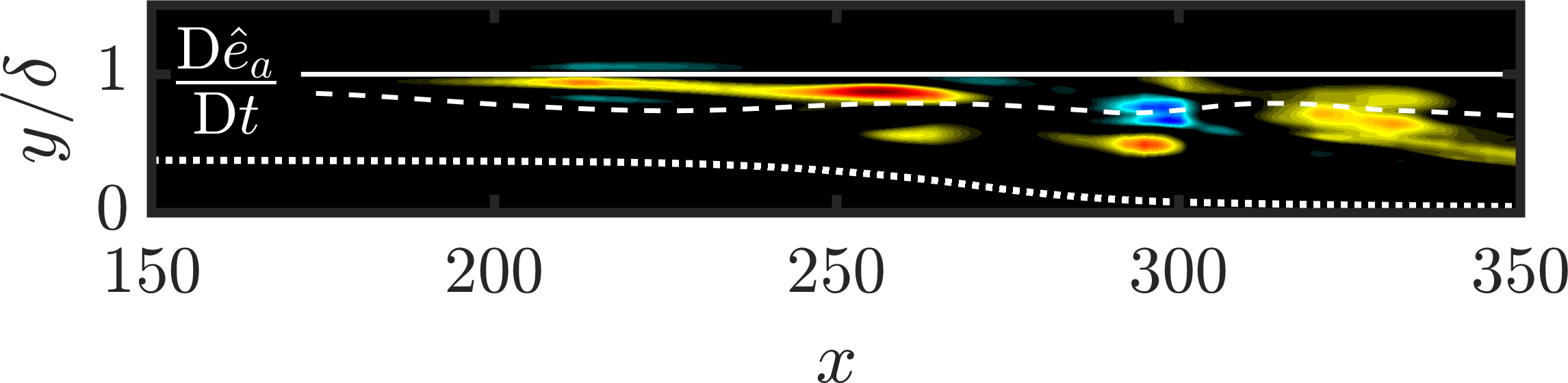}
  \end{subfigure}\vspace{-0cm}
  \caption{Contours of normalized (\textit{a}) linear transfer, (\textit{b}) net nonlinear transfer, (\textit{c}) nonlinear transfer through $(-10,0)+(20,0)$ and (\textit{d}) nonlinear transfer through $(3,1)+(7,-1)$ to the acoustic energy of the primary second mode P2--$(10,0)$. The energy transfers are normalized by the maximum value between the linear and net nonlinear terms at each streamwise location. The white solid, dashed and dotted lines in the plots correspond with figure \ref{fig:mode}. Contour bar ({\color[rgb]{0.8 0 0}$\blacksquare$}{\color[rgb]{0 0 0}$\blacksquare$}{\color[rgb]{0 0 0.8}$\blacksquare$}) ranges between $\pm1$.}
  \label{fig:Ta_contour}
\end{figure}
We further check the spatial locations with intense acoustic energy transfer. Figure \ref{fig:Ta_contour} gives the normalized linear and nonlinear transfer contour. The net nonlinear transfer and the transfers via crucial processes $(-10,0)+(20,0)$ as well as $(3,1)+(-7,1)$ are shown. Clearly, the most pronounced linear acoustic energy transfer is initially located below the sonic line (see figure \ref{fig:Ta_contour}(\textit{a})). In this early stage, the harmonic triad $(10,0)+(10,0)\rightarrow(20,0)$ extracts the acoustic energy from $(10,0)$ around the GIP and the wall (see figure \ref{fig:Ta_contour}(\textit{b})). The final attenuation of the acoustic signature of this primary wave is still associated with the negative linear transfer in $220<x<250$, albeit the net nonlinear transfer supplies acoustic energy. Downstream of $x=300$, the secondary growth of acoustic energy is more attributed to $(3,1)+(7,-1)\rightarrow(10,0)$, interacting outside the sonic line. Hence, in the nonlinear stage, the associated acoustic feature of P2 differs generally from that of the linear stage.\par

\section{Discussions and conclusions}\label{sec:conclusion}
Dynamics involving the mean-flow and instability waves of multiple spatio/temporal scales are crucial in the laminar-turbulent transition of boundary layers. The realistic transition is subject to broadband forcings under flight or wind-tunnel conditions. The resulting co-existence of multiple primary and high-order modes highlights the complexity and importance of nonlinearity. In this research, we utilize the DNS database of a Mach 6 boundary layer flow, which is initiated by optimal forcing terms to trigger multiple primary instabilities. Efforts are made to extract and quantify the complex mechanism of interactions among multiple primary and high-order modes. To elucidate the role of the nonlinear interaction between two optimal primary waves, we do not employ a broadband forcing.\par
Mode decomposition first reveals that, near the transition onset, the primary waves start to deviate from the linear solution due to nonlinearity. The nonlinear effect results in the sustaining growth of the first primary instability in conjunction with the saturation and subsequent secondary growth of the second primary instability. Meanwhile, the energy is re-distributed into multiple scales, causing the broadening of spectra. The generated higher-order waves may inherit the physical properties of associated primary instability waves due to the role of nonlinear forcings.\par
The nonlinear dynamics are then analyzed based on a linear input--output formulation, which describes how different forcings contribute to the response patterns. In the generation stage of each higher-order wave, a specific leading nonlinear triad is identified to contribute to each forcing. These higher-order waves manifest solely in response to the identified dominant forcing during their generation. At the moderate and late transitional stages, the MFD-associated background forcing is prominent for all the waves. At this stage, the forcings are, however, not equally transferred to the response via the resolvent operator. In other words, the base-flow-associated resolvent operator exerts different levels of `leverage' to transfer different forcings. For the first mode, the reconstructed response to the 20 leading triadic forcings can well represent the perturbation pattern in a relatively early stage. Furthermore, the secondary growth of the first mode is related to the MFD- and cubic-associated forcings, and can be suppressed by the summed triadic forcing.\par
\begin{figure}
  \centering
  \includegraphics[width=1\textwidth]{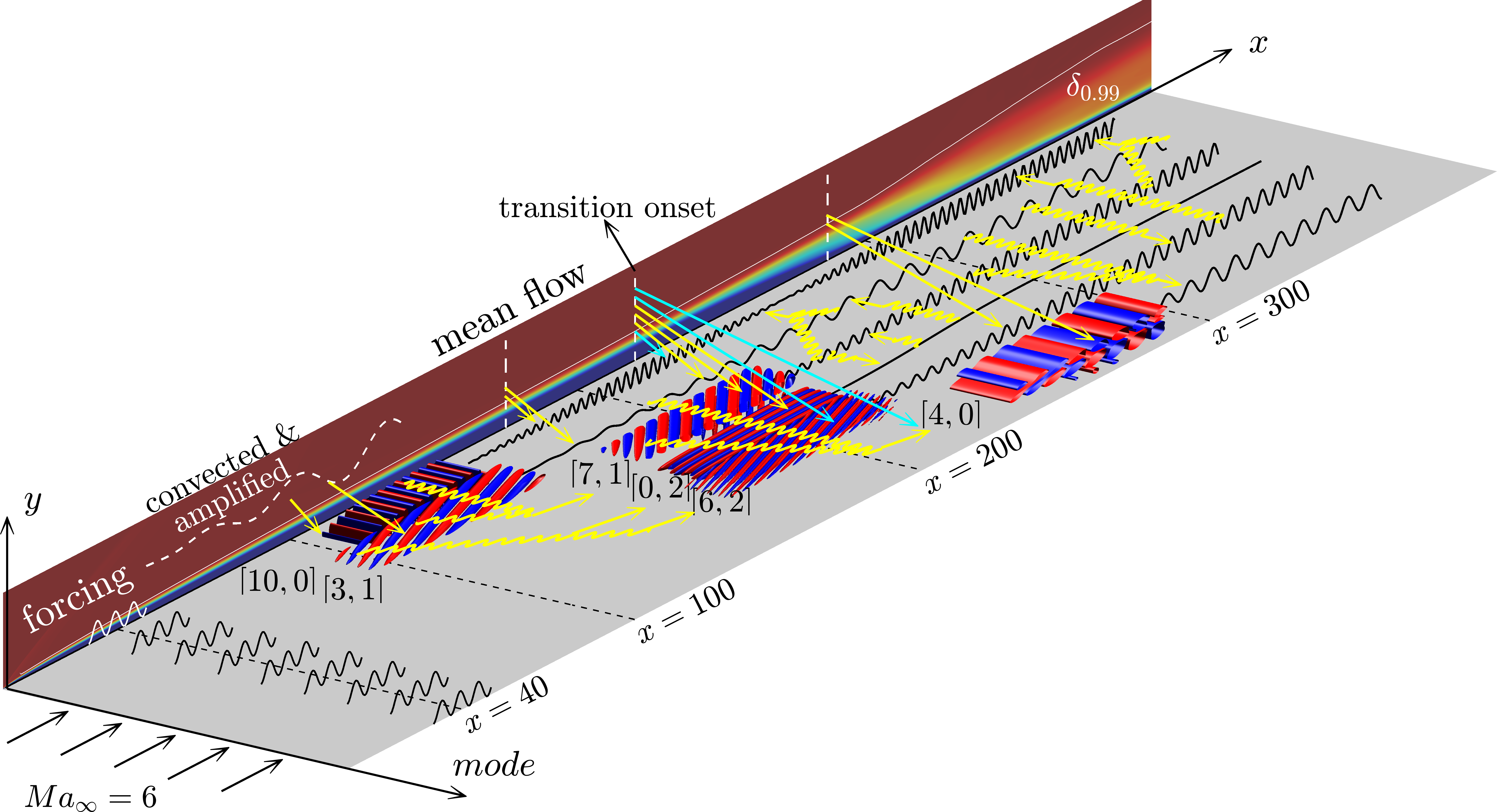}
  \caption{Energy transfer between the mean flow, the primary modes and important high-order Fourier modes. The light blue line represents the loss of energy of corresponding mode through linear process, and the yellow line denotes energy gain of corresponding mode through linear transfer or a nonlinear triadic interaction.}
  \label{fig:summary}
\end{figure}

By quantifying the linear and nonlinear effects via the spectral energy equation, some important energy pathways among different waves are depicted in figure \ref{fig:summary}. Initiated by the upstream forcing, the optimal oblique first mode and Mack second mode are triggered, and grow continuously through the amplification of the base flow. As these primary waves are amplified to certain magnitude, nonlinear effects give rise to the generation of secondary waves, including the difference mode, the harmonic mode and the stationary streak. In the initial growth stage, these secondary waves also interact with the mean flow. The streak and the difference mode gain energy, as illustrated by the yellow arrows pointed to modes $(7,1)$ and $(0,2)$. By contrast, the first-mode harmonic loses energy through this linear transfer mechanism at the transition onset, as shown by the blue arrow points from the mean flow to $(6,2)$. Downstream of the transition onset, the increasing deviation of the mean flow from the laminar solution starts to suppress the growth of the Mack second mode, as shown by the blue arrow points from the mean flow to $(10,0)$. This causes the `quiet zone' of this primary wave, which has also been captured by wind-tunnel measurements. Meanwhile, the continuous growth of existing primary and secondary waves further results in more intense nonlinear triadic interactions. They facilitate the growth of higher-order waves, e.g. the tertiary wave $(4,0)$. The initial energy supplier for $(4,0)$ is from nonlinear interactions instead of the mean flow. Next, the mean flow begins to supply energy to $(4,0)$, as marked by the yellow arrow to this mode. Further downstream at $x>300$, the growing nonlinearity of the flow leads to complex energy transfer pathways among all the aforementioned waves. Notably, the secondary growth of the Mack second mode is a consequence of nonlinear energy transfer through multiple triads. In this stage, the mean flow supports the growth of most waves through linear transfer. The energy is then re-distributed into more spatio-temporal scales through nonlinear interactions as the flow develops, which eventually leads to the breakdown into turbulence.\par
Compared to the controlled transition with one primary wave, the co-existence of two primary waves in our DNS also leads to an enriched growth process of streaks. The emergence of the steady streak is attributed to the first-mode oblique interaction as well as the secondary difference wave. Then the streak grows subject to the linear mean-flow effect, and meanwhile, feeds energy back to the first mode and the difference mode. To conclude, the self-sustaining cycle of the streak involves multiple important waves under the considered circumstance.\par
The higher-order waves, such as the difference mode P1D, the harmonic mode P1S and a tertiary wave, are also observed to inherit physical signature from the associated lower-order wave. For instance, both the first mode and its harmonic are amplified along the GIP due to their close interaction. The difference mode P12D owns acoustic signature owing to the impact of the primary second mode. In addition, the breakdown and secondary growth of the acoustic signature are also investigated for the second mode. In detail, the mean-flow variation downstream of the transition onset leads to a linear drain of acoustic energy beneath the sonic line. This effect is the main cause of the breakdown of trapped acoustic waves. The secondary growth of the acoustic energy in this mode is attributed mostly to the transfer via nonlinear forcings. These transfer regions are located outside the sonic line, and thus the associated physical nature is  different from that of the linear second mode.\par
The main contributions of this research can be summarized from two aspects. Methodologically, a generic framework is constructed (i) for input--output analysis of nonlinear systems with periodic perturbations, and (ii) to quantify the transfer of energy components under different definitions of norms as well as among different frequencies and wavenumbers. Physically, energy transfer pathways and the leading nonlinear interactions are identified via this framework. Several findings are reported for the mutual interactions between two primary instabilities in hypersonic transitional boundary layers. For more complicated natural forcings, the methodological framework may also reveal leading mechanisms upon valid mode decomposition of data.\par

\vspace{5mm}
\noindent\textbf{Funding.} This work is supported by the Hong Kong Research Grants Council (no. 15203724). 

\vspace{5mm}
\noindent\textbf{Acknowledgements.} The open-source solver for DNS provided by Prof. Xinliang Li and the resolvent-analysis solver by Prof. Jiaao Hao are sincerely appreciated.

\vspace{5mm}
\noindent{\bf Declaration of Interests.} The authors report no conflict of interest.

\bibliographystyle{jfm}
\bibliography{jfm}

\end{document}